\documentclass[10pt]{article}
\pdfoutput=1

\usepackage{preamb}

\pgfdeclarelayer{back}
\pgfdeclarelayer{front}
\pgfsetlayers{back,main,front}

\tikzset{->-/.style={decoration={
			markings,
			mark=at position #1 with {\arrow{Triangle[width=3.5pt, length=3.5pt]}}},postaction={decorate}}}

\tikzset{-<-/.style={decoration={
			markings,
			mark=at position #1 with {\arrowreversed{Triangle[width=3.5pt, length=3.5pt]}}},postaction={decorate}}}

\tikzset{
    every picture/.append style={
        line join=round,
        line width = 0.7pt,
        line cap=round
    }
}

\tikzset{mpoint/.style={ transform shape, sloped, 
minimum width=20pt, minimum height=4pt, inner sep=0pt, ultra thin, font=\tiny}}

\tikzset{
    clip even odd rule/.code={\pgfseteorule},
    invclip/.style={clip,insert path=[clip even odd rule]{
    [reset cm](-\maxdimen,-\maxdimen)rectangle(\maxdimen,\maxdimen)
    }}} 

\pgfdeclaredecoration{simple line}{initial}{
    \state{initial}[width=\pgfdecoratedpathlength-1sp]{\pgfmoveto{\pgfpointorigin}}
    \state{final}{\pgflineto{\pgfpointorigin}}
}
\tikzset{
    shift left/.style={decorate,decoration={simple line,raise=#1}},
    shift right/.style={decorate,decoration={simple line,raise=-1*#1}},
}
    
\newcommand\clipIntersection[1]{
    (#1.north west) -- (#1.north east) -- (#1.south east) -- (#1.south west) -- (#1.north west)
}

\tikzstyle{dot}=[dash pattern= on 0.6pt off 2.1pt]
\tikzstyle{obj}=[line width = 0.7pt, line join=round, line cap=round, shorten >= 1pt, shorten <= 1pt]
\tikzstyle{edge}=[line width = 0.7pt, line join=round, line cap=round]
\tikzstyle{fat}=[line width = 1.1pt, line join=round, line cap=round]
\tikzstyle{speObj}=[line width = 0.7pt, line join=round, line cap=round, shorten >= 1pt, shorten <= 1pt, color=BrickRed]
\tikzstyle{mod}=[color=BlueViolet,line width = 0.7pt, line join=round, line cap=round]
\tikzstyle{dotmod}=[dash pattern= on 0.6pt off 2.1pt,color=BlueViolet,line width = 0.7pt, line join=round, line cap=round]
\tikzstyle{mor}=[draw=none, fill=gray!22]
\tikzstyle{wig}=[snake=zigzag, segment amplitude=0.7pt, segment length=3.5pt, rounded corners=0.6pt]

\newcommand\clipAroundNode[2]{
    \begin{pgfscope}
        \begin{scope}[overlay]
            \path[invclip] \clipIntersection{#1};#2
        \end{scope}
    \end{pgfscope}
}

\newcommand\clipAroundTwoNodes[3]{
    \begin{pgfscope}
        \begin{scope}[overlay]
            \path[invclip] \clipIntersection{#1} \clipIntersection{#2};#3
        \end{scope}
    \end{pgfscope}
}

\newcommand\clipAroundFourNodes[5]{
    \begin{pgfscope}
        \begin{scope}[overlay]
            \path[invclip] \clipIntersection{#1} \clipIntersection{#2} \clipIntersection{#3} \clipIntersection{#4};#5
        \end{scope}
    \end{pgfscope}
}

\newcommand{\middleCoo}[6]{
    \path (#1) --coordinate[pos=0.5](#3) (#2);
    \path (#1) --coordinate[pos={0.5-#6}](#4) (#2);
    \path (#1) --coordinate[pos={0.5+#6}](#5) (#2);
}

\newcommand{\barycentre}[8]{
    \middleCoo{#1}{#2}{A}{}{}{#8};
    \middleCoo{#2}{#3}{B}{}{}{#8};
    \middleCoo{#1}{#3}{C}{}{}{#8};
    \path[name path=PA#3] (A) -- (#3);
    \path[name path=PB#1] (B) -- (#1);
    \path[name path=PC#2] (C) -- (#2);
    \path [name intersections={of=PA#3 and PB#1,by={#4}}];
    \path (#1) --coordinate[pos=0.84](#5) (#4);
    \path (#2) --coordinate[pos=0.84](#6) (#4);
    \path (#3) --coordinate[pos=0.84](#7) (#4);
}

\newcommand{\barycentreSq}[9]{
    \middleCoo{#1}{#2}{A1}{A2}{A3}{\sp};
    \middleCoo{#3}{#4}{B1}{B2}{B3}{\sp};
    \middleCoo{#1}{#3}{C1}{C2}{C3}{\sp};
    \middleCoo{#2}{#4}{D1}{D2}{D3}{\sp};
    \middleCoo{A1}{B1}{#5}{}{}{\sp};
    \path[name path=PA2B2] (A2) -- (B2);
    \path[name path=PA3B3] (A3) -- (B3);
    \path[name path=PC2D2] (C2) -- (D2);
    \path[name path=PC3D3] (C3) -- (D3);       
    \path [name intersections={of=PA2B2 and PC2D2,by={#6}}];
    \path [name intersections={of=PA2B2 and PC3D3,by={#8}}];
    \path [name intersections={of=PA3B3 and PC2D2,by={#7}}];
    \path [name intersections={of=PA3B3 and PC3D3,by={#9}}];
}

\newcommand{\intersec}[5]{
    \path[name path=P12] (#1) -- (#2);
    \path[name path=P34] (#3) -- (#4);
    \path [name intersections={of=P12 and P34,by={#5}}];
}

\newcommand{\mor}[2]{
    \node[mor, circle, inner sep=3.3pt] at (#1) {};
    \node at (#1) {${\sss #2}$};
}

\newcommand{\centerarc}[5]{
    \draw[#1] ($(#2)+({#5*cos(#3)},{#5*sin(#3)})$) arc (#3:#4:#5);
}

\newcommand{\roundArrow}[4]{
    \path[name path=P12A, shift #4=3pt] (#1) --++ ($1.2*(#2)-1.2*(#1)$);
    \path[name path=P12B, shift #4=8pt] (#1) --++ ($1.2*(#2)-1.2*(#1)$);
    \path[name path=P23A, shift #4=-3pt] (#3) --++ ($1.2*(#2)-1.2*(#3)$);
    \path[name path=P23B, shift #4=-8pt] (#3) --++ ($1.2*(#2)-1.2*(#3)$);
    \path [name intersections={of=P12A and P23B,by={p1}}];
    \path [name intersections={of=P12A and P23A,by={p2}}];
    \path [name intersections={of=P12B and P23A,by={p3}}];
    \draw[rounded corners=4pt, line width=.5pt, -{Triangle[scale=.5]}] (p1) -- (p2) -- (p3);
}

\newcommand{\monoidalAssociator}[1]{
	\begin{tikzpicture}[scale=0.6,baseline={([yshift=-1ex]current bounding box.center)}]
	    \def\l{1.5};
        \ifthenelse{#1 = 1}{
            \draw[obj] (0,0) -- (0,1);
            \draw[obj] (0,1) -- (-\l,3);
            \draw[obj] (0,1) -- (\l,3);
            \draw[obj] (-\l/2,2) -- (0,3);
            \node[mor, circle, inner sep=3.3pt] at (-\l/2,2) {};
            \node at (-\l/2,2) {${\sss i}$};
            \node[mor, circle, inner sep=3.3pt] at (0,1) {};
            \node at (0,1) {${\sss j}$};
            \node at (-0.65,1.3) {${\sss X_5}$};
        }{}
        \ifthenelse{#1 = 2}{
            \draw[obj] (0,0) -- (0,1);
            \draw[obj] (0,1) -- (-\l,3);
            \draw[obj] (0,1) -- (\l,3);
            \draw[obj] (\l/2,2) -- (0,3);
            \node[mor, circle, inner sep=3.3pt] at (0,1) {};
            \node at (0,1) {${\sss l}$};
            \node[mor, circle, inner sep=3.3pt] at (\l/2,2) {};
            \node at (\l/2,2) {${\sss k}$};
            \node at (0.7,1.3) {${\sss X_6}$};
        }{}
        \node[above] at (-\l,3) {${\sss X_1}$};
        \node[above] at (0,3) {${\sss X_2}$};
        \node[above] at (\l,3) {${\sss X_3}$};
        \node[below] at (0,0) {${\sss X_4}$};
	\end{tikzpicture}
}

\newcommand{\innerProduct}[1]{
    \begin{tikzpicture}[scale=0.6,baseline={([yshift=-1ex]current bounding box.center)}]
        \def\l{1.5};
        \ifthenelse{#1 = 1}{
            \draw[obj] (0,0) -- (0,1);
            \draw[obj] (0,2) -- (0,3);
            \draw[rounded corners=1pt] (0,1) -- (-\l/4,1.5) -- (0,2);
            \draw[rounded corners=1pt] (0,1) -- (\l/4,1.5) -- (0,2);
            \node[mor, circle, inner sep=3.3pt] at (0,1) {};
            \node at (0,1) {${\sss i}$};
            \node[mor, circle, inner sep=3.3pt] at (0,2) {};
            \node at (0,2) {${\sss j}$};
            \node[above, yshift=-1pt] at (0,3) {${\sss X_3'}$};
            \node[below] at (0,0) {${\sss X_3}$};
            \node[left, xshift=1pt] at (-\l/4,1.5) {${\sss X_1}$};
            \node[right, xshift=-1pt] at (\l/4,1.5) {${\sss X_2}$};
        }{}
        \ifthenelse{#1 = 2}{
            \draw[obj] (0,0) -- (0,3);
            \node[below] at (0,0) {${\sss X_3}$};
            \node[above, yshift=-1pt] at (0,3) {${\phantom{\sss X_3'}}$};
        }{}
        \ifthenelse{#1 = 3}{
            \draw[obj] (0,1) to[out=-90, in=-90, looseness=2] (1.5,1.5) to[out=90, in=90, looseness=2] (0,2);
            \draw[rounded corners=1pt] (0,1) -- (-\l/4,1.5) -- (0,2);
            \draw[rounded corners=1pt] (0,1) -- (\l/4,1.5) -- (0,2);
            \node[mor, circle, inner sep=3.3pt] at (0,1) {};
            \node at (0,1) {${\sss i}$};
            \node[mor, circle, inner sep=3.3pt] at (0,2) {};
            \node at (0,2) {${\sss j}$};
            \node[] at (.7,3.1) {${\sss X_3}$};
            \node[] at (.7,-.1) {${\sss X_3}$};
            \node[left, xshift=1pt] at (-\l/4,1.5) {${\sss X_1}$};
            \node[right, xshift=-1pt] at (\l/4,1.5) {${\sss X_2}$};
        }{}
    \end{tikzpicture}
}

\newcommand{\leftModuleAssociator}[1]{
    \begin{tikzpicture}[scale=0.6,baseline={([yshift=-1ex]current bounding box.center)}]
        \ifthenelse{#1 = 1}{
            \draw[mod] (0,0) -- (0,1);
            \draw[mod, dot] (0,1) -- (0,3);
            \draw[speObj] (-1,3) -- (0,2);
            \draw[speObj] (-2,3) -- (0,1);
            \node[mor, circle, inner sep=3.3pt] at (0,2) {};
            \node at (0,2) {${\sss v}$};
            \node[mor, circle, inner sep=3.3pt] at (0,1) {};
            \node at (0,1) {${\sss m_1}$};
        }{}
        \ifthenelse{#1 = 2}{
            \draw[mod, dot] (0,1) -- (0,3);
            \draw[mod] (0,0) -- (0,1);
            \draw[speObj, name path=P1] (-2,3) -- (0,1);
            \path[name path=P2] (-1,3) -- (-1,0);
            \path [name intersections={of=P1 and P2,by={sp}}];
            \draw[speObj] (-1,3) -- (sp);
            \node[mor, circle, inner sep=3.3pt] at (sp) {};
            \node at (sp) {${\sss i}$};
            \node[mor, circle, inner sep=3.3pt] at (0,1) {};
            \node at (0,1) {${\sss m_2}$};
            \node at (-0.75,1.2) {${\sss M_2}$};
        }{}
        \node[above] at (-1,3) {${\sss V}$};
        \node[above] at (-2,3) {${\sss M_1}$};
        \node[below] at (0,0) {${\sss \mathbb C_{As}}$};
    \end{tikzpicture}
}

\newcommand{\rightModuleAssociator}[1]{
	\begin{tikzpicture}[scale=0.6,baseline={([yshift=-1ex]current bounding box.center)}]
        \ifthenelse{#1 = 1}{
            \draw[mod] (0,0) -- (0,3);
            \draw[obj] (1,3) -- (0,2);
            \draw[obj] (2,3) -- (0,1);
            \node[mor, circle, inner sep=3.3pt] at (0,2) {};
            \node at (0,2) {${\sss i}$};
            \node[mor, circle, inner sep=3.3pt] at (0,1) {};
            \node at (0,1) {${\sss j}$};
            \node[left] at (0,1.5) {${\sss M_3}$};
        }{}
        \ifthenelse{#1 = 2}{
            \draw[mod] (0,0) -- (0,3);
            \draw[obj, name path=P1] (2,3) -- (0,1);
            \path[name path=P2] (1,3) -- (1,0);
            \path [name intersections={of=P1 and P2,by={sp}}];
            \draw[obj] (1,3) -- (sp);
            \node[mor, circle, inner sep=3.3pt] at (sp) {};
            \node at (sp) {${\sss k}$};
            \node[mor, circle, inner sep=3.3pt] at (0,1) {};
            \node at (0,1) {${\sss l}$};
            \node at (0.7,1.2) {${\sss X_3}$};
        }{}
        \node[above] at (0,3) {${\sss M_1}$};
        \node[above] at (1,3) {${\sss X_1}$};
        \node[above] at (2,3) {${\sss X_2}$};
        \node[below] at (0,0) {${\sss M_2}$};
	\end{tikzpicture}
}

\newcommand{\moduleFunctor}[1]{
    \begin{tikzpicture}[scale=0.6,baseline={([yshift=-1ex]current bounding box.center)}]
       \ifthenelse{#1 = 1}{
            \draw[mod] (0,2) -- (0,3);
            \draw[mod] (0,0) -- (0,2);
            \draw[speObj] (-1,3) -- (0,2);
            \draw[obj] (1,3) -- (0,1);
            \node[mor, circle, inner sep=3.3pt] at (0,2) {};
            \node at (0,2) {${\sss i}$};
            \node[mor, circle, inner sep=3.3pt] at (0,1) {};
            \node at (0,1) {${\sss j}$};
            \node[left] at (0,1.5) {${\sss M_3}$};
            \node[above] at (0,3) {${\sss M_1}$};
            \node[above] at (1,3) {${\sss X}$};
            \node[above] at (-1,3) {${\sss Y}$};
            \node[below] at (0,0) {${\sss M_2}$};
        }{}
        \ifthenelse{#1 = 2}{
            \draw[mod] (0,1) -- (0,3);
            \draw[mod] (0,0) -- (0,1);
            \draw[speObj] (-1,3) -- (0,1);
            \draw[obj] (1,3) -- (0,2);
            \node[mor, circle, inner sep=3.3pt] at (0,2) {};
            \node at (0,2) {${\sss k}$};
            \node[mor, circle, inner sep=3.3pt] at (0,1) {};
            \node at (0,1) {${\sss l}$};
            \node[right] at (0,1.5) {${\sss M_4}$};
            \node[above] at (0,3) {${\sss M_1}$};
            \node[above] at (1,3) {${\sss X}$};
            \node[above] at (-1,3) {${\sss Y}$};
            \node[below] at (0,0) {${\sss M_2}$};
        }{}
        \ifthenelse{#1 = 3}{
            \draw[mod] (0,2) -- (0,3);
            \draw[mod] (0,0) -- (0,2);
            \draw[obj] (1,3) -- (0,2);
            \draw[speObj] (-1,3) -- (0,1);
            \node[mor, circle, inner sep=3.3pt] at (0,2) {};
            \node at (0,2) {${\sss 1}$};
            \node[mor, circle, inner sep=3.3pt] at (0,1) {};
            \node at (0,1) {${\sss w}$};
            \node[right] at (0,1.5) {${\sss \mathbb C_{Ar \cat g}}$};
            \node[above] at (0,3) {${\sss \mathbb C_{Ar}}$};
            \node[above] at (1,3) {${\sss \mathbb C_g}$};
            \node[above] at (-1,3) {${\sss V_x}$};
            \node[below] at (0,0) {${\sss \mathbb C_{As \cat g}}$};
        }{}
        \ifthenelse{#1 = 4}{
            \draw[mod] (0,1) -- (0,3);
            \draw[mod] (0,0) -- (0,1);
            \draw[obj] (1,3) -- (0,1);
            \draw[speObj] (-1,3) -- (0,2);
            \node[mor, circle, inner sep=3.3pt] at (0,2) {};
            \node at (0,2) {${\sss v}$};
            \node[mor, circle, inner sep=3.3pt] at (0,1) {};
            \node at (0,1) {${\sss 1}$};
            \node[left] at (0,1.5) {${\sss \mathbb C_{As}}$};
            \node[above] at (0,3) {${\sss \mathbb C_{Ar}}$};
            \node[above] at (1,3) {${\sss \mathbb C_g}$};
            \node[above] at (-1,3) {${\sss V_x}$};
            \node[below] at (0,0) {${\sss \mathbb C_{As \cat g}}$};
        }{}
        \ifthenelse{#1 = 5}{
            \draw[mod, dot] (0,1) -- (0,3);
            \draw[mod] (0,0) -- (0,1);
            \draw[obj] (1,3) -- (0,2);
            \draw[speObj] (-1,3) -- (0,1);
            \node[mor, circle, inner sep=3.3pt] at (0,2) {};
            \node at (0,2) {${\sss 1}$};
            \node[mor, circle, inner sep=3.3pt] at (0,1) {};
            \node at (0,1) {${\sss n}$};
            \node[above] at (1,3) {${\sss \mathbb C_g}$};
            \node[above] at (-1,3) {${\sss M}$};
            \node[below] at (0,0) {${\sss \mathbb C_{As \cat g}}$};
        }{}
        \ifthenelse{#1 = 6}{
            \draw[mod, dot] (0,2) -- (0,3);
            \draw[mod] (0,0) -- (0,2);
            \draw[obj] (1,3) -- (0,1);
            \draw[speObj] (-1,3) -- (0,2);
            \node[mor, circle, inner sep=3.3pt] at (0,2) {};
            \node at (0,2) {${\sss m}$};
            \node[mor, circle, inner sep=3.3pt] at (0,1) {};
            \node at (0,1) {${\sss 1}$};
            \node[left] at (0,1.5) {${\sss \mathbb C_{As}}$};
            \node[above] at (1,3) {${\sss \mathbb C_g}$};
            \node[above] at (-1,3) {${\sss M}$};
            \node[below] at (0,0) {${\sss \mathbb C_{As \cat g}}$};
        }{}
        \ifthenelse{#1 = 7}{
            \draw[mod] (0,2) -- (0,3);
            \draw[mod] (0,0) -- (0,2);
            \draw[obj] (1,3) -- (0,2);
            \draw[speObj] (-1,3) -- (0,1);
            \node[mor, circle, inner sep=3.3pt] at (0,2) {};
            \node at (0,2) {${\sss k}$};
            \node[mor, circle, inner sep=3.3pt] at (0,1) {};
            \node at (0,1) {${\sss l}$};
            \node[right] at (0,1.5) {${\sss M_4}$};
            \node[above] at (0,3) {${\sss M_1}$};
            \node[above] at (1,3) {${\sss  W}$};
            \node[above] at (-1,3) {${\sss V_x}$};
            \node[below] at (0,0) {${\sss M_2}$};
        }{}
        \ifthenelse{#1 = 8}{
            \draw[mod] (0,1) -- (0,3);
            \draw[mod] (0,0) -- (0,1);
            \draw[obj] (1,3) -- (0,1);
            \draw[speObj] (-1,3) -- (0,2);
            \node[mor, circle, inner sep=3.3pt] at (0,2) {};
            \node at (0,2) {${\sss i}$};
            \node[mor, circle, inner sep=3.3pt] at (0,1) {};
            \node at (0,1) {${\sss j}$};
            \node[left] at (0,1.5) {${\sss M_3}$};
            \node[above] at (0,3) {${\sss M_1}$};
            \node[above] at (1,3) {${\sss W}$};
            \node[above] at (-1,3) {${\sss V_x}$};
            \node[below] at (0,0) {${\sss M_2}$};
        }{}
    \end{tikzpicture}
}

\newcommand{\PEPS}[5]{
    \def\sty{#1}
    \def\morI{#2}
    \def\morII{#3}
    \def\morIII{#4}
    \def\morIV{#5}
    \PEPScont
}

\newcommand{\PEPScont}[7]{
    \begin{tikzpicture}[scale=1.15,baseline={([yshift=-.5ex]current bounding box.center)}]
        \def\r{8pt};
        \def\l{3.5};
        \def\sp{0.08};
        
        \ifthenelse{#7 = 1 \OR #7 = 3}{\def\sgn{1}}{}
        \ifthenelse{#7 = 2}{\def\sgn{-1}}{}
        
        \coordinate (a1) at (0,0);
        \coordinate (a2) at (\l,0);
        \coordinate (a3) at (\l/2,{\sgn*sqrt(3)*\l/2});
        
        \clip ({(0.5-\sp)*\l*cos(60)-0.05},-\sgn*0.05) rectangle ({\l-(0.5-\sp)*\l*cos(60)+0.05},{\sgn*(\l-(0.5+\sp)*\l*sin(60))+\sgn*0.05});
        
        \middleCoo{a1}{a2}{m12}{m12-1}{m12-2}{\sp};
        \middleCoo{a2}{a3}{m23}{m23-2}{m23-3}{\sp};
        \middleCoo{a1}{a3}{m13}{m13-1}{m13-3}{\sp};
        \barycentre{a1}{a2}{a3}{b123}{b123-1}{b123-2}{b123-3}{\sp};
        
        \ifthenelse{#7 = 1}{
            \draw[obj,->-=0.77] (m12) --node[pos=0.45,fill=white,rectangle,inner sep=1pt] {${\sss #6}$} (b123);
            \draw[obj,->-=0.71] (b123) --node[pos=0.38,sloped,fill=white,rectangle,inner sep=0pt] {${\sss #4}$} (m13);
            \draw[obj,->-=0.71] (b123) --node[pos=0.38,sloped,fill=white,rectangle,inner sep=0pt] {${\sss #5}$} (m23);
        }{}
        \ifthenelse{#7 = 2}{
            \draw[obj,->-=0.7] (b123) --node[pos=0.38,fill=white,rectangle,inner sep=1pt] {${\sss #6}$} (m12);
            \draw[obj,->-=0.77] (m13) --node[pos=0.44,sloped,fill=white,rectangle,inner sep=0pt] {${\sss #4}$} (b123);
            \draw[obj,->-=0.77] (m23) --node[pos=0.44,sloped,fill=white,rectangle,inner sep=0pt] {${\sss #5}$} (b123);
        }{}
        
        \draw[mor] 
            (m12-1) to[out=\sgn*90, in=\sgn*90, looseness=1.5] (m12-2)
            (m13-1) to[out=-\sgn*30, in=-\sgn*30, looseness=1.5] (m13-3)
            (m23-2) to[out=-\sgn*150, in=-\sgn*150, looseness=1.5] (m23-3);
        \node[mor, circle, inner sep=4pt] at (b123) {};
        
        \draw[rounded corners=\r, \sty] 
        (m12-1) -- (b123-1) -- (m13-1)
        (m12-2) -- (b123-2) -- (m23-2)
        (m23-3) -- (b123-3) -- (m13-3);
        \node[yshift=-\sgn*5pt, xshift=-8pt] at (b123-1) {${\sss #1}$};
        \node[yshift=-\sgn*5pt, xshift=8pt] at (b123-2) {${\sss #2}$};
        \node[yshift=\sgn*9pt] at (b123-3) {${\sss #3}$}; 
        \node[] at (b123) {${\sss \morII}$};
        \node[yshift=\sgn*3.5pt] at (m12) {${\sss \morIII}$};
        \node[yshift=-\sgn*1.5pt, xshift=2.5pt] at (m13) {${\sss \morI}$};
        \node[yshift=-\sgn*1.5pt, xshift=-2.5pt] at (m23) {${\sss \morIV}$};
    \end{tikzpicture}
}

\newcommand{\PEPSHex}[7]{
    \def\sty{#1}
    \def\modI{#2}
    \def\modII{#3}
    \def\modIII{#4}
    \def\modIV{#5}
    \def\modV{#6}
    \def\modVI{#7}
    \PEPSHexcont
}

\newcommand{\PEPSHexcont}[6]{
    \begin{tikzpicture}[scale=1.15,baseline={([yshift=-.5ex]current bounding box.center)}]
        \def\r{8pt};
        \def\l{3.5};
        \def\sp{0.08};

        \coordinate (a1) at (0,0);
        \coordinate (a2) at ({\l/2*cos(30)},{-\l/2*sin(30)});
        \coordinate (a3) at ({2*\l/2*cos(30)},{0});
        \coordinate (a4) at ({2*\l/2*cos(30)},{\l/2});
        \coordinate (a5) at ({\l/2*cos(30)},{\l/2+\l/2*sin(30)});
        \coordinate (a6) at ({0},{\l/2});
        
        \middleCoo{a1}{a2}{m12}{m12-1}{m12-2}{2*\sp};
        \middleCoo{a2}{a3}{m23}{m23-2}{m23-3}{2*\sp};
        \middleCoo{a3}{a4}{m34}{m34-3}{m34-4}{2*\sp};
        \middleCoo{a4}{a5}{m45}{m45-4}{m45-5}{2*\sp};
        \middleCoo{a5}{a6}{m56}{m56-5}{m56-6}{2*\sp};
        \middleCoo{a1}{a6}{m16}{m16-1}{m16-6}{2*\sp};

        \draw[obj, ->-=.36, ->-=.84] (m12) --node[pos=0.2,fill=white,rectangle,inner sep=1pt, sloped](){${\sss #1}$} node[pos=0.68,fill=white,rectangle,inner sep=1pt, sloped](){${\sss #2}$}  (m45);
        \draw[obj, ->-=.36, ->-=.84] (m56) -- node[pos=0.2,fill=white,rectangle,inner sep=1pt, sloped](){${\sss #3}$} node[pos=0.68,fill=white,rectangle,inner sep=1pt, sloped](){${\sss #4}$} (m23);
        \draw[obj, ->-=.36, ->-=.84] (m16) -- node[pos=0.2,fill=white,rectangle,inner sep=1pt, sloped](){${\sss #5}$} node[pos=0.68,fill=white,rectangle,inner sep=1pt, sloped](){${\sss #6}$} (m34);

        \draw[mor] 
            (m12-1) to[out=60, in=60, looseness=1.5] (m12-2)
            (m56-5) to[out=-60, in=-60, looseness=1.5] (m56-6)
            (m23-2) to[out=120, in=120, looseness=1.5] (m23-3)
            (m45-4) to[out=-120, in=-120, looseness=1.5] (m45-5)
            (m16-1) to[out=0, in=0, looseness=1.5] (m16-6)
            (m34-3) to[out=180, in=180, looseness=1.5] (m34-4);

        \node[mor, circle, inner sep=4pt] at ({\l/2*cos(30)},{\l/4}) {};
        \node[] at ({\l/2*cos(30)},{\l/4}) {${\sss 1}$};
        \node[xshift=2pt, yshift=3pt] at (m12) {${\sss 1}$};
        \node[xshift=-2pt, yshift=3pt] at (m23) {${\sss 1}$};
        \node[xshift=2pt, yshift=-3pt] at (m56) {${\sss 1}$};
        \node[xshift=-2pt, yshift=-3pt] at (m45) {${\sss 1}$};
        \node[xshift=3.5pt] at (m16) {${\sss 1}$};
        \node[xshift=-3.5pt] at (m34) {${\sss 1}$};
        
        \intersec{m12-1}{m45-5}{m16-1}{m34-3}{b1};
        \intersec{m12-2}{m45-4}{m23-2}{m56-6}{b2};
        \intersec{m23-3}{m56-5}{m34-3}{m16-1}{b3};
        \intersec{m34-4}{m16-6}{m45-4}{m12-2}{b4};
        \intersec{m45-5}{m12-1}{m56-5}{m23-3}{b5};
        \intersec{m16-6}{m34-4}{m56-6}{m23-2}{b6};

        \draw[mod, rounded corners=\r] 
            (m12-2) -- (b2) -- (m23-2)
            (m23-3) -- (b3) -- (m34-3) 
            (m34-4) -- (b4) -- (m45-4)
            (m45-5) -- (b5) -- (m56-5)
            (m56-6) -- (b6) -- (m16-6)
            (m16-1) -- (b1) -- (m12-1);

        \node[yshift=-22pt] at (b2) {${\sss \modII}$};
        \node[yshift=22pt] at (b5) {${\sss \modV}$};
        \node[yshift=-7pt, xshift=-17pt] at (b1) {${\sss \modI}$};
        \node[yshift=-7pt, xshift=17pt] at (b3) {${\sss \modIII}$};
        \node[yshift=7pt, xshift=17pt] at (b4) {${\sss \modIV}$};
        \node[yshift=7pt, xshift=-17pt] at (b6) {${\sss \modVI}$};
    \end{tikzpicture}
}

\newcommand{\MPO}[6]{
    \def\styA{#1}
    \def\styB{#2}
    \def\morI{#3}
    \def\morII{#4}
    \def\morIII{#5}
    \def\morIV{#6}
    \MPOcont
}

\newcommand{\MPOcont}[7]{
	\begin{tikzpicture}[scale=1.15,baseline={([yshift=-.5ex]current bounding box.center)}]
        \def\r{8pt};
        \def\l{3.5};
        \def\sp{0.16};
        \coordinate (a1) at (0,0);
        \coordinate (a2) at (\l/2,0);
        \coordinate (a3) at (0,\l/2);
        \coordinate (a4) at (\l/2,\l/2);
        
        \middleCoo{a1}{a2}{m12}{m12-1}{m12-2}{\sp};
        \middleCoo{a1}{a3}{m13}{m13-1}{m13-3}{\sp};
        \middleCoo{a2}{a4}{m24}{m24-2}{m24-4}{\sp};
        \middleCoo{a3}{a4}{m34}{m34-3}{m34-4}{\sp};
        \barycentreSq{a1}{a2}{a3}{a4}{b1234}{b1234-1}{b1234-2}{b1234-3}{b1234-4};
        
        \ifthenelse{#7 = 2}{
            \draw[speObj,->-=0.7,shorten <= 1pt, shorten >= 1pt] (m13) --node[pos=0.35,fill=white,rectangle,inner sep=0pt,text=black] {${\sss #6}$} node[mpoint, pos=0.5](mk1){}  (m24);
        }{}
        \ifthenelse{#7 = 1}{
            \draw[speObj,->-=0.7] (m24) --node[pos=0.35,fill=white,rectangle,inner sep=0pt,text=black] {${\sss #6}$} node[mpoint, pos=0.5](mk1){}  (m13);
        }{}

        \clipAroundNode{mk1}{
            \ifthenelse{#7=5}{
                \draw[speObj,->-=0.75] (m12) --node[pos=0.3,fill=white,rectangle,inner sep=1pt,text=black] {${\sss #5}$} (m34);
            }{
                \draw[obj,->-=0.75] (m12) --node[pos=0.3,fill=white,rectangle,inner sep=1pt] {${\sss #5}$} (m34);
            }
        }  

        \draw[mor] 
        (m12-1) to[out=90, in=90, looseness=1.5] (m12-2)
        (m13-1) to[out=0, in=0, looseness=1.5] (m13-3)
        (m24-2) to[out=-180, in=-180, looseness=1.5] (m24-4)
        (m34-3) to[out=-90, in=-90, looseness=1.5] (m34-4);
        
        \draw[obj, rounded corners=\r, \styA] 
        (m12-1) -- (b1234-1) -- (m13-1)
        (m12-2) -- (b1234-2) -- (m24-2);
        \draw[obj, rounded corners=\r, \styB] 
        (m13-3) -- (b1234-3) -- (m34-3)
        (m34-4) -- (b1234-4) -- (m24-4);
        
        \node[yshift=3pt] at (m12) {${\sss \morIV}$};
        \node[yshift=-3.5pt] at (m34) {${\sss \morIII}$};
        \node[xshift=3pt] at (m13) {${\sss \morI}$};   
        \node[xshift=-3pt] at (m24) {${\sss \morII}$};   
        \node[xshift=-8pt,yshift=8pt] at (b1234-3) {${\sss #1}$};
        \node[xshift=8pt,yshift=8pt] at (b1234-4) {${\sss #2}$};
        \node[xshift=-8pt,yshift=-8pt] at (b1234-1) {${\sss #4}$};
        \node[xshift=8pt,yshift=-8pt] at (b1234-2) {${\sss #3}$}; 
	\end{tikzpicture}
}

\newcommand{\pulling}[1]{
    \begin{tikzpicture}[scale=1.15,baseline={([yshift=-.5ex]current bounding box.center)}]
        \def\r{8pt};
        \def\l{3.5};
        \def\sp{0.08};
        \def\red{0.5};
        
        \ifthenelse{#1=1}{
            \clip (-0.4,-0.05) rectangle (3.9,3.5);
            \coordinate (a1) at (0,0);
            \coordinate (a2) at (\l,0);
            \coordinate (a3) at (\l/2,{sqrt(3)*\l/2});	    	    	
            \middleCoo{a1}{a2}{m12}{m12-1}{m12-2}{\sp};
            \middleCoo{a2}{a3}{m23}{m23-2}{m23-3}{\sp};
            \middleCoo{a1}{a3}{m13}{m13-1}{m13-3}{\sp};
            \barycentre{a1}{a2}{a3}{b123}{b123-1}{b123-2}{b123-3}{\sp};
            \foreach \x in {m13,m13-1,m13-3}{
                \coordinate (\x) at ($(\x)+({cos(30)*\red},{-sin(30)*\red})$);
            }
            \foreach \x in {m23,m23-2,m23-3}{
                \coordinate (\x) at ($(\x)+({-cos(30)*\red},{-sin(30)*\red})$);
            }
            \draw[obj,->-=0.77] (m12) --node[pos=0.45,fill=white,rectangle,inner sep=1pt] {${\sss X_3}$} (b123);       
            \draw[mor] (m12-1) to[out=90, in=90, looseness=1.5] (m12-2);
            \draw[mod, rounded corners=\r] 
                (m12-1) -- (b123-1) -- (m13-1)
                (m12-2) -- (b123-2) -- (m23-2)
                (m23-3) -- (b123-3) -- (m13-3);
            
            \def\sp{0.16};
            \coordinate (b1) at ($({\l*cos(60)/4},{\l*sin(60)/4})+({cos(30)*\red},{-sin(30)*\red})$);
            \coordinate (b2) at ($({3*\l*cos(60)/4},{3*\l*sin(60)/4})+({cos(30)*\red},{-sin(30)*\red})$);
            \coordinate (b3) at ($(b1)+({-\l*cos(30)/2},{\l*sin(30)/2})$);
            \coordinate (b4) at ($(b2)+({-\l*cos(30)/2},{\l*sin(30)/2})$);
            \middleCoo{b1}{b3}{n13}{n13-1}{n13-3}{\sp};
            \middleCoo{b2}{b4}{n24}{n24-2}{n24-4}{\sp};
            \middleCoo{b3}{b4}{n34}{n34-3}{n34-4}{\sp};
            \barycentreSq{b1}{b2}{b3}{b4}{b1234}{b1234-1}{b1234-2}{b1234-3}{b1234-4};
            \draw[speObj, shorten <=-1pt,->-=0.7] (n24) -- node[mpoint, pos=0.5](mk1){} (n13);
            \clipAroundNode{mk1}{
                \draw[obj,->-=0.5] (b123) --node[pos=0.25,sloped,fill=white,rectangle,inner sep=0pt] {${\sss X_1}$} (n34);
            } 
            \draw[mor] 
                (n34-3) to[out=-30, in=-30, looseness=1.5] (n34-4)
                (n13-1) to[out=60, in=60, looseness=1.5] (n13-3);
            \draw[mod, rounded corners=\r]
                (m13-1) -- (b1234-1) -- (n13-1)
                (m13-3) -- (b1234-2) -- (n24-2)
                (n13-3) -- (b1234-3) -- (n34-3)
                (n34-4) -- (b1234-4) -- (n24-4);
                
            \coordinate (c1) at ($(a2)+({-3*\l*cos(60)/4},{3*\l*sin(60)/4})+({-cos(30)*\red},{-sin(30)*\red})$);
            \coordinate (c2) at ($(a2)+({-\l*cos(60)/4},{\l*sin(60)/4})+({-cos(30)*\red},{-sin(30)*\red})$);
            \coordinate (c3) at ($(c1)+({\l*cos(30)/2},{\l*sin(30)/2})$);
            \coordinate (c4) at ($(c2)+({\l*cos(30)/2},{\l*sin(30)/2})$);
            \middleCoo{c1}{c3}{o13}{o13-1}{o13-3}{\sp};
            \middleCoo{c2}{c4}{o24}{o24-2}{o24-4}{\sp};
            \middleCoo{c3}{c4}{o34}{o34-3}{o34-4}{\sp};
            \barycentreSq{c1}{c2}{c3}{c4}{c1234}{c1234-1}{c1234-2}{c1234-3}{c1234-4};
            \draw[speObj,shorten >= -1pt,->-=0.7] (o24) -- node[mpoint, pos=0.5](mk1){} (o13);
            \clipAroundNode{mk1}{
                \draw[obj,->-=0.5] (b123) --node[pos=0.25,sloped,fill=white,rectangle,inner sep=0pt] {${\sss X_2}$} (o34);
            } 
            \draw[mor] 
                (o34-3) to[out=-150, in=-150, looseness=1.5] (o34-4)
                (o24-2) to[out=120, in=120, looseness=1.5] (o24-4);
            \draw[mod, rounded corners=\r]
                (m23-2) -- (c1234-2) -- (o24-2)
                (m23-3) -- (c1234-1) -- (o13-1)
                (o13-3) -- (c1234-3) -- (o34-3)
                (o34-4) -- (c1234-4) -- (o24-4);
            
            \draw[line width = 0.7pt, line join=round, line cap=join, color=BrickRed] (o13) to[out=120,in=60,looseness=1] node[pos=0.5,fill=white,rectangle,inner sep=0pt,text=black](){${\sss Y}$} (n24);
            \draw[mod] (n24-2) to[out=60,in=120,looseness=1] (o13-1);
            \draw[mod] (n24-4) to[out=60,in=120,looseness=1] node[pos=0.5,yshift=6pt,text=black](){${\sss M_4}$} (o13-3);
            \node[mor, circle, inner sep=4pt] at (b123) {};    
            \node at (b123) {${\sss m}$};
            \node[yshift=9pt] at (b123-3) {${\sss }$};
            \node[yshift=-6pt,xshift=-4pt] at (m13-1) {${\sss M_1}$};
            \node[yshift=10pt] at (b123-3) {${\sss M_3}$};
            \node[yshift=-6pt,xshift=4pt] at (m23-2) {${\sss M_2}$};
            \node[xshift=-10pt,yshift=-3pt] at (b1234-3) {${\sss M_5}$};
            \node[xshift=10pt,yshift=-3pt] at (c1234-4) {${\sss M_6}$};
            \node[yshift=3pt] at (m12) {${\sss n}$};
            \node[xshift=1.5pt,yshift=2.6pt] at (n13) {${\sss i}$};
            \node[xshift=2.5pt,yshift=-1.5pt] at (n34) {${\sss k}$};
            \node[xshift=-1.5pt,yshift=2.6pt] at (o24) {${\sss j}$};
            \node[xshift=-2.5pt,yshift=-1.5pt] at (o34) {${\sss l}$};
        }{}
        
        \ifthenelse{#1=2}{
            \clip (0.65,-1.35) rectangle (2.85,1.85);
            \coordinate (a1) at (0,0);
            \coordinate (a2) at (\l,0);
            \coordinate (a3) at (\l/2,{sqrt(3)*\l/2});	    	    	
        	\middleCoo{a1}{a2}{m12}{m12-1}{m12-2}{\sp};
        	\middleCoo{a2}{a3}{m23}{m23-2}{m23-3}{\sp};
        	\middleCoo{a1}{a3}{m13}{m13-1}{m13-3}{\sp};
        	\barycentre{a1}{a2}{a3}{b123}{b123-1}{b123-2}{b123-3}{\sp};
        	\foreach \x in {m12,m12-1,m12-2}{
        	    \coordinate (\x) at ($(\x)+(0,\red)$);
        	}
            \draw[obj,->-=0.71] (b123) --node[pos=0.38,sloped,fill=white,rectangle,inner sep=0pt] {${\sss X_1}$} (m13);
            \draw[obj,->-=0.71] (b123) --node[pos=0.38,sloped,fill=white,rectangle,inner sep=0pt] {${\sss X_2}$} (m23);
            \draw[mor] 
                (m13-1) to[out=-30, in=-30, looseness=1.5] (m13-3)
                (m23-2) to[out=-150, in=-150, looseness=1.5] (m23-3);
            \draw[mod, rounded corners=\r] 
                (m12-1) -- (b123-1) -- (m13-1)
                (m12-2) -- (b123-2) -- (m23-2)
                (m23-3) -- (b123-3) -- (m13-3);
                
            \def\sp{0.16};
            \coordinate (b1) at (\l/4,-\l/2+\red);
            \coordinate (b2) at (3*\l/4,-\l/2+\red);
            \coordinate (b3) at (\l/4,\red);
            \coordinate (b4) at (3*\l/4,\red);
            \middleCoo{b1}{b3}{n13}{n13-1}{n13-3}{\sp};
            \middleCoo{b2}{b4}{n24}{n24-2}{n24-4}{\sp};
            \middleCoo{b1}{b2}{n12}{n12-1}{n12-2}{\sp};
            \barycentreSq{b1}{b2}{b3}{b4}{b1234}{b1234-1}{b1234-2}{b1234-3}{b1234-4};
            \draw[speObj,->-=0.7] (n24) --node[pos=0.35,fill=white,rectangle,inner sep=0pt,text=black] {${\sss Y}$} node[mpoint, pos=0.5](mk1){}  (n13);
            \clipAroundNode{mk1}{
                \draw[obj,->-=0.8] (n12) --node[pos=0.6,fill=white,rectangle,inner sep=0pt] {${\sss X_3}$} (b123);  
            } 
            \draw[mor] 
                (n12-1) to[out=90, in=90, looseness=1.5] (n12-2)
                (n13-1) to[out=0, in=0, looseness=1.5] (n13-3)
                (n24-2) to[out=-180, in=-180, looseness=1.5] (n24-4);
            \draw[mod, rounded corners=\r]
                (m12-1) -- (b1234-3) -- (n13-3)
                (m12-2) -- (b1234-4) -- (n24-4)
                (n12-1) -- (b1234-1) -- (n13-1)
                (n12-2) -- (b1234-2) -- (n24-2);
            \node[mor, circle, inner sep=4pt] at (b123) {};    
            \node at (b123) {${\sss m}$};
            \node[xshift=-8pt,yshift=-8pt] at (b1234-1) {${\sss M_1}$};
            \node[xshift=8pt,yshift=-8pt] at (b1234-2) {${\sss M_2}$};
            \node[xshift=-8pt,yshift=14pt] at (b1234-3) {${\sss M_5}$};
            \node[xshift=8pt,yshift=14pt] at (b1234-4) {${\sss M_6}$};
            \node[yshift=9pt] at (b123-3) {${\sss M_4}$};
            \node[yshift=3pt] at (n12) {${\sss n}$};
            \node[xshift=2pt,yshift=-1.5pt] at (m13) {${\sss k}$};
            \node[xshift=-2pt,yshift=-1.5pt] at (m23) {${\sss l}$};
            \node[xshift=3pt] at (n13) {${\sss i}$};
            \node[xshift=-3pt] at (n24) {${\sss j}$};
        }{}
	\end{tikzpicture}
}

\newcommand{\deltaT}[7]{
    \begin{tikzpicture}[scale=1,baseline={([yshift=-.5ex]current bounding box.center)}]
        \def\l{.9};
        \coordinate (a) at (0,0);
        \draw[obj, ->-=.68] (0,0) -- node[pos=1.25]{${\sss #4}$} ({\l*cos(30)},{\l*sin(30)});
        \draw[obj, ->-=.68] (0,0) -- node[pos=1.25]{${\sss #3}$} ({\l*cos(120)},{\l*sin(120)});
        \draw[obj, ->-=.68] (0,0) -- node[pos=1.25]{${\sss #2}$} ({\l*cos(150)},{\l*sin(150)});
        \centerarc{dash pattern= on 0pt off 4.5pt, line width=1pt}{a}{58}{100}{1.3*\l/2};                      
        \draw[obj, -<-=.5] (0,0) -- node[pos=1.25]{${\sss #7}$} ({\l*cos(30)},{-\l*sin(30)});
        \draw[obj, -<-=.5] (0,0) -- node[pos=1.25]{${\sss #6}$} ({\l*cos(120)},{-\l*sin(120)});
        \draw[obj, -<-=.5] (0,0) -- node[pos=1.25]{${\sss #5}$} ({\l*cos(150)},{-\l*sin(150)});
        \centerarc{dash pattern= on 0pt off 4.5pt, line width=1pt}{a}{-58}{-100}{1.3*\l/2}; 
        \ifthenelse{#1=1}{
            \node[line width=.7pt, draw=black, fill=black!40, circle, inner sep=2.3pt] at (0,0) {};
        }{
            \node[line width=.7pt, draw=black, fill=white, circle, inner sep=2.3pt] at (0,0) {};
        }
    \end{tikzpicture}
}

\newcommand{\MPOA}[1]{
    \begin{tikzpicture}[scale=1,baseline={([yshift=-.57ex]current bounding box.center)}]
        \def\l{.8};
        \draw[->-=.83] (0,-\l) -- (0,\l);
        \draw[speObj, ->-=.83] (\l,0) -- node[pos=.25,above,color=black](){${\sss #1}$} (-\l,0);
        )
        \node[rectangle, inner sep=3.2pt, draw=BrickRed, fill=white] at (0,0) {};
    \end{tikzpicture}
}

\newcommand{\PEPSA}[1]{
    \begin{tikzpicture}[scale=1,baseline={([yshift=-.5ex]current bounding box.center)},z={(0,1)},y={(-0.3,0.3)}]
        \def\l{3.7};
        \def\h{.9};
        \coordinate (a) at (0,0,0);
        \coordinate (b) at (\l,0,0);
        \coordinate (c) at ({\l+cos(60)*\l},{sin(60)*\l},0);
        \coordinate (d) at (\l,{2*sin(60)*\l},0);
        \coordinate (e) at (0,{2*sin(60)*\l},0);
        \coordinate (f) at ({-cos(60)*\l},{sin(60)*\l},0);
        \coordinate (p) at (\l/2,{\l*sin(60)},0);

        \foreach \x/\y in {a/b,b/c,c/d,d/e,e/f,f/a}{
            \middleCoo{\x}{\y}{\x\y}{}{}{0};
            \draw[->-=.65] (\x\y) --node[mpoint, pos=.5](mp\x\y){} node[mpoint, pos=.8](mpp\x\y){} ($(\x\y)+(0,0,\h)$);
        }
        \ifthenelse{#1=1 \OR #1=2}{
            \clipAroundTwoNodes{mpbc}{mpef}{    
                \draw[obj, ->-=.27, ->-=.77] (b) -- (a);
                \draw[obj, ->-=.27, ->-=.77] (d) -- (e);
                \draw[obj, ->-=.28, ->-=.78] (a) -- (f);
                \draw[obj, ->-=.35, ->-=.78] (c) -- (d);
                \draw[obj, ->-=.3, ->-=.8] (f) -- (e);
                \draw[obj, ->-=.3, ->-=.8] (b) -- (c);
                \draw[obj] (a) -- ($(a)+({-\l/3*cos(60)},{-\l/3*sin(60)},0)$);
                \draw[obj] (b) -- ($(b)+({\l/3*cos(60)},{-\l/3*sin(60)},0)$);
                \draw[obj] (c) -- ($(c)+({\l/4},0,0)$);
                \draw[obj] (d) -- ($(d)+({\l/3*cos(60)},{\l/3*sin(60)},0)$);
                \draw[obj] (e) -- ($(e)+({-\l/3*cos(60)},{\l/3*sin(60)},0)$);
                \draw[obj] (f) -- ($(f)+(-\l/4,0,0)$);
            }
    
            \foreach \x in {a,b,c,d,e,f}{
                \node[line width=.7pt, draw=black, fill=black!40, circle, inner sep=2.3pt] at (\x) {};
            }
            \foreach \x in {ab,bc,cd,de,ef,fa}{
                \node[line width=.7pt, draw=black, fill=white, circle, inner sep=2.3pt] at (\x) {};
            }
    
            \ifthenelse{#1=2}{
                \middleCoo{a}{fa}{afa}{}{}{0};
                \middleCoo{d}{de}{dde}{}{}{0};
                \draw[speObj, -<-=.2, -<-=.5, -<-=.92] (-2.5,{-\l/3*sin(60)},0) to[out=90, in=-120, looseness=1.5] (afa) to[out=60, in=-90, out looseness=2, in looseness=1] node[pos=.4, above, color=black]{${\sss x}$} (dde) to[out=90, in=-10, looseness=1.3] (.65*\l,{2*\l*sin(60)+\l/3*sin(60)},0);  
                \node[rectangle, inner sep=3.2pt, draw=BrickRed, fill=white, rotate=72] at (afa) {};
                \node[rectangle, inner sep=3.2pt, draw=BrickRed, fill=white] at (dde) {};
            }{}
        }{}
        
        \ifthenelse{#1=3 \OR #1=4}{
            \clipAroundTwoNodes{mpbc}{mpef}{    
                \draw[dot] (b) -- (a);
                \draw[dot] (d) -- (e);
                \draw[dot] (a) -- (f);
                \draw[dot] (c) -- (d);
                \draw[dot] (f) -- (e);
                \draw[dot] (b) -- (c);
                \draw[dot] (a) -- ($(a)+({-\l/3*cos(60)},{-\l/3*sin(60)},0)$);
                \draw[dot] (b) -- ($(b)+({\l/3*cos(60)},{-\l/3*sin(60)},0)$);
                \draw[dot] (c) -- ($(c)+({\l/4},0,0)$);
                \draw[dot] (d) -- ($(d)+({\l/3*cos(60)},{\l/3*sin(60)},0)$);
                \draw[dot] (e) -- ($(e)+({-\l/3*cos(60)},{\l/3*sin(60)},0)$);
                \draw[dot] (f) -- ($(f)+(-\l/4,0,0)$);
            }
            \clipAroundTwoNodes{mppab}{mppfa}{
                \draw[-<-=.45, shorten <=-13pt] (ab) -- (p);
                \draw[-<-=.45, shorten >=-13pt] (p) --node[pos=.7, inner sep=0pt](sp3){} (de);
                \draw[->-=.55, shorten <=-26pt] (bc) -- (p);
                \draw[->-=.55, shorten <=-21pt] (cd) -- (p);
                \draw[->-=.55, shorten >=-26pt] (p) --node[pos=.5, inner sep=0pt](sp2){} (ef);
                \draw[->-=.55, shorten >=-21pt] (p) --node[pos=.72, inner sep=0pt](sp1){} (fa);
            }
            \node[line width=.7pt, draw=black, fill=white, circle, inner sep=2.3pt] at (p) {};
            \foreach \x in {ab,bc,cd,de,ef,fa}{
                \node[line width=.7pt, draw=black, fill=black!40, circle, inner sep=2.3pt] at (\x) {};
            }
    
            \ifthenelse{#1=4}{
                \draw[speObj, -<-=.2, -<-=.7] (-2.5,{-\l/3*sin(60)},0) to[out=90,in=-80, looseness=1] (sp1) to[out=100, in=-100, looseness=1] (sp2) to[out=80, in=-140, looseness=1] (sp3) to[out=50, in=-10, looseness=1] (.65*\l,{2*\l*sin(60)+\l/3*sin(60)},0);  
                \node[rectangle, inner sep=3.2pt, draw=BrickRed, fill=white, sloped, rotate=45] at (sp3) {};
                \node[rectangle, inner sep=3.2pt, draw=BrickRed, fill=white, sloped, rotate=11] at (sp1) {};
                \node[rectangle, inner sep=3.2pt, draw=BrickRed, fill=white, rotate=-10] at (sp2) {};
                \path[->-=.3] (p) -- (ef);
                \node[] at (1,5.5) {${\sss \chi}$};
            }{}
        }{}
    \end{tikzpicture}
}

\newcommand{\cellDecomposition}[0]{
    \begin{tikzpicture}[scale=1,baseline={([yshift=-.5ex]current bounding box.center)},z={(0,1)},y={(-0.25,0.25)}]
        \def\l{1};
        \def\h{2.7};
        \coordinate (a) at (0,0,0);
        \coordinate (b) at (1.8*\l,1.8*\l,0);
        \coordinate (c) at (1.5*\l,4*\l,0);
        \coordinate (d) at (0,5*\l,0);
        \coordinate (e) at (-2*\l,2.5*\l,0);
        \coordinate (f) at (-1.8*\l,0,0);
        \coordinate (a1) at (0.5*\l,-1*\l,0);
        \coordinate (b1) at (2.8*\l,1.8*\l,0);
        \coordinate (c1) at (2.5*\l,4*\l,0);
        \coordinate (d1) at (-.18*\l,5.9*\l,0);
        \coordinate (e1) at (-3*\l,2.5*\l,0);
        \coordinate (f1) at (-1.8*\l,-1.5*\l,0);
        \coordinate (f2) at (-2.8*\l,0.6*\l,0);
        \coordinate (ft) at ($(f)+(0,0,\h)$);
        \coordinate (at) at ($(a)+(0,0,\h)$);
        \coordinate (a1t) at ($(a1)+(0,0,\h)$);
        \coordinate (bt) at ($(b)+(0,0,\h)$);

        \fill[color=BlueViolet, opacity=.1] (-4*\l,7*\l,0) -- (-4*\l,-2.2*\l,0) -- (3.5*\l,-2.2*\l,0) -- (3.5*\l,7*\l,0) --cycle;
        \draw[edge, dot] (-4*\l,7*\l,0) -- (-4*\l,-2.2*\l,0) -- (3.5*\l,-2.2*\l,0) -- (3.5*\l,7*\l,0);
        \node[right] at (3.5*\l,-2*\l,0) {${\Sigma_\Upsilon \times \{0\}}$};
        \node[right] at (3.5*\l,-2*\l,\h) {${\Sigma_\Upsilon \times \{1\}}$};
        \draw[edge, dot] (-4*\l,7*\l,\h) -- (-4*\l,-2.2*\l,\h) -- (3.5*\l,-2.2*\l,\h) -- (3.5*\l,7*\l,\h);

        \draw[edge] (f) -- (f2);
        \foreach \x in {a,b,c,d,e,f}{
            \draw[edge] (\x) -- ($(\x)+(0,0,\h)$);
        }
        \path (a) --node[mpoint, pos=.35](mp1){} ($(a)+(0,0,\h)$);
        \path (f) --node[mpoint, pos=.15](mp3){} node[mpoint, pos=.45](mp4){} ($(f)+(0,0,\h)$);
        \path (b) --node[mpoint, pos=.2](mp2){} ($(b)+(0,0,\h)$);
        \clipAroundFourNodes{mp1}{mp2}{mp3}{mp4}{
            \foreach \x in {a,b,c,d,e,f}{
                \draw (\x) -- (\x1);
            }
            \draw[edge] (a) -- (b) -- (c) -- (d) -- (e) -- (f) --cycle;
            \draw[edge] (c) -- (e);
        }
        \foreach \x/\y in {d/d1,c/c1,e/e1,c/d,d/e,c/e,b/c,e/f,a/b,f/a,b/b1,a/a1,f/f1,f/f2}{
            \fill[gray, opacity=.08] (\x) -- (\y) -- ($(\y)+(0,0,\h)$) -- ($(\x)+(0,0,\h)$) -- cycle;  
        }
        \path[->-=.52] (a) -- ($(a)+(0,0,\h)$);
        \path[->-=.6] (a) -- (b);
        \path[->-=.6] (f) -- (a);
        \path[->-=.6] (a) -- (a1);
        \node[fill=black, circle, inner sep=1.4pt] at (a) {};
        \roundArrow{a}{f}{f1}{left};
        \roundArrow{c}{e}{f}{left};
        \roundArrow{b}{a}{a1}{left};
        \roundArrow{f}{ft}{at}{right};
        \roundArrow{a}{at}{a1t}{right};
        \roundArrow{at}{bt}{b}{right};
    \end{tikzpicture}
}

\newcommand{\evaluation}[0]{
    \begin{tikzpicture}[scale=1,baseline={([yshift=-.5ex]current bounding box.center)},z={(0,1)},y={(0,0.33)}]
        \def\l{3.5};
        \def\h{2.7};
        \def\tb{360};
        \def\tc{130};
        \def\td{180};
        \def\te{290};       
        \coordinate (a) at (0,0,0);
        \coordinate (b) at ({\l*cos(\tb)},{\l*sin(\tb)},0);
        \coordinate (b1) at ({\l*cos(\te+(\tb-\te)/2)},{\l*sin(\te+(\tb-\te)/2)},0);
        \coordinate (c) at ({\l*cos(\tc)},{\l*sin(\tc)},0);
        \coordinate (c1) at ({\l*cos(\tc/4)},{\l*sin(\tc/4)},0);
        \coordinate (c2) at ({\l*cos(\tc/2)},{\l*sin(\tc/2)},0);
        \coordinate (c3) at ({\l*cos(3*\tc/4)},{\l*sin(3*\tc/4)},0);
        \coordinate (d) at ({\l*cos(\td)},{\l*sin(\td)},0);
        \coordinate (d1) at ({\l*cos(\tc+0.5*(\td-\tc))},{\l*sin(\tc+0.5*(\td-\tc))},0);
        \coordinate (e) at ({\l*cos(\te)},{\l*sin(\te)},0);
        \coordinate (e1) at ({\l*cos(\td+(\te-\td)/3)},{\l*sin(\td+(\te-\td)/3)},0);
        \coordinate (e2) at ({\l*cos(\td+2*(\te-\td)/3)},{\l*sin(\td+2*(\te-\td)/3)},0);
        \coordinate (at) at (0,0,\h/1.543);
        \coordinate (ah) at (0,0,\h);
        \coordinate (bt) at ({.5*\l*cos(\tb)},{.5*\l*sin(\tb)},0);
        \coordinate (ct) at ({.5*\l*cos(\tc)},{.5*\l*sin(\tc)},0);
        \coordinate (dt) at ({.5*\l*cos(\td)},{.5*\l*sin(\td)},0);
        \coordinate (et) at ({.5*\l*cos(\te)},{.5*\l*sin(\te)},0);
        \coordinate (bh) at ({\l*cos(\tb)},{\l*sin(\tb)},\h);
        \coordinate (ch) at ({\l*cos(\tc)},{\l*sin(\tc)},\h);
        \coordinate (dh) at ({\l*cos(\td)},{\l*sin(\td)},\h);
        \coordinate (eh) at ({\l*cos(\te)},{\l*sin(\te)},\h);

        \clip (-3.6,-4.4,0) rectangle (3.6,11,0);
        \fill[color=BlueViolet, opacity=.1] (b) -- (c1) -- (c2) -- (c3) -- (c) -- (d1) -- (d) -- (e1) -- (e2) -- (e) -- (b1) -- cycle;

        \node at (-2,-1.5,0) {${\sss M_1}$};
        \node at (2,-1.5,0) {${\sss M_2}$};
        \node at (-2.5,1.,0) {${\sss M_4}$};
        \node at (2.4,1,0) {${\sss M_3}$};
        \node at (-1,1,1.6) {${\sss X_2}$};
        
        \draw[edge,->-=.4] (a) --node[mpoint, pos=.2](mp1){} (ah);
        \draw[fat, ->-=.5] (bt) to[out=90,in=0,looseness=1] (at);
        \draw[fat, ->-=.6] (at) to[out=180,in=90,looseness=1] node[mpoint, pos=.7](mp3){} (dt);
        \draw[fat, ->-=.6] (at) to[out=180,in=90,looseness=.9] (ct);
        \draw[fat, ->-=.55] (et) to[out=90,in=-30,looseness=.8] node[mpoint, pos=.5](mp2){} node[mpoint, pos=.2](mp4){} (at);
        
        \clipAroundFourNodes{mp1}{mp2}{mp3}{mp4}{
            \draw[fat, ->-=.1, ->-=.35, ->-=.57, ->-=.8] ({\l/2*cos(0)},{\l/2*sin(0)},0) arc(0:-360:\l/2);
            \draw[edge, ->-=.35] (a) -- (b);
            \draw[edge, ->-=.75] (c) -- (a);
            \draw[edge, ->-=.75] (d) -- (a);
            \draw[edge, ->-=.3] (a) -- (e);
        }
        \foreach \x in {b,c,d,e}{
            \fill[gray, opacity=.08] (a) -- (\x) -- ($(\x)+(0,0,\h)$) -- (ah) -- cycle;  
        }
        \node[fill=black, circle, inner sep=1.4pt] at (a) {};
        \fill[color=LimeGreen, opacity=.1] (bt) arc(0:-180:\l/2) to[out=90,in=180,looseness=1] (at) to[out=0, in=90, looseness=1] (bt);
        \fill[color=LimeGreen, opacity=.1] (bt) arc(0:-180:\l/2) to[out=-90,in=180,looseness=1] ($(a)+(0,0,-\h/2)$) to[out=0, in=-90, looseness=1] (bt);

        \node[xshift=4pt, yshift=4pt] at (0,0,0) {${\sss \msf v}$};
        \node at (-1.8,0,1.6) {${\sss X_1}$};
        \node at (1.8,0,1.6) {${\sss X_3}$};
        \node at (.9,-2,1.6) {${\sss X_4}$};
        \roundArrow{d}{dh}{ah}{right};
        \roundArrow{c}{ch}{ah}{right};
        \roundArrow{ah}{bh}{b}{right};
        \roundArrow{a}{ah}{eh}{right};
        \roundArrow{a}{d}{e1}{left};
        \roundArrow{a}{b}{c1}{left};
        \roundArrow{a}{c}{d1}{left};
        \roundArrow{a}{e}{b1}{left};
    \end{tikzpicture}
}

\newcommand{\evaluationLine}[0]{
    \begin{tikzpicture}[scale=1,baseline={([yshift=-.5ex]current bounding box.center)},z={(0,1)},y={(0,0.3)}]
        \def\l{3.5};
        \def\h{2.7};
        \def\tb{360};
        \def\tc{120};
        \def\td{180};
        \def\te{300};       
        \coordinate (a) at (0,0,0);
        \coordinate (b) at ({\l*cos(\tb)},{\l*sin(\tb)},0);
        \coordinate (b1) at ({\l*cos(\te+(\tb-\te)/2)},{\l*sin(\te+(\tb-\te)/2)},0);
        \coordinate (c) at ({\l*cos(\tc)},{\l*sin(\tc)},0);
        \coordinate (c1) at ({\l*cos(\tc/2)},{\l*sin(\tc/2)},0);
        \coordinate (d) at ({\l*cos(\td)},{\l*sin(\td)},0);
        \coordinate (d1) at ({\l*cos(\tc+0.5*(\td-\tc))},{\l*sin(\tc+0.5*(\td-\tc))},0);
        \coordinate (e) at ({\l*cos(\te)},{\l*sin(\te)},0);
        \coordinate (e1) at ({\l*cos(\td+(\te-\td)/2)},{\l*sin(\td+(\te-\td)/2)},0);
        \coordinate (at) at (0,0,\h/1.543);
        \coordinate (ah) at (0,0,\h);
        \coordinate (bt) at ({.5*\l*cos(\tb)},{.5*\l*sin(\tb)},0);
        \coordinate (ct) at ({.5*\l*cos(\tc)},{.5*\l*sin(\tc)},0);
        \coordinate (dt) at ({.5*\l*cos(\td)},{.5*\l*sin(\td)},0);
        \coordinate (et) at ({.5*\l*cos(\te)},{.5*\l*sin(\te)},0);
        \coordinate (ch) at ({\l*cos(\tc)},{\l*sin(\tc)},\h);
        \coordinate (eh) at ({\l*cos(\te)},{\l*sin(\te)},\h);

        \clip (-3.6,-4.8,0) rectangle (3.6,12.3,0);
        \fill[color=BlueViolet, opacity=.1] (b) -- (c1) -- (c) -- (d1) -- (d) -- (e1) -- (e) -- (b1) -- cycle;

        \node at (-2,-1.5,0) {${\sss M_1}$};
        \node at (2,-1.5,0) {${\sss M_4}$};
        \node at (-2.5,1.1,0) {${\sss M_3}$};
        \node at (2.3,1.1,0) {${\sss M_2}$};
        \node at (-1.4,1,2.6) {${\sss X}$};

        \draw[fat, ->-=.4] (et) to[out=90, in=88,in looseness=1.8, out looseness=4.3] node[mpoint, pos=.05](mp1){} node[mpoint, pos=.1](mp2){} (ct);
        
        \clipAroundTwoNodes{mp1}{mp2}{
            \draw[fat, color=BrickRed] (a) -- (bt);
            \draw[edge, color=BrickRed] (bt) -- (b);
            \draw[edge, ->-=.75] (c) -- (a);
            \draw[edge, color=BrickRed] (d) -- (dt);
            \draw[fat, color=BrickRed, ->-=.55] (a) --node[pos=.25, below,text=black](){${\sss Y}$} (dt);
            \draw[fat, ->-=.07, ->-=.35, ->-=.57, ->-=.8] ({\l/2*cos(0)},{\l/2*sin(0)},0) arc(0:-360:\l/2);
            \draw[edge] (a) -- (e);
        }
        \foreach \x in {c,e}{
            \fill[gray, opacity=.08] (a) -- (\x) -- ($(\x)+(0,0,\h)$) -- (ah) -- cycle; 
        }
        \fill[color=LimeGreen, opacity=.1] (bt) arc(0:-180:\l/2) to[out=90,in=180,looseness=1] (at) to[out=0, in=90, looseness=1] (bt);
        \fill[color=LimeGreen, opacity=.1] (bt) arc(0:-180:\l/2) to[out=-90,in=180,looseness=1] ($(a)+(0,0,-\h/2)$) to[out=0, in=-90, looseness=1] (bt);

        \node[xshift=4pt, yshift=4pt] at (0,0,0) {${\sss \msf c}$};
        \roundArrow{c}{ch}{ah}{right};
        \roundArrow{a}{d}{e1}{left};
        \roundArrow{a}{b}{c1}{left};
        \roundArrow{a}{c}{d1}{left};
        \roundArrow{a}{e}{b1}{left};
    \end{tikzpicture}
}

\newcommand{\planarDiagram}[1]{
    \begin{tikzpicture}[scale=.95,baseline={([yshift=-.5ex]current bounding box.center)}]
        \ifthenelse{#1=1}{
            \def\l{3.5};
            \def\tc{130};
            \def\te{290}; 
            \fill[color=LimeGreen, opacity=.1, rounded corners=2pt] (0,0) rectangle (\l,\l);
            \draw[obj, ->-=.11, ->-=.37, ->-=.58, ->-=.85] (5*\l/6,\l/2) arc(0:-360:\l/3);
            \draw[obj, ->-=.3, ->-=.8] (5*\l/6,\l/2) --node[above, pos=.8](){${\sss X_1}$} node[below,pos=.3](){${\sss X_3}$} (\l/6,\l/2);
            \draw[obj, ->-=.55]  (\l/2,\l/2) --node[pos=.6, right](){${\sss X_2}$} ($(\l/2,\l/2)+({\l/3*cos(\tc)},{\l/3*sin(\tc)})$);
            \draw[obj, ->-=.55] ($(\l/2,\l/2)+({\l/3*cos(\te)},{\l/3*sin(\te)})$) --node[pos=.5, left](){${\sss X_4}$} (\l/2,\l/2);
            \node[mor, circle, inner sep=3.3pt] at (\l/6,\l/2) {};
            \node[mor, circle, inner sep=3.3pt] at (5*\l/6,\l/2) {};
            \node[mor, circle, inner sep=3.3pt] at (\l/2,\l/2) {};
            \node[xshift=5pt, yshift=5pt] at (\l/2,\l/2) {${\sss \msf v}$};
            \node[mor, circle, inner sep=3.3pt] at ($(\l/2,\l/2)+({\l/3*cos(\tc)},{\l/3*sin(\tc)})$) {};
            \node[mor, circle, inner sep=3.3pt] at ($(\l/2,\l/2)+({\l/3*cos(\te)},{\l/3*sin(\te)})$) {};
            \node at ($(\l/2,\l/2)+({1.3*\l/3*cos(65)},{1.2*\l/3*sin(65)})$) {${\sss M_3}$};
            \node at ($(\l/2,\l/2)+({1.3*\l/3*cos(155)},{1.3*\l/3*sin(155)})$) {${\sss M_4}$};
            \node at ($(\l/2,\l/2)+({1.3*\l/3*cos(235)},{1.3*\l/3*sin(235)})$) {${\sss M_1}$};
            \node at ($(\l/2,\l/2)+({1.3*\l/3*cos(325)},{1.3*\l/3*sin(325)})$) {${\sss M_2}$};
        }{}
        \ifthenelse{#1=2}{
            \def\l{3.5};
            \def\tb{30}
            \def\tc{150};
            \def\td{270}; 
            \fill[color=LimeGreen, opacity=.1, rounded corners=2pt] (0,0) rectangle (\l,\l);
            \draw[obj, ->-=.10, ->-=.43, ->-=.76] (5*\l/6,\l/2) arc(0:-360:\l/3);
            \draw[obj, ->-=.55]  (\l/2,\l/2) --node[pos=.6, yshift=-8pt](){${\sss V_2}$} ($(\l/2,\l/2)+({\l/3*cos(\tb)},{\l/3*sin(\tc)})$);
            \draw[obj, ->-=.55]  (\l/2,\l/2) --node[pos=.6, yshift=-8pt](){${\sss V_1}$} ($(\l/2,\l/2)+({\l/3*cos(\tc)},{\l/3*sin(\tc)})$);
            \draw[obj, ->-=.55] ($(\l/2,\l/2)+({\l/3*cos(\td)},{\l/3*sin(\td)})$) --node[pos=.5, left](){${\sss V_3}$} (\l/2,\l/2);
            \node[mor, circle, inner sep=3.3pt] at (\l/2,\l/2) {};
            \node[yshift=8pt] at (\l/2,\l/2) {${\sss \msf v}$};
            \node[mor, circle, inner sep=3.3pt] at ($(\l/2,\l/2)+({\l/3*cos(\tb)},{\l/3*sin(\tb)})$) {};
            \node[mor, circle, inner sep=3.3pt] at ($(\l/2,\l/2)+({\l/3*cos(\tc)},{\l/3*sin(\tc)})$) {};
            \node[mor, circle, inner sep=3.3pt] at ($(\l/2,\l/2)+({\l/3*cos(\td)},{\l/3*sin(\td)})$) {};
            \node at ($(\l/2,\l/2)+({1.3*\l/3*cos(90)},{1.2*\l/3*sin(90)})$) {${\sss M_3}$};
            \node at ($(\l/2,\l/2)+({1.3*\l/3*cos(-30)},{1.3*\l/3*sin(-30)})$) {${\sss M_2}$};
            \node at ($(\l/2,\l/2)+({1.3*\l/3*cos(210)},{1.3*\l/3*sin(210)})$) {${\sss M_1}$};
        }{}
        \ifthenelse{#1=3}{
            \def\l{3.5};
            \def\tb{90}
            \def\tc{180};
            \def\td{270}; 
            \fill[color=LimeGreen, opacity=.1, rounded corners=2pt] (0,0) rectangle (\l,\l);
            \draw[obj, ->-=.14, ->-=.39, ->-=.64, ->-=.89] (5*\l/6,\l/2) arc(0:-360:\l/3);
            \draw[speObj, ->-=.75] ($(\l/2,\l/2)+({\l/3*cos(0)},{\l/3*sin(0)})$) --node[mpoint, pos=.5](mp1){} node[pos=.3,below,text=black](){${\sss Y}$} ($(\l/2,\l/2)+({\l/3*cos(\tc)},{\l/3*sin(\tc)})$); 
            \clipAroundNode{mp1}{
                \draw[obj, ->-=.3] ($(\l/2,\l/2)+({\l/3*cos(\td)},{\l/3*sin(\td)})$) -- ($(\l/2,\l/2)+({\l/3*cos(\tb)},{\l/3*sin(\tb)})$) node[pos=.75, left](){${\sss X}$}; 
            }

            \node[xshift=5pt, yshift=5pt] at (\l/2,\l/2) {${\sss \msf c}$};
            \foreach \t in {0,\tb,\tc,\td}{
                \node[mor, circle, inner sep=3.3pt] at ($(\l/2,\l/2)+({\l/3*cos(\t)},{\l/3*sin(\t)})$) {};
            }
            \node at ($(\l/2,\l/2)+({1.3*\l/3*cos(45)},{1.2*\l/3*sin(45)})$) {${\sss M_2}$};
            \node at ($(\l/2,\l/2)+({1.3*\l/3*cos(135)},{1.3*\l/3*sin(135)})$) {${\sss M_3}$};
            \node at ($(\l/2,\l/2)+({1.3*\l/3*cos(225)},{1.3*\l/3*sin(225)})$) {${\sss M_1}$};
            \node at ($(\l/2,\l/2)+({1.3*\l/3*cos(315)},{1.3*\l/3*sin(315)})$) {${\sss M_4}$};
        }{}
    \end{tikzpicture}
}

\newcommand{\halfBraiding}[2]{
	\begin{tikzpicture}[scale=1,baseline={([yshift=-.5ex]current bounding box.center)}]
    	\def\dx{1.1};
    	\def\dy{1.5};
    	\begin{pgfonlayer}{front}
    	    \draw[speObj,wig] (0,0) --node[mpoint, pos=0.5](mk){} (-\dx,\dy);
    	\end{pgfonlayer}
    	\clipAroundNode{mk}{
    	    \draw[speObj] (0,\dy) -- (-\dx,0);
    	}
    	\node[below] at (0,0) {${\sss #1}$};
    	\node[above] at (-\dx,\dy) {${\sss #1}$};
    	\node[below] at (-\dx,0) {${\sss #2}$};
    	\node[above] at (0,\dy) {${\sss #2}$};
	\end{tikzpicture}
}

\newcommand{\resolution}[6]{
    \begin{tikzpicture}[scale=1,baseline={([yshift=-.5ex]current bounding box.center)}]
        \def\dx{1.1};
        \def\dy{1.5};
        \def\rr{0.3};
        \coordinate (a) at ($(0.5*\dx,0.5*\dy)+(\rr*0.5*\dx,-\rr*0.5*\dy)$);
        \coordinate (b) at ($(0.5*\dx,0.5*\dy)+(-\rr*0.5*\dx,\rr*0.5*\dy)$);
        \draw[speObj] (\rr*\dx,0) -- (a);
        \draw[speObj] ({(1-\rr)*\dx},\dy) -- (b);
        \draw[speObj] (0,\dy) -- (\dx,0);
        \node[below] at (\rr*\dx,0) {${\sss #3}$};
        \node[above] at ({(1-\rr)*\dx},\dy) {${\sss #1}$};
        \node[xshift=6pt, yshift=4pt] at (0.5*\dx,0.5*\dy) {${\sss #2}$};
        \node[below] at (\dx,0) {${\sss #4}$};
        \node[above] at (0,\dy) {${\sss #4}$};
        \mor{a}{#6};
        \mor{b}{#5};
    \end{tikzpicture}
} 

\newcommand{\networkLines}[2]{
    \begin{tikzpicture}[scale=1,baseline={([yshift=-.5ex]current bounding box.center)}]
        \fill[color=BlueViolet, opacity=.1] (0,0) rectangle (2,2);

        \ifthenelse{\equal{#1}{}}{}{
           \draw[speObj,wig] (0,1) --node[mpoint, pos=0.5](mk){} node[pos=.25,above](){${\sss #1}$}(2,1); 
        }
        \ifthenelse{\equal{#2}{}}{}{
            \ifthenelse{\equal{#1}{}}{
                \draw[speObj] (1,0) --node[right, pos=.25](){${\sss #2}$} (1,2);
            }{
                \clipAroundNode{mk}{
                    \draw[speObj] (1,0) --node[right, pos=.25](){${\sss #2}$} (1,2);
                }   
            }  
        }
        \path[->-=.25, color=BlueViolet!50] (0,0) -- (2,0);
        \path[->-=.25, color=BlueViolet!50] (0,2) -- (2,2);
        \path[->-=.24, ->-=.29, color=BlueViolet!50] (0,0) -- (0,2);
        \path[->-=.24, ->-=.29, color=BlueViolet!50] (2,0) -- (2,2);
    \end{tikzpicture}
} 

\newcommand{\cellDecompositionW}[0]{
	\begin{tikzpicture}[scale=1,baseline={([yshift=-.5ex]current bounding box.center)},z={(0,1)},y={(-0.25,0.25)}]
        \def\l{1};
        \def\h{-2.7};
        \coordinate (a) at (0,0,0);
        \coordinate (b) at (1.8*\l,1.8*\l,0);
        \coordinate (c) at (1.5*\l,4*\l,0);
        \coordinate (d) at (0,5*\l,0);
        \coordinate (e) at (-2*\l,2.5*\l,0);
        \coordinate (f) at (-1.8*\l,0,0);
        \coordinate (a1) at (0.5*\l,-1*\l,0);
        \coordinate (b1) at (2.8*\l,1.8*\l,0);
        \coordinate (c1) at (2.5*\l,4*\l,0);
        \coordinate (d1) at (-.18*\l,5.9*\l,0);
        \coordinate (e1) at (-3*\l,2.5*\l,0);
        \coordinate (f1) at (-1.8*\l,-1.5*\l,0);
        \coordinate (f2) at (-2.8*\l,0.6*\l,0);
        \coordinate (a1t) at ($(a1)+(0,0,\h)$);

        \foreach \p/\q in {a/at, b/bt, c/ct, d/dt, e/et, f/ft} {
			\coordinate (\q) at ($(\p)+(0,0,\h)$);
        }
        
        \draw[edge, dot] (-4*\l,7*\l,0) -- (-4*\l,-2.2*\l,0) -- (3.5*\l,-2.2*\l,0) -- (3.5*\l,7*\l,0);
        \node[right] at (3.5*\l,-2*\l,0) {${\Sigma_\Upsilon \times \{2\}}$};
        \node[right] at (3.5*\l,-2*\l,\h) {${\Sigma_\Upsilon \times \{1\}}$};
        \draw[edge, dot] (-4*\l,7*\l,\h) -- (-4*\l,-2.2*\l,\h) -- (3.5*\l,-2.2*\l,\h) -- (3.5*\l,7*\l,\h);

        \draw[edge] (f) -- (f2);

        \draw[edge] (a) -- (b) -- (c) -- (d) -- (e) -- (f) --cycle;
        \draw[edge] (c) -- (e);
        
        \foreach \x in {a,b,c,d,e,f}{
        	\draw (\x) -- (\x1);
        }
        
        \node[name=mp1] at (intersection of d--dt and c--e) {};
        \node[name=mp2] at (intersection of d--dt and f--a) {};
        \node[name=mp3] at (intersection of c--ct and a--b) {};
        \node[name=mp4] at (intersection of c--ct and a--a1) {};
        \node[name=mp5] at (intersection of e--et and f--f2) {};
        
        \clipAroundFourNodes{mp1}{mp2}{mp3}{mp4}{
        	\foreach \x/\y in {a/at, b/bt, c/ct, d/dt, f/ft}{
        		\draw[edge] (\x) -- (\y);
        	}
        }
        
        \clipAroundNode{mp5}{
        	\draw[edge] (e) -- (et);
        }

        \node[draw, fill=white, circle, inner sep=.5pt] at ($(c)!0.6!(ct)$) {{\small $\vartheta$}};
        \node[draw, fill=white, circle, inner sep=.5pt] at ($(d)!0.7!(dt)$) {{\small $\vartheta$}};
        \node[draw, fill=white, circle, inner sep=.5pt] at ($(e)!0.5!(et)$) {{\small $\vartheta$}};

        \foreach \x/\y in {d/d1,c/c1,e/e1,c/d,d/e,c/e,b/c,e/f,a/b,f/a,b/b1,a/a1,f/f1,f/f2}{
            \fill[gray, opacity=.08] (\x) -- (\y) -- ($(\y)+(0,0,\h)$) -- ($(\x)+(0,0,\h)$) -- cycle;  
        }
        \node[draw, fill=white, circle, inner sep=.5pt] at ($(a)!0.5!(at)$) {{\small $\vartheta$}};
        \node[draw, fill=white, circle, inner sep=.5pt] at ($(b)!0.5!(bt)$) {{\small $\vartheta$}};
        \node[draw, fill=white, circle, inner sep=.5pt] at ($(f)!0.5!(ft)$) {{\small $\vartheta$}};

        \node[draw, fill=white, circle, inner sep=.5pt] at ($(a)!0.6!(b)$) {{\small $\tht$}};
        \node[draw, fill=white, circle, inner sep=.5pt] at ($(b)!0.5!(c)$) {{\small $\tht$}};
        \node[draw, fill=white, circle, inner sep=.5pt] at ($(c)!0.5!(d)$) {{\small $\tht$}};
        \node[draw, fill=white, circle, inner sep=.5pt] at ($(d)!0.5!(e)$) {{\small $\tht$}};
        \node[draw, fill=white, circle, inner sep=.5pt] at ($(e)!0.5!(f)$) {{\small $\tht$}};
        \node[draw, fill=white, circle, inner sep=.5pt] at ($(f)!0.5!(a)$) {{\small $\tht$}};
        \node[draw, fill=white, circle, inner sep=.5pt] at ($(c)!0.45!(e)$) {{\small $\tht$}};
        
        \fill[color=BurntOrange, opacity=.1] (-4*\l,7*\l,0) -- (-4*\l,-2.2*\l,0) -- (3.5*\l,-2.2*\l,0) -- (3.5*\l,7*\l,0) --cycle;

    \end{tikzpicture}
}

\newcommand{\algebraObj}[1]{
    \begin{tikzpicture}[scale=0.6,baseline={([yshift=-1ex]current bounding box.center)}]
        \def\l{1.5};
        \ifthenelse{#1 = 1}{
            \draw[obj] (0,-1) -- (0,1);
            \draw[obj] (0,1) -- (-\l,3);
            \draw[obj] (0,1) -- (\l,3);
            \node[mor, circle, inner sep=3.3pt] at (-\l/2,2) {};
            \node[mor, circle, inner sep=3.3pt] at (\l/2,2) {};
            \node[mor, circle, inner sep=3.3pt] at (0,-.12) {};
            \node at (0,-.12) {${\sss l}$};
            \node at (\l/2,2) {${\sss k}$};
            \node at (-\l/2,2) {${\sss j}$};
            \node[draw, rectangle, fill=white, inner sep=2pt, rounded corners=2pt] at (0,1) {${\sss \mu}$};
            \node at (-0.65,1.3) {${\sss \mc A}$};
            \node at (0.65,1.3) {${\sss \mc A}$};
            \node[right] at (0,0.4) {${\sss \mc A}$};
        }{}
        \ifthenelse{#1 = 2}{
            \draw[obj] (0,-1) -- (0,1);
            \draw[obj] (0,1) -- (-\l,3);
            \draw[obj] (0,1) -- (\l,3);
            \node[mor, circle, inner sep=3.3pt] at (0,1) {};
            \node at (0,1) {${\sss i}$};
        }{}
        \node[above] at (-\l,3) {${\sss X_1}$};
        \node[above] at (\l,3) {${\sss X_2}$};
        \node[below] at (0,-1) {${\sss X_3}$};
    \end{tikzpicture}
}

\newcommand{\sandwich}[0]{
    \begin{tikzpicture}[scale=.85,baseline={([yshift=-.5ex]current bounding box.center)},z={(0,1)},y={(-0.25,0.25)}]
        \def\l{1};
        \def\h{2.7};
        \coordinate (a) at (0,0,0);
        \coordinate (b) at (1.8*\l,1.8*\l,0);
        \coordinate (c) at (1.5*\l,4*\l,0);
        \coordinate (d) at (0,5*\l,0);
        \coordinate (e) at (-2*\l,2.5*\l,0);
        \coordinate (f) at (-1.8*\l,0,0);
        \coordinate (a1) at (0.5*\l,-1*\l,0);
        \coordinate (b1) at (2.8*\l,1.8*\l,0);
        \coordinate (c1) at (2.5*\l,4*\l,0);
        \coordinate (d1) at (-.18*\l,5.9*\l,0);
        \coordinate (e1) at (-3*\l,2.5*\l,0);
        \coordinate (f1) at (-1.8*\l,-1.5*\l,0);
        \coordinate (f2) at (-2.8*\l,0.6*\l,0);
        \coordinate (ft) at ($(f)+(0,0,\h)$);
        \coordinate (at) at ($(a)+(0,0,\h)$);
        \coordinate (a1t) at ($(a1)+(0,0,\h)$);
        \coordinate (bt) at ($(b)+(0,0,\h)$);

        \fill[color=BlueViolet, opacity=.1] (-4*\l,7*\l,0) -- (-4*\l,-2.2*\l,0) -- (3.5*\l,-2.2*\l,0) -- (3.5*\l,7*\l,0) --cycle;
        \draw[edge, dot] (-4*\l,7*\l,0) -- (-4*\l,-2.2*\l,0) -- (3.5*\l,-2.2*\l,0) -- (3.5*\l,7*\l,0);
        \node[right] at (3.5*\l,-2*\l,0) {${\Sigma_\Upsilon \times \{0\}}$};
        \node[right] at (3.5*\l,-2*\l,\h) {${\Sigma_\Upsilon \times \{1\}}$};
        \draw[edge, dot] (-4*\l,7*\l,\h) -- (-4*\l,-2.2*\l,\h) -- (3.5*\l,-2.2*\l,\h) -- (3.5*\l,7*\l,\h);

        \draw[edge] (f) -- (f2);
        \foreach \x in {a,b,c,d,e,f}{
            \draw[edge] (\x) -- ($(\x)+(0,0,\h)$);
        }
        \path (a) --node[mpoint, pos=.35](mp1){} ($(a)+(0,0,\h)$);
        \path (f) --node[mpoint, pos=.15](mp3){} node[mpoint, pos=.45](mp4){} ($(f)+(0,0,\h)$);
        \path (b) --node[mpoint, pos=.2](mp2){} ($(b)+(0,0,\h)$);
        \clipAroundFourNodes{mp1}{mp2}{mp3}{mp4}{
            \foreach \x in {a,b,c,d,e,f}{
                \draw (\x) -- (\x1);
            }
            \draw[edge] (a) -- (b) -- (c) -- (d) -- (e) -- (f) --cycle;
            \draw[edge] (c) -- (e);
        }
        \foreach \x/\y in {d/d1,c/c1,e/e1,c/d,d/e,c/e,b/c,e/f,a/b,f/a,b/b1,a/a1,f/f1,f/f2}{
            \fill[gray, opacity=.08] (\x) -- (\y) -- ($(\y)+(0,0,\h)$) -- ($(\x)+(0,0,\h)$) -- cycle;  
        }

        \def\sh{7.3};
        \begin{scope}
            \def\h{-2.7};
            \foreach \x in {a,b,c,d,e,f,a1,b1,c1,d1,e1,f1,f2}{
               \coordinate (\x) at ($(\x)+(0,0,\sh)$); 
            }
            \coordinate (a1t) at ($(a1)+(0,0,\h)$);
    
            \foreach \p/\q in {a/at, b/bt, c/ct, d/dt, e/et, f/ft} {
                \coordinate (\q) at ($(\p)+(0,0,\h)$);
            }
            
            \draw[edge, dot] (-4*\l,7*\l,\sh) -- (-4*\l,-2.2*\l,\sh) -- (3.5*\l,-2.2*\l,\sh) -- (3.5*\l,7*\l,\sh);
            \node[right] at (3.5*\l,-2*\l,\sh) {${\Sigma_\Upsilon \times \{2\}}$};
            \node[right] at (3.5*\l,-2*\l,\h+\sh) {${\Sigma_\Upsilon \times \{1\}}$};
            \draw[edge, dot] (-4*\l,7*\l,\h+\sh) -- (-4*\l,-2.2*\l,\h+\sh) -- (3.5*\l,-2.2*\l,\h+\sh) -- (3.5*\l,7*\l,\h+\sh);
    
            \draw[edge] (f) -- (f2);
            \draw[edge] (a) -- (b) -- (c) -- (d) -- (e) -- (f) --cycle;
            \draw[edge] (c) -- (e);
            
            \foreach \x in {a,b,c,d,e,f}{
                \draw (\x) -- (\x1);
            }
            
            \node[name=mp1] at (intersection of d--dt and c--e) {};
            \node[name=mp2] at (intersection of d--dt and f--a) {};
            \node[name=mp3] at (intersection of c--ct and a--b) {};
            \node[name=mp4] at (intersection of c--ct and a--a1) {};
            \node[name=mp5] at (intersection of e--et and f--f2) {};
            
            \clipAroundFourNodes{mp1}{mp2}{mp3}{mp4}{
                \foreach \x/\y in {a/at, b/bt, c/ct, d/dt, f/ft}{
                    \draw[edge] (\x) -- (\y);
                }
            }
            
            \clipAroundNode{mp5}{
                \draw[edge] (e) -- (et);
            }
    
            \node[draw, fill=white, circle, inner sep=.5pt] at ($(c)!0.6!(ct)$) {${\sss \vartheta}$};
            \node[draw, fill=white, circle, inner sep=.5pt] at ($(d)!0.7!(dt)$) {${\sss \vartheta}$};
            \node[draw, fill=white, circle, inner sep=.5pt] at ($(e)!0.5!(et)$) {${\sss \vartheta}$};
    
            \foreach \x/\y in {d/d1,c/c1,e/e1,c/d,d/e,c/e,b/c,e/f,a/b,f/a,b/b1,a/a1,f/f1,f/f2}{
                \fill[gray, opacity=.08] (\x) -- (\y) -- ($(\y)+(0,0,\h)$) -- ($(\x)+(0,0,\h)$) -- cycle;  
            }
            \node[draw, fill=white, circle, inner sep=.5pt] at ($(a)!0.5!(at)$) {${\sss \vartheta}$};
            \node[draw, fill=white, circle, inner sep=.5pt] at ($(b)!0.5!(bt)$) {${\sss \vartheta}$};
            \node[draw, fill=white, circle, inner sep=.5pt] at ($(f)!0.5!(ft)$) {${\sss \vartheta}$};

            \node[draw, fill=white, circle, inner sep=.5pt] at ($(a)!0.6!(b)$) {${\sss \tht}$};
            \node[draw, fill=white, circle, inner sep=.5pt] at ($(b)!0.5!(c)$) {${\sss \tht}$};
            \node[draw, fill=white, circle, inner sep=.5pt] at ($(c)!0.5!(d)$) {${\sss \tht}$};
            \node[draw, fill=white, circle, inner sep=.5pt] at ($(d)!0.5!(e)$) {${\sss \tht}$};
            \node[draw, fill=white, circle, inner sep=.5pt] at ($(e)!0.5!(f)$) {${\sss \tht}$};
            \node[draw, fill=white, circle, inner sep=.5pt] at ($(f)!0.5!(a)$) {${\sss \tht}$};
            \node[draw, fill=white, circle, inner sep=.5pt] at ($(c)!0.45!(e)$) {${\sss \tht}$};
            
            \fill[color=BurntOrange, opacity=.1] (-4*\l,7*\l,\sh) -- (-4*\l,-2.2*\l,\sh) -- (3.5*\l,-2.2*\l,\sh) -- (3.5*\l,7*\l,\sh) --cycle;
        \end{scope}
        \draw [decorate,decoration={brace,amplitude=4pt},xshift=0pt,yshift=0pt]
            (-4*\l,2.4*\l,-0.3) -- (-4*\l,2.4*\l,\h+0.3) node [midway,left,xshift=-0.5em,yshift=0.3em] {${\la \text{topological} |}$}; 
        \draw [decorate,decoration={brace,amplitude=4pt},xshift=0pt,yshift=0pt]
            (-4*\l,2.4*\l,-\h+\sh-0.3) -- (-4*\l,2.4*\l,\sh+0.3) node [midway,left,xshift=-0.5em,yshift=0.3em] {${| \text{boundary} \ra}$};         
    \end{tikzpicture}
}

\title{\boldmath\smaller Symmetry topological field theory  
and non-abelian Kramers--Wannier dualities  of generalised Ising models}

\author[]{Clement Delcamp,*}
\author[]{Nafiz Ishtiaque}

\affiliation[]{Institut des Hautes \'Etudes Scientifiques, Bures-sur-Yvette, France}

\emailAdd{delcamp@ihes.fr}
\emailAdd{ishtiaque@ihes.fr}

\abstract{\\~\\ For a class of two-dimensional Euclidean lattice field theories admitting topological lines encoded into a spherical fusion category, we explore aspects of their realisations as boundary theories of a three-dimensional topological quantum field theory. After providing a general framework for explicitly constructing such realisations, we specialise to non-abelian generalisations of the Ising model and consider two operations: gauging an arbitrary subsymmetry and performing Fourier transforms of the local weights encoding the dynamics of the theory. These are carried out both in a traditional way and in terms of the three-dimensional topological quantum field theory. Whenever the whole symmetry is gauged, combining both operations recovers the non-abelian Kramers--Wannier duals \`a la Freed and Teleman of the generalised Ising models. Moreover, we discuss the interplay between renormalisation group fixed points of gapped symmetric phases and these generalised Kramers--Wannier dualities.  
}

\begin{document} 
	\vspace*{-2em}
	\maketitle
	\flushbottom
	\newpage
	
\section{Introduction}

Noether's theorem assigns to a global \emph{continuous invertible 0-form symmetry} in a quantum field theory a collection of \emph{one-codimensional} \emph{topological} defects, such that correlation functions involving networks of these defects are insensitive to deformations preserving the topology of the underlying submanifold, unless they are pulled through charged operators. Conversely, a modern perspective identifies the existence of a collection of topological defects in a quantum field theory as the defining property of a (global) symmetry. This viewpoint has led to various generalisations of the ordinary notion of symmetry---typically referred to as \emph{generalised symmetries}---whereby defects are not necessarily one-codimensional and the accompanied transformations of the fields are not necessarily invertible or continuous \cite{Gaiotto:2014kfa,Freed:2022qnc}.

While ordinary global symmetries are encoded into groups, generalised symmetries are encoded into higher mathematical structures. In (1+1) dimensions, the requirement that we must be able to construct junctions of topological defects, together with finiteness and semisimplicity assumptions, suggests an axiomatisation in terms of \emph{(spherical) fusion categories} \cite{Bhardwaj:2017xup,Chang:2018iay,Thorngren:2019iar,PhysRevResearch.2.043086,Schafer-Nameki:2023jdn,Shao:2023gho}. Though somewhat exotic, such generalised symmetries are not uncommon in (1+1)-dimensional physical systems. For instance, \emph{Verlinde lines} in a (1+1)d rational Conformal Field Theory (CFT) \cite{Verlinde:1988sn, PETKOVA2001157}, which fuse according to the representation theory of the underlying chiral vertex algebra, provide examples of typically non-invertible topological defects. A particularly celebrated example is that of the \emph{Kramers--Wannier} duality line in the diagonal \emph{Ising} CFT \cite{Oshikawa:1996dj,Oshikawa:1996ww,Frohlich:2004ef,Frohlich:2006ch}. Another common source of examples are ordinary global symmetries with respect to \emph{non-abelian} groups, which upon \emph{gauging} produce non-invertible topological lines labelled by representations of the group \cite{Tachikawa:2017gyf, Bhardwaj:2017xup, Delcamp:2021szr,Bhardwaj:2022lsg,Bartsch:2022mpm,Bhardwaj:2022yxj}.

Within this paradigm, a framework has recently emerged enabling a calculus of topological defects that leverages well-established methods from Topological Quantum Field Theory (TFQT). Fundamentally, the idea is that a theory with a given finite symmetry can be realised as a \emph{boundary theory} of a higher-dimensional \emph{fully extended} TQFT hosting a gauged version of this symmetry. An early incarnation of this mechanism within the context of two-dimensional rational CFTs was developed in ref.~\cite{Fuchs:2002cm, FUCHS2004277, FUCHS2004511, FUCHS2005539}. More recently, it has either been referred to as `topological holography' \cite{KONG201762,PhysRevResearch.2.043086,PhysRevResearch.2.033417,Ji:2021esj,Chatterjee:2022kxb,Huang:2023pyk,Huang:2024ror}, `strange correlators' \cite{PhysRevLett.112.247202,PhysRevLett.121.177203}, `symmetry Topological Field Theory (SymTFT)' \cite{Apruzzi:2021nmk,Kaidi:2022cpf,Kaidi:2023maf}, or `sandwich construction' \cite{Freed:2022qnc}, depending on the physical context. The upshot is that all the symmetry aspects of a theory can be understood in terms of this higher-dimensional TQFT, e.g. 
anomalies, generalised charges, twisted sectors, gapped symmetric phases, order/disorder operators, phase transitions, etc. Most relevant to the present work, it makes gauging non-invertible (but also invertible) symmetries particularly convenient \cite{Gaiotto:2020iye,Lootens:2021tet,Delcamp:2021szr,Diatlyk:2023fwf}.

The exploitation of this calculus of topological defects has mainly focused on the continuous setting.  Nevertheless, there have been remarkable developments within the context of one-dimensional quantum lattice models, where topological defects act as linear operators on some microscopic Hilbert space. A particularly convenient way of approaching (1+1)d lattice models with non-invertible symmetries is to employ the \emph{anyonic chain} construction, which has the merit of clearly distinguishing an \emph{abstract symmetry}---defined in a completley intrinsic way---from its concrete realisation in a given physical system. First introduced for the case of topological lines encoded into the \emph{Fibonacci} modular tensor category \cite{Feiguin:2006ydp}, it was subsequently extended to ${\rm su}(2)_k$ anyonic theories in ref.~\cite{PhysRevB.87.235120}, and later to the case of an arbitrary fusion category in ref.~\cite{Buican:2017rxc}. This framework can be used for instance to realise fixed point Hamiltonians of (1+1)d gapped phases preserving (non-anomalous) symmetries encoded into fusion categories \cite{Inamura:2021szw}. Recently, a generalisation of the anyonic chain construction was introduced in ref.~\cite{Lootens:2021tet,Lootens:2022avn,Lootens:2023wnl}, realising many aspects of the SymTFT philosophy in one-dimensional quantum lattice models. Amongst others, this generalisation presents two overlapping advantages over the original construction: (i) It is more versatile, facilitating the task of writing a given quantum model as an anyonic chain, and thus enabling a systematic study of its symmetry. (ii) It makes the implementation of generalised gauging procedures particularly straightforward, which upon a careful treatment of boundary conditions and charge sectors can be promoted to \emph{dualities}. Further developments were presented in ref.~\cite{Bhardwaj:2024kvy}, and preliminary results regarding the higher-dimensional scenario were already obtained in ref.~\cite{Delcamp:2023kew,Inamura:2023qzl,Moradi:2023dan,Choi:2024rjm}. 

\bigskip\noindent
The goal of the present manuscript is to apply the philosophy of ref.~\cite{Lootens:2021tet, Lootens:2022avn,Bhardwaj:2024kvy} to \emph{2d Euclidean lattice field theories}. Many aspects of the SymTFT philosophy have already been implemented within this context, see e.g. ref.~\cite{Aasen:2016dop,Freed:2018cec,PhysRevLett.121.177203,Aasen:2020jwb,Vanhove:2021nav}, reminiscing of ideas that originated in the study of \emph{integrable} systems
\cite{Temperley:1971iq,Pasquier1988,Roche1990}. Our work complements previous constructions in the following ways: (i) We introduce a systematic and concrete way of realising a large class of physical models as boundary theory of a fully extended TQFT. (ii) We explicitly perform the coupling of the resulting partition functions with topological lines encoding the symmetry. (iii) We compute the effective field theory obtained after gauging any subsymmetry, as well as the resulting topological lines. (iv) We discuss the interplay between the aforementioned aspects and the notion that the same symmetric theory can be realised as a boundary theory of different TQFTs.

Concretely, consider a cell decomposition $\Sigma_\Upsilon$ of an oriented two-dimensional surface $\Sigma$ and a spherical fusion category $\mc C$. The spherical fusion category $\mc C$ serves as input datum of a state sum TQFT via the Turaev--Viro--Barrett--Westbury construction \cite{Turaev:1992hq,Barrett:1993ab}. After picking a cell decomposition of the manifold $\Sigma_\Upsilon \times [0,1]$, we impose a topological brane boundary condition on $\Sigma_\Upsilon \times \{0\}$, and a (typically non-topological) `physical' boundary condition on $\Sigma_\Upsilon \times \{1\}$. Generally, topological brane boundary conditions are solely encoded into \emph{module categories} over $\mc C$ \cite{kongBdries,Fuchs:2012dt}. For now, let us suppose that we impose the so-called \emph{Dirichlet} brane boundary condition associated with the module category $\mc C$ over itself, which is the only choice that can always be made regardless of $\mc C$. Amongst the bulk topological line operators of the TQFT, only those encoded into $\mc C$ itself do not condense along $\Sigma_\Upsilon \times \{0\}$. These topological lines can be thought of as the analogues of \emph{'t Hooft lines} in a gauge theory. The interval compactification along $[0,1]$ of the partition function the (2+1)d state sum TQFT assigned to $\Sigma_\Upsilon \times [0,1]$ realises the partition function of a 2d $\mc C$-symmetric theory. The symmetry follows from the ability to couple the theory with topological lines in $\mc C$, which are images of the topological lines on $\Si_\Up \times \{0\}$ under the compactification. Within this context, choosing a different topological brane boundary condition amounts to gauging a subset of lines, resulting in a distinct symmetry; whereas choosing a different physical boundary condition amounts to picking a different physical theory altogether. This is the content of the SymTFT proposal. Interestingly, there are typically different choices of spherical fusion categories which, given appropriate boundary conditions on $\Si_\Up \times \{0\}$ and $\Si_\Up \times \{1\}$, result in dual descriptions of the same 2d symmetric partition function.

How does the procedure sketched above actually produce a partition function? In order to make the construction more precise, it is convenient to first invoke topological invariance so as to `stretch' $\Si_\Up \times [0,1]$ into $\Sigma_\Upsilon \times [0,2]$; we then cut $\Si_\Up \times [0,2]$ transversely along $\Sigma_\Upsilon \times \{1\}$ so that $\Sigma_\Upsilon \times [0,2] \cong (\Sigma_\Upsilon \times [0,1]) \cup_{\Sigma_\Upsilon \times \{1\}} (\Sigma_\Upsilon \times [1,2])$, and think of $\Sigma_\Upsilon \times \{1\}$ as a `gluing' boundary, i.e. a boundary along which manifolds can be glued. By definition, the TQFT assigns a topologically invariant Hilbert space to $\Sigma_\Upsilon \times \{1\}$ and a topological state in this Hilbert space to the manifold $\Si_\Up \times [0,1]$. Importantly, the Hilbert space assigned to $\Si_\Up \times \{1\}$ is a subspace of a `microscopic' Hilbert space $\mc H_\mc C(\Sigma_\Upsilon)$. The so-called physical boundary condition actually amounts to a certain choice of state in $\mc H_\mc C(\Sigma_\Up)$. This state can be obtained by imposing another brane boundary condition on $\Si \times \{2\}$ and applying the state-sum construction to $\Si_\Up \times [1,2]$ in the presence of Boltzmann weights that encode the dynamics of the theory. The inner product between the topological state associated with $\Si_\Up \times [0,1]$ and the (typically not topological) state associated with $\Si_\Up \times [1,2]$ produces the 2d partition function of the symmetric theory. We schematically depict this construction below:\footnote{We duplicated the spacelike surface $\Si_\Up \times \{1\}$ for visual clarity, but the inner product is obtained when identifying both copies of $\Sigma_\Up \times \{1\}$.}
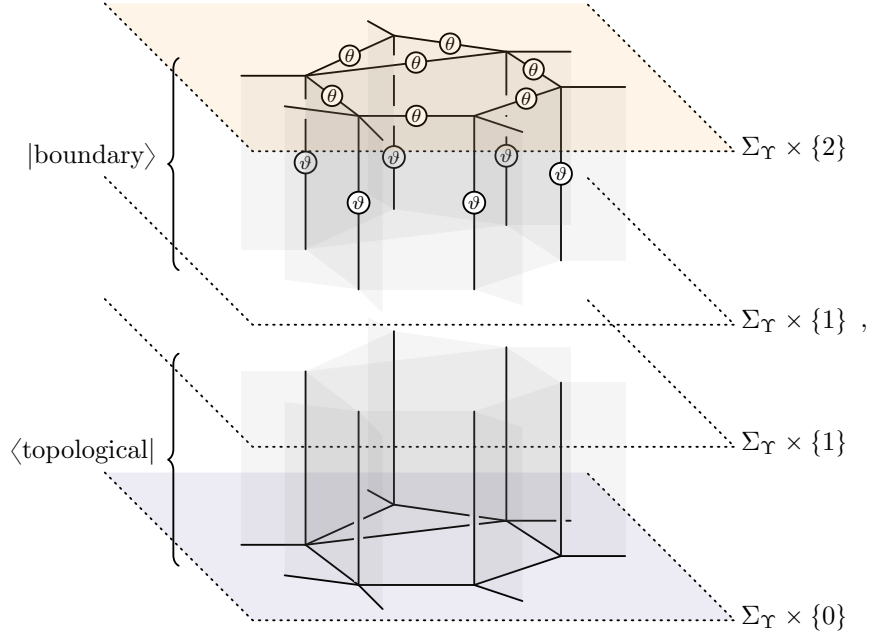
\begin{equation*}
    \sandwich \, ,
\end{equation*}
where the blobs labelled by $\theta$ and $\vartheta$ embody the Boltzmann weights encoding the dynamics of the theory. The same derivation can be repeated in the presence of a network of topological lines on $\Si_\Up \times \{0\}$. Whenever a non-contractible network is inserted, a different topological state is obtained, resulting in a partition function in a different \emph{topological sector}. A basis of topological states then lifts to a basis of topological sectors. All other things being equal, gauging via a change of topological brane boundary condition simply amounts to a change of basis for the invariant Hilbert space assigned by the TQFT to $\Si_\Up \times \{1\}$. This implies that finding out the topological sectors for which the partition functions of the initial theory and that resulting from gauging any subsymmetry can be equated, amounts to working out the corresponding basis transformation. These constructions can be made especially explicit invoking the calculus of topological \emph{tensor network} states and operators  \cite{PhysRevB.79.085119,SCHUCH20102153,Sahinoglu:2014upb,Williamson:2017uzx,Lootens:2020mso}, the relation between tensor networks and state sums having been clarified in ref.~\cite{PhysRevB.96.035101,Lootens:2020mso}. Importantly, the philosophy pursued in this manuscript systematically extends to higher dimensions, relying in particular on the corresponding calculus of tensor network states and operators \cite{Williamson:2020hxw,Delcamp:2020rds,Delcamp:2021szr}.

We illustrate our construction with finite group generalisations of the celebrated 2d Ising model. These are naturally realised as boundary theories of the state sum TQFT with input datum the spherical fusion category $\Vect_G$ of $G$-graded vector spaces for the Dirichlet topological boundary condition. Importantly, this realisation is by no means unique. In particular, it follows from \emph{electromagnetic duality} of the TQFT \cite{Buerschaper_2009, Hu:2020ghf} that the same models can also be realised as boundary theories of the TQFT with input datum the category $\Rep(G)$ of finite-dimensional representations of $G$ for the Neumann topological boundary condition. One explicitly checks that the resulting partition functions are related by Fourier transform on finite groups, as expected, using techniques in the same vein as ref.~\cite{Oeckl:2000hs,Barrett:2008wh,Bahr:2011yc}. For both formulations, one explicitly shows that gauging any subsymmetry amounts to changing the topological boundary condition. More specifically, one relates in the former formulation such a change to summing over all possible insertions of a collection of topological lines; one then verifies, using the Fourier transform on finite groups, that the resulting partition function coincides with that obtained in the latter formulation. Whenever the whole symmetry is gauged,  the partition functions after Fourier transform recover those of the non-abelian \emph{Kramers--Wannier} duals of the generalised Ising models \cite{Kramers:1941kn, 10.1063/1.1665530,Savit:1979ny, RevModPhys.51.659}, as defined by Freed and Teleman in ref.~\cite{Freed:2018cec} (see also \cite{Liu:2021hhy}). Addressing the topological sectors allows to equate the partition function of a generalised Ising model with that of its Kramers--Wannier dual. Along the way, we further comment on constructing partition functions of renormalisation group fixed points of gapped symmetric phases, and their interplay with the generalised Kramers--Wannier duality.

\bigskip \bigskip \noindent
\begin{center}
    \textbf{Organisation of the manuscript}
\end{center}
We begin in sec.~\ref{sec:abelian} by reviewing the Kramers--Wannier duality of generalised Ising models in the abelian case, closely following the exposition in ref.~\cite{Freed:2018cec}. Significantly simpler than its non-abelian analogue, we utilise  this example to motivate our construction. In sec.~\ref{sec:topo}, we review how to construct families of topological states in the form of tensor networks via the Turaev--Viro--Barrett--Westbury state sum construction, and compute the action of topological lines on such states. In sec.~\ref{sec:symTh}, we extend the state sum construction so as to define families of boundary states within the so-called microscopic Hilbert space, to which the topological states also belong. Partition functions of symmetric theories are then constructed as inner products between topological states and boundary states. Specialising to models with invertible symmetries, we explain in sec.~\ref{sec:VecG-FT} how the same theory can be realised as boundary theories of dual TQFTs. In sec.~\ref{sec:gauging}, we discuss the gauging of subsymmetries and the action of the resulting topological lines. We bring everything together in sec.~\ref{sec:RepG-FT} equating partition functions of generalised Ising models and those of their Kramers--Wannier duals.

\vfill \noindent
\begin{center}
	\textbf{Acknowledgements}
\end{center}

\noindent
CD would like to thank Hank Chen, Laurens Lootens, and Frank Verstraete for numerous discussions on closely related topics. NI thanks Shota Komatsu for helpful discussions on the topic.

\newpage
\section{Motivation: abelian Kramers--Wannier duality\label{sec:abelian}}

\emph{In order to motivate our framework, let us first examine the abelian Kramers--Wannier of generalised Ising models in close analogy with ref.~\cite{Freed:2018cec}.}

\subsection{Partition function and topological lines\label{sec:abelian_lines}}

Consider an oriented lattice $\TSi$ embedded in the two-torus $\mathbb T^2$. We denote by $\msf V(\TSi)$, $\msf E(\TSi)$ and $\msf P(\TSi)$ the sets of vertices, edges and plaquettes, respectively. Given an \emph{abelian} group $A$ and a collection $\theta: \msf E(\TSi) \to \mathbb C^A$ of \emph{even} functions $\theta_\msf e \equiv \theta(\msf e) : A \to \mathbb C$, we define a theory on $\TSi$ with `Boltzmann weight' $\theta$ via its partition function\footnote{Notice that we do not impose $\theta$ to be positive, so that the resulting theory typically does not possess a statistical mechanical interpretation. Therefore, in general, $\theta$ does not encode a Boltzmann weight in the strictest sense. }
\begin{equation}
    \label{eq:ZIsingA}
    \mc Z^A(\TSi;\theta)
    :=  
    \sum_{\sigma \in A^{\msf V(\TSi)}} 
    \prod_{\msf e \in \msf E(\TSi)} \tht_\msf e \big(\sigma_{\pme}^{-1} \sigma_{\ppe}^{\phantom{-1}}  \big) \, ,
\end{equation}
where $\pme$ and $\ppe$ refer to the source and target vertices of $\msf e$, respectively, and $\sigma_\msf v \equiv \sigma(\msf v)$. We define a differential on the spaces of $A$-valued forms on the lattice  
\begin{equation}
    A^{\msf V(\TSi)}
    \xrightarrow{{\rm d}^{(0)}}
    A^{\msf E(\TSi)}
    \xrightarrow{{\rm d}^{(1)}}
    A^{\msf P(\TSi)} \, ,
\end{equation}
such that
\begin{equation}
    (\dd^{(0)} k)_{\msf e} := k^{-1}_{\pme} k_{\ppe}^{\phantom{-1}}
    \q \text{and} \q
    (\dd^{(1)} a)_{\msf p} := \prod_{\msf e \subset \partial \msf p} a_{\msf e} \, ,
\end{equation}
where $\prod_{\msf e \subset \partial \msf p}$ is over the edges $\msf e$ on the boundary of the plaquette $\msf p \in \msf P(\TSi)$, with their orientations assumed to induce that of $\msf p$. Then, noticing that $A^{\msf V(\TSi)} \cong \im \dd^{(0)} \times \ker \dd^{(0)}$, one can rewrite the partition function as $\mc Z^A(\TSi;\theta) = |A|\sum_{a \subseteq \im {\rm d}^{(0)}} \prod_{\msf e \in \msf E(\TSi)} \theta_\msf e(a_{\msf e})$.

For any group element $x \in A$, transforming the field $\si$ as $\si(\mathsf v) \mapsto x \si(\mathsf v)$ for all vertices $\mathsf v \in \mathsf V(\TSi)$ keeps the Boltzmann weights invariant, and thus $A$ is a global (0-form) symmetry of the theory. Crucially, having a 0-form global symmetry $A$ implies that the theory can be coupled to a background (1-form) flat gauge field. Such a gauge field (or connection) is stipulated by a function $a \in A^{\msf E(\TSi)}$, the flatness condition imposing that $\dd^{(1)}a = \mathbb 1$. i.e., $a$ belongs to $\ker {\rm d}^{(1)}$. Coupling the theory to $a \in \ker {\rm d}^{(1)}$, the partition function is modified in the following way:
\begin{equation}
    \label{eq:ZIsingA_a}
    \mc Z^A(\TSi;\theta)(a)
    :=  
    \sum_{\sigma \in A^{\msf V(\TSi)}} 
    \prod_{\msf e \in \msf E(\TSi)} \tht_\msf e \big(a_\msf e \sigma_{\pme}^{-1} \sigma_{\ppe}^{\phantom{-1}}  \big) \, .
\end{equation}
Given $k \in A^{\msf V(\TSi)}$, the Boltzmann weights are individually invariant under the \emph{gauge} transformation
\begin{equation}
    \begin{split}
    \sigma_\msf v &\mapsto k_\msf v \sigma_\msf v \, , \q \forall \, \msf v \in \msf V(\TSi) \, ,
    \\
    a_\msf e &\mapsto k^{\phantom{-1}}_{\pme}a_\msf e k_{\ppe}^{-1} = a_\msf e ({\rm d}^{(0)}k)_\msf e \, , \q \forall \, \msf e \in \msf E(\TSi) \, .    
    \end{split}
\end{equation}
This implies that the partition function $\mc Z^A(\TSi;\theta)(a)$ is invariant under the gauge transformation $a \mapsto a \, {\rm d}^{(0)}k$, and the corresponding connections are \emph{gauge equivalent}. Putting everything together, this is the statement that the \emph{moduli space} of flat connections is given by the cohomology group $H^1(\TSi,A) := \frac{\ker \rd^{(1)}}{\im \rd^{(0)}}$. Cohomology classes in $H^1(\TSi, A)$ are labelled by the holonomies of the connections around the non-contractible 1-cycles of $\TSi$, i.e. $H^1(\TSi,A) \cong \Hom(\pi_1(\mathbb T^2),A)$. But $\pi_1(\mathbb T^2) \cong \mathbb Z \times \mathbb Z$, hence $H^1(\TSi,A) \cong A \times A$. In other words, coupling the theory to flat gauge fields related by gauge transformations results in physically indistinguishable theories. On the other hand, coupling the theory to flat gauge fields that are not related by gauge transformations results in theories that are locally indistinguishable but differ globally. These theories are said to be in distinct \emph{topological sectors} with respect to their global symmetry, which correspond to different points on the moduli space of flat connections. By convention, we shall refer to the sector corresponding to $(\mathbb 1,\mathbb 1) \in A \times A$ as the `trivial' sector, and all other sectors as being `twisted'.

With the different topological sectors defined, we observe that the partition function \eqref{eq:ZIsingA} corresponds to the partition function of the theory in the trivial sector. Starting from the trivial gauge field, consider performing the following gauge transformation: Let $\ell$ be a closed contractible loop along the Poincar\'e dual ${\TSi}^{\! \vee}$ of $\TSi$. We define $k \in A^{\msf V(\TSi)}$ such that $k(\msf v) := x \in A$, for every vertex $\msf v \in \msf V(\TSi)$ inside the region $\Omega_\ell$ bounded by $\ell$, and $k(\msf v) :=\mathbb 1$, for every other vertex. Performing the gauge transformation with gauge parameters $k$ on the trivial gauge field results in $a \in A^{\msf E(\TSi)}$ such that
\begin{equation}
    a(\msf e) = 
    \left\{ \begin{array}{ll}
    x & \text{if } \iota(\msf e) \subset \ell \\
    x^{-1} & \text{if } \iota(-\msf e) \subset \ell 
    \\
    \mathbb 1 & \text{otherwise}
    \end{array} \right. ,
\end{equation}
for every oriented edge $\msf e \in \msf E(\TSi)$, where $\iota : \msf E(\TSi) \xrightarrow{\sim} \msf E({\TSi}^{\! \vee})$.
By construction, the information of the gauge field $a$ is encoded into the \emph{defect line} labelled by $x \in A$ with support $\ell$. Crucially, this defect line is \emph{topological} as any continuous deformation of $\ell$ amounts to performing a specific gauge transformation. Within this context, the contractibility of $\ell$ implies that the gauge field $a$ is still in the trivial sector. In standard gauge theoretic language, such topological one-codimensional defects along the dual lattice are referred to as \emph{'t Hooft lines}, since crossing them along the primal lattice causes a shift of holonomies. Generic gauge transformations would then result in a network of topological lines, in such a way that the oriented product of group elements labelling lines meeting at a given junction must vanish, thereby encoding the flatness of the corresponding gauge field.

More generally, the various topological sectors of the theory are obtained as follows: Let $(h_1,h_2) \in \Hom(\pi_1(\mathbb T^2),A)$. A representative of the corresponding cohomology class in $H^1(\TSi,A)$ can easily be obtained by imposing that a given gauge field in $A^{\msf E(\TSi )}$ has the correct holonomies. Consider two inequivalent non-contractible 1-cycles $(\gamma_1,\gamma_2)$ along the one-skeleton of the Poincar\'e dual of $\TSi$. These two cycles are taken to be the supports of two topological lines labelled by $h_1$ and $h_2$, respectively. One then defines the representative gauge field $a(h_1,h_2)$ as
\begin{equation}
    \label{eq:gaugeFieldA}
    a(h_1,h_2)_\msf e 
    =
    \left\{
    \begin{array}{ll}
        h_i  & \text{if $\iota(\msf e) \subset \gamma_i$}
        \\
        h_i^{-1}  & \text{if $\iota(-\msf e) \subset \gamma_i$}
        \\
        \mathbb 1 & \text{otherwise}
    \end{array}
    \right. ,
\end{equation}
for every oriented edge $\msf e \in \msf E(\TSi)$. Any other representative is related to this one via gauge transformation, and as such one can unambiguously define $\mc Z^A(\TSi;\theta)(h_1,h_2):= \mc Z^A(\TSi;\theta)(a(h_1,h_2))$. Bringing everything together recovers the notion that the theory assigns to the symmetry a collection of one-dimensional topological lines in $A$, and that the different sectors are accessed by inserting topological lines along the non-contractible cycles of the Poincar\'e dual of $\TSi$. 

\subsection{Quantum double model \label{sec:QD}}

The main feature of the partition function \eqref{eq:ZIsingA} that motivates the present work is that it can be rewritten as the inner product between a topological state and a product state. Specifically, the topological state is found to be one of the ground states of a finite abelian group generalisation of the celebrated (2+1)d \emph{toric code} \cite{KITAEV20032}, which is a prototypical example of a physical system exhibiting \emph{topological order} \cite{doi:10.1142/S0217979290000139}. These finite group generalisations, which are typically referred to \emph{quantum double models}, can be formulated as Hamiltonian realisations of 3d BF theory or equivalently untwisted 3d \emph{Dijkgraaf--Witten} theory \cite{Dijkgraaf1990}.  In finite volume, we distinguish $|A \times A|$ such ground states spanning a subspace of the microscopic Hilbert space $\bigotimes_{\msf e \in \msf E(\TSi)} \mathbb C[A]$. Writing $\mathbb C[A] = {\rm Span}_\mathbb C\{ | a \ra \, | \, a \in A\}$ such that $\la a_1 | a_2 \ra = |A| \delta_{a_1,a_2}$, one particular normalised ground state reads
\begin{equation}
    |\TSi,A \ra := |A|^{-\frac{|\msf V(\TSi)|+|\msf E(\TSi)|+1}{2}}
    \!\!\!\sum_{\si \in A^{\msf V(\TSi)}} \bigotimes_{\msf e \in \msf E(\TSi)} | \si^{-1}_{\pme} \si_{\ppe}^{\phantom{-1}} \ra \, .
\end{equation}
Let us now define the product state. Given a function $\theta_\msf e : A \to \mathbb C$, we write $| \theta_\msf e \ra := \frac{1}{|A|}\sum_{a \in A}\theta_\msf e(a) |a \ra \in \mathbb C[A]$ so that $\la a | \theta_\msf e \ra = \theta_\msf e(a)$. We then construct the following product state in $\bigotimes_{\msf e \in \msf E(\TSi)} \mathbb C[A]$ containing the information of the Boltzmann weights:
\begin{equation}
    | \TSi; \theta \ra := |A|^\frac{|\msf V(\TSi)|+|\msf E(\TSi)|+1}{2} \bigotimes_{\msf e \in \msf E(\TSi)} | \theta_\msf e \ra \, .
\end{equation}
By inspection, one finally finds that
\begin{equation}
    \label{eq:ZIsingA_fact}
    \mc Z^A(\TSi;\theta) = \la \TSi, A \, | \, \TSi ;\theta \ra \, ,
\end{equation}
as desired.
In the following, we shall refer to such a construction as the realisation of the 2d  symmetric theory as a boundary theory of a (2+1)d topological field theory.
As commented above, we chose one particular topological state in order to recover the partition function in the trivial topological sector. What about the other ground states? These are obtained by acting with operators that amounts to nucleating pairs of dual magnetic excitations, moving one of the excitations along a non-contractible cycle, and annihilating them back. It follows from topological invariance that any continuous deformations of the paths followed by such excitations yield the same ground state. The results are states of the form
\begin{equation}
    \label{eq:QD_GS_a}
    |\TSi,A,(h_1,h_2) \ra := |A|^{-\frac{|\msf V(\TSi)|+|\msf E(\TSi)|+1}{2}}
    \!\!\!\sum_{\si \in A^{\msf V(\TSi)}} \bigotimes_{\msf e \in \msf E(\TSi)} | a(h_1,h_2)_\msf e \, \si^{-1}_{\pme} \si_{\ppe}^{\phantom{-1}} \ra \, , 
\end{equation}
where $a(h_1,h_2)_\msf e$ was defined in eq.~\eqref{eq:gaugeFieldA}. It readily follows from the definitions that
\begin{equation}
    \label{eq:ZIsingA_fact_h}
    \mc Z^{A}(\TSi;\theta)(h_1,h_2) = \la \TSi,A,(h_1,h_2) \, | \, \TSi;\theta \ra \, .
\end{equation}
This is merely the statement that the moduli space of flat connections is isomorphic to the ground state subspace of the (2+1)d topological model. More generally, the implications of the above expression is that all the symmetry aspects of our 2d theory can be understood in terms of the (2+1)d topological model. Although the benefits of such an expression are marginal in the abelian case, it turns out to be a precious tool when dealing with more intricate symmetry structures. 

\bigskip \noindent
In general, making full use of an expression such as \eqref{eq:ZIsingA_fact} for a given partition function will require a specific parametrisation of topological states in terms of \emph{tensor networks}. Specifically, in order to realise the ground state $| \TSi,A\ra$ as a tensor network, one requires two types of rank-$(p+q)$ tensors in $\mathbb C[A]^{\otimes p} \to \mathbb C[A]^{\otimes q}$. The first one is given by $\sum_{\{a_i\}_i} \big(\prod_{i=1}^{p+q-1}\delta_{a_i,a_{i+1}}\big)\bigotimes_{i=p+1}^{p+q}|a_i\ra \bigotimes_{j=1}^{p}\la a_j |$,  and the second one by $\sum_{\{a_i\}_i} \delta_{\prod_{i=p+1}^{p+q}a_i ,\prod_{j=1}^p a_j}\bigotimes_{i=p+1}^{p+q}|a_i\ra \bigotimes_{j=1}^{p}\la a_j | $. As customary with tensor networks, let us introduce a graphical notation for these two types of tensors:\footnote{Whenever considering more delicate operations or going beyond finite groups, a more sophisticated graphical calculus needs to be employed (c.f. sec.~\ref{sec:state-sum}).}
\begin{equation*}
\begin{split}
    \!\!\! \deltaT{1}{}{}{}{}{}{} \!\!\!\!\!
    &\equiv 
    \sum_{\{a_i\}_{i=1}^{p+q}} 
    \!\!
    \deltaT{1}{a_{p+1}}{a_{p+2}}{a_{p+q}}{a_1}{a_2}{a_p} 
    \!
    \bigotimes_{i=p+1}^{p+q} \! |a_i \ra \, \bigotimes_{j=1}^p \, \la a_j |
    \equiv
    \!\!
    \sum_{\{a_i\}_{i=1}^{p+q}} 
    \!\!
    \bigg(\prod_{i=1}^{p+q-1}\delta_{a_i,a_{i+1}} \bigg) 
    \bigotimes_{i=p+1}^{p+q} \! |a_i \ra \, \bigotimes_{j=1}^p \, \la a_j | \, ,
    \\ 
    \!\!\! \deltaT{2}{}{}{}{}{}{} \!\!\!\!\!
    &\equiv 
    \sum_{\{a_i\}_{i=1}^{p+q}} 
    \!\!
    \deltaT{2}{a_{p+1}}{a_{p+2}}{a_{p+q}}{a_1}{a_2}{a_p} 
    \!
    \bigotimes_{i=p+1}^{p+q} \! |a_i \ra \, \bigotimes_{j=1}^p \, \la a_j |
    \equiv
    \!\!
    \sum_{\{a_i\}_{i=1}^{p+q}} 
    \!\!
    \delta_{\prod_{i=p+1}^{p+q}a_i ,\prod_{j=1}^p a_j}
    \bigotimes_{i=p+1}^{p+q} \! |a_i \ra \, \bigotimes_{j=1}^p \, \la a_j | \, .
\end{split}
\end{equation*}
In order to construct the state $| \TSi,A \ra$ it suffices to assign a rank-$(p+q)$ `black' tensor to every $(p+q)$-valent vertex $\msf v \in \msf V(\TSi)$ such that the orientation of the indices is backwards compared to those of the incident edges, a rank-3 `white' tensor to every oriented edge $\msf e \in \msf E(\TSi)$, and contract neighbouring tensors according to the pattern dictated by $\TSi$. Supposing for simplicity that $\TSi$ is the \emph{hexagonal lattice}, one obtains a tensor network of the form
\begin{equation}
    \label{eq:PEPSA_1}
    \raisebox{12pt}{\PEPSA{1}} \hspace{2.2em}  \, ,
\end{equation}
where we notice that some indices are left uncontracted so that the resulting tensor network belongs to $\bigotimes_{\msf e \in \msf E(\TSi)} \mathbb C[A]$, as desired. One should think of the black tensors as performing the summation over configuration variables assigned to the vertices, whereas the white tensors encode the nearest-neighbour interactions. One can readily check that up to normalisation this state reproduces $| \TSi,A \ra$. This parametrisation gives a unique perspective on the topological lines in $A$. Indeed, the operation of coupling a topological line labelled by $x \in A$ and supported on the dual lattice can now be performed in the following way:
\begin{equation}
    \raisebox{6pt}{\PEPSA{2}}  \hspace{2.3em} , \q \MPOA{x} \equiv \sum_{a \in A}|xa \ra \la a | \, . \hspace{-2em}
\end{equation}
The definitions of the various tensors guarantee that such a line is indeed topological and lifts to 't Hooft lines, i.e. it amounts to inserting the gauge field defined in eq.~\eqref{eq:gaugeFieldA}.\footnote{In the terminology of the introduction, this line lives on a topological brane boundary, and the `physical' topological line defined previously is its image under the identification \eqref{eq:ZIsingA_fact_h}.}
Although this approach is hardly beneficial in the finite abelian group case, it greatly simplifies defining the action of topological lines in the general case. 

\subsection{Dynamical gauge fields\label{sec:dyn_gauge}}

By summing over background connections in eq.~\eqref{eq:ZIsingA_a}, one promotes the gauge field to be dynamical, resulting in the partition function of the theory where the symmetry is gauged:\footnote{
We are using $\widetilde{\mc Z}$ for the partition function of theory resulting from gauging the symmetry $A$, and the use of $\widehat A$ foreshadows the appearance of a new global symmetry, as clarified below. 
}
\begin{equation}
\label{eq:ZIsingA_gauged}
\begin{split}
    \widetilde{\mc Z}^{\widehat A}(\TSi;\theta)
    &:= 
    \frac{1}{|\ker \rd^{(1)}|}
    \sum_{a \in \ker \rd^{(1)}} \mc Z^A(\TSi;\theta)(a) \
    \\
    &= 
    \frac{1}{|\ker \rd^{(1)}|}
    \sum_{\sigma \in A^{\msf V(\TSi)}} 
    \sum_{a \in \ker \rd^{(1)}}
    \prod_{\msf e \in \msf E(\TSi)} \tht_\msf e \big(a_\msf e \, (\rd^{(0)} \sigma^{-1})_\msf e  \big) \, .
\end{split}
\end{equation}
Performing the shift of summation variable $a \mapsto a \rd^{(0)}\sigma$, one obtains
\begin{equation}
    \widetilde{\mc Z}^{\what A}(\TSi;\theta)
    = \frac{|A^{\msf V(\TSi)}|}{|\ker \rd^{(1)}|}
     \sum_{a \in \ker \rd^{(1)}}
    \prod_{\msf e \in \msf E(\TSi)} \tht_\msf e \big(a_\msf e  \big) \, .
\end{equation}
Since $A^{\msf V(\TSi)} \cong \im \rd^{(0)} \times \ker \rd^{(0)}$, $| \ker \rd^{(0)} | = |A|$ and $\frac{|\ker \rd^{(1)}|}{|\im \rd^{(0)}|} = |A| \times |A|$, one finally finds that
\begin{equation}
    \label{eq:ZIsingA_gaugedBis}
    \widetilde{\mc Z}^{\widehat A}(\TSi;\theta)
    = 
    \frac{1}{|A|}
    \sum_{a \in \ker \rd^{(1)}}
    \prod_{\msf e \in \msf E(\TSi)} \tht_\msf e \big(a_\msf e  \big) \, .
\end{equation}
Invoking $\mc Z^A(\TSi;\theta) = |A| \sum_{a \in \im \rd^{(0)}} \prod_{\msf e \in \msf E(\TSi)} \theta_\msf e(a_\msf e)$ and decomposing the sum over $a \in \ker \rd^{(1)}$ into a sum over holonomies $(h_1,h_2) \in A \times A$ and a sum over gauge transformations in $\im \rd^{(0)}$, one immediately finds the following relation:\footnote{On a simply connected surface $\Sigma$, where there is a unique topological sector, one rather has $\mc Z^{\what A}(\Sigma_\Upsilon;\theta) = \mc Z^A(\Sigma_\Upsilon;\theta)$.}
\begin{equation}
    \widetilde{\mc Z}^{\what A}(\TSi;\theta) = \frac{1}{|A|^2}\sum_{h_1,h_2 \in A} \mc Z^A(\TSi;\theta)(h_1,h_2) \, .
\end{equation}
When gauging the symmetry $A$ in eq.~\eqref{eq:ZIsingA_gauged}, one summed over all gauge fields indiscriminately, resulting in a gauge invariant measure over the space of gauge fields. However, this measure is by no means unique. In particular, the sum can be weighted by any functional on the moduli space, the space of functionals being spanned by characters $(\chi_1,\chi_2) : A \times A \to \mathbb C^{\times}$. As a consequence, the most general partition function for the theory after gauging is of the form
\begin{equation}
\begin{split}
    \label{eq:ZIsingAd_chi}
    \widetilde{\mc Z}^{\what A}(\TSi;\theta)(\chi_1,\chi_2) &:= \frac{1}{|A|^2} \sum_{h_1,h_2 \in A} \chi_1(h_1)\chi_2(h_2) \, \mc Z^A(\TSi;\theta)(h_1,h_2) 
    \\
    &=
    \frac{1}{|A|} \sum_{a \ker \rd^{(1)}} 
    \chi_1 \bigg(\prod_{\msf e \subset \gamma_1^\vee} a_\msf e\bigg)
    \chi_2\bigg(\prod_{\msf e \subset \gamma_2^\vee}a_\msf e\bigg)
    \prod_{\msf e \in \msf E(\TSi)} 
    \theta_\msf e(a_\msf e)
    \, ,
\end{split}
\end{equation}
which we recognise as the Fourier transform of the partition function $\mc Z^A(\TSi;\theta) : A \times A \to \mathbb C$. While $\mc Z^A(\TSi;\theta)$ is a function on the moduli space $H^1(\TSi,A)$ of flat $A$-connections, $\widetilde{\mc Z}^{\what A}(\TSi;\theta)$ is a function of the moduli space $H^1(\TSi,\what A)$ of flat $\what A$-connections, where $\what A = \Hom(A,\mathbb C^\times)$ is the Pontrjagin dual of $A$. This implies that after gauging the symmetry $A$, the resulting theory acquires a dual symmetry $\what A$, and $\widetilde{\mc Z}^{\what A}(\TSi;\theta)(\chi_1,\chi_2)$ precisely corresponds to the partition function in the twisted topological sector $(\chi_1,\chi_2) \in \what A \times \what A$. In terms of eq.~\eqref{eq:ZIsingA_gaugedBis}, the corresponding topological lines are supported on non-contractible cycles $(\gamma_1^\vee,\gamma_2^\vee)$ along ${\TSi}$, respectively, and act by modifying the Boltzmann weights $\theta_\msf e$ as $\theta_\msf e \cdot \chi_i$ if $\msf e \subset \gamma_i^\vee$ and $\theta_\msf e \cdot \chi_i^\vee$ if $-\msf e \subset \gamma_i^\vee$, with $i \in \{1,2\}$, for every $\msf e \in \msf E({\TSi}^{\! \vee})$. More general topological lines supported on arbitrary closed loops act analogously. In standard gauge theoretic language, these are referred to as \emph{Wilson lines}.

\bigskip \noindent
As before, the partition function $\widetilde{\mc Z}^{\what A}(\TSi;\theta)$ can be rewritten as an inner product between a topological state and a product state. While the product state is the same as before, namely $|\TSi;\theta \ra$, the topological state is now given by
\begin{equation}
    \widetilde{|\TSi ,A \ra} 
    := |A|^{-\frac{|\msf V(\TSi)|+|\msf E(\TSi)|+3}{2}}
    \!\!\! \sum_{a \in \ker \rd^{(1)}} \bigotimes_{\msf e \in \msf E(\TSi)} |a_\msf e \ra 
\end{equation}
so that $\widetilde{\mc Z}^{\what A}(\TSi;\theta) = \widetilde{\la \TSi , A \, |}\, \TSi; \theta \ra$. Importantly, $\widetilde{| \TSi,A \ra}$ is still in the state space spanned by states $| \TSi,A,(h_1,h_2) \ra$ over $(h_1,h_2) \in A \times A$. Indeed, one can check that 
\begin{equation}
    \widetilde{|\TSi,A \ra} = \frac{1}{|A|^2} \sum_{h_1,h_2 \in A} | \TSi,A,(h_1,h_2) \ra \, ,
\end{equation}
as expected. More generally, we can construct a different basis of the same state space that is spanned by vectors
\begin{equation}\label{eq:tildechi1chi2}
    \widetilde{|\TSi,A,}(\chi_1,\chi_2) \ra := \frac{1}{|A|^2} \sum_{h_1,h_2 \in A} \chi_1^\vee(h_1) \chi_2^\vee(h_2) \, | \TSi,A,(h_1,h_2) \ra \, ,
\end{equation}
over $(\chi_1,\chi_2) \in \what A \times \what A$. One can verify that $\widetilde{\mc Z}^{\what A}(\TSi;\theta)(\chi_1,\chi_2) = \widetilde{\la \TSi,A,}(\chi_1,\chi_2) \, | \, \TSi;\theta \ra$. As expected, these new basis vectors also admit a tensor network parametrisation. In order to construct $\widetilde{|\TSi,A \ra}$, it suffices to assign a rank-$(p+q)$ `white' tensor to every $(p+q)$-valent vertex $\msf v \in \msf V({\TSi}^{\! \vee})$ of the Poincar\'e dual of $\TSi$, a rank-3 `black' tensor to every oriented edge $\msf e \in \msf E({\TSi}^{\! \vee})$, and contract neighbouring tensors according to the pattern dictated by ${\TSi}^{\! \vee}$. As earlier, supposing for simplicity that $\TSi$ is the hexagonal lattice, one obtain a tensor network of the form
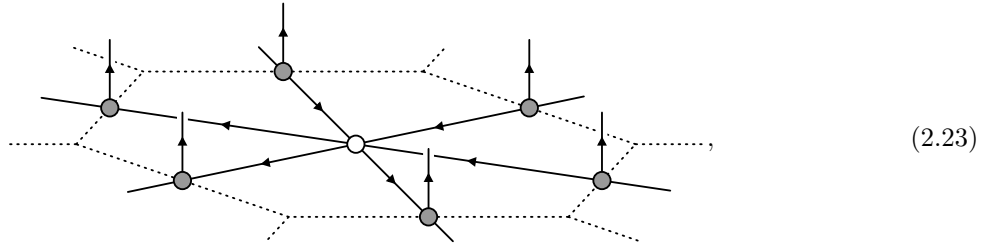
\begin{equation}
    \label{eq:PEPSA_2}
    \raisebox{12pt}{\PEPSA{3}} \hspace{2.2em}  \, ,
\end{equation}
where the dotted lines here represent the primal lattice $\TSi$. These derivations thus relate the notion of gauging to the existence of distinct tensor network parametrisations of the same family of topological states \cite{Delcamp:2020rds,Williamson:2020hxw,Lootens:2020mso}. In sec~\ref{sec:state-sum}, we will trace back the origin of these distinct parametrisations to a choice of topological brane boundary condition. As for the original theory, this gives us access to a different implementation of topological lines, which are now labelled by characters $\chi \in \what A$:
\begin{equation}
    \raisebox{10pt}{\PEPSA{4}}  \hspace{2.5em} , 
    \q \MPOA{\chi} \equiv \sum_{a \in A}\chi(a)\, |a  \ra \la a | \, . \hspace{-2em}
\end{equation}
It follows from the definitions of the various tensors that these lines are indeed topological and lift to Wilson lines as previously defined.\footnote{As for the initial theory, we should think of these lines as living on a brane boundary that is imposed to be topological. Previously, the choice of boundary condition was Dirichlet, while after gauging it is Neumann. One can then understand the swap of topological lines operated upon gauging as the statement that amongst the bulk topological lines of 3d BF theory, only `t Hooft or Wilson lines survive on a Dirichlet or Neumann topological boundary, respectively.}

\subsection{Fourier transform on finite abelian groups \label{sec:FTA}}

So far, we have only considered partition functions whose configuration variables are valued in the abelian group $A$. Given that the theory after gauging admits topological lines valued in $\what A$, one would expect an alternative formulation where configuration variables are also valued in $\what A$. Starting from the partition function of the theory in the topological sector $(h_1,h_2) \in A \times A$, performing an inverse Fourier transform of all the Boltzmann weights $\theta_\msf e : A \to \mathbb C$ yields
\begin{equation}
    \mc Z^A(\TSi;\theta)(h_1,h_2)
    =  
    \sum_{\chi \in {\what A}^{\msf E(\TSi)}}
    \sum_{\sigma \in A^{\msf V(\TSi)}} 
    \prod_{\msf e \in \msf E(\TSi)} \tht^\vee_\msf e (\chi_\msf e) \, \chi_\msf e^\vee \big(a(h_1,h_2)_\msf e \,  \si^{-1}_{\pme} \si_{\ppe}^{\phantom{-1}} \big) \, .
\end{equation}
After using the fact that $\chi_\msf e : A \to \mathbb C^\times$ are group homomorphisms, we can reorganise the product over edges so as to obtain 
\begin{equation}
    \mc Z^A(\TSi;\theta)(h_1,h_2) 
    =  \!\!\!
    \! \sum_{\chi \in {\what A}^{\msf E(\TSi)}} \!\!
    \bigg(\prod_{\msf e \in \msf E(\TSi)}\theta_\msf e^\vee(\chi_\msf e) 
    \, \chi_\msf e^\vee \big(a(h_1,h_2 \big)_\msf e)\bigg) \!\! 
    \prod_{\msf v \in \msf V(\TSi)} \sum_{\sigma_\msf v \in A}
    \bigg( \prod_{\msf e \leftarrow \msf v} \chi_\msf e(\sigma_\msf v) \bigg)
    \bigg( \prod_{\msf e \to \msf v}\chi_\msf e^\vee(\sigma_\msf v) \bigg) ,
\end{equation}
where the products $\prod_{\msf e \leftarrow \msf v}$ and $\prod_{\msf e \to \msf v}$ are over edges $\msf e \in \msf E(\TSi)$ such that $\pme = \msf v$ and $\ppe = \msf v$, respectively. Using the property $\chi_1(\sigma) \, \chi_2(\sigma) = (\chi_1 \cdot \chi_2)(\sigma)$, together with the fact that $\sum_{a \in A}\chi(a) = \delta_{1,\chi}|A|$ for any $\chi \in \what A$, the previous expression simplifies:
\begin{equation}
    \mc Z^A(\TSi;\theta)(h_1,h_2) 
    = |A|^{|\msf V(\TSi)|} \!\!  
     \sum_{\chi \in {\what A}^{\msf E(\TSi)}} \!\!
    \bigg(\prod_{\msf e \in \msf E(\TSi)}\theta_\msf e^\vee(\chi_\msf e) 
    \, \chi_\msf e^\vee \big(a(h_1,h_2 \big)_\msf e)\bigg)  
    \prod_{\msf v \in \msf V(\TSi)} \!\!
    \delta_{1,\chi_\msf v} \, ,
\end{equation}
where $\chi_\msf v := \prod_{\msf e \leftarrow \msf v} \chi_\msf e \prod_{\msf e \to \msf v}\chi_\msf e^\vee$. At this stage, it is convenient to rewrite the partition function in terms of ${\TSi}^{\! \vee}$:
\begin{equation}
    \mc Z^A(\TSi;\theta)(h_1,h_2) 
    =  |\what A|^{|\msf P({\TSi}^{\! \vee})|} \!\!\!
    \sum_{\chi \in {\what A}^{\msf E({\TSi}^{\! \vee})}} \!\!
    \bigg(\prod_{\msf e \in \msf E({\TSi}^{\! \vee})}\theta_\msf e^\vee(\chi_\msf e) 
    \, \chi_\msf e^\vee \big((\iota^{-1})^* a(h_1,h_2)_\msf e \big)\bigg) \!\!  
    \prod_{\msf p \in \msf P({\TSi}^{\! \vee})} \!\!\!
    \delta_{1,(\rd^{(1)}\chi^\vee)_\msf p} \, ,
\end{equation}
where we recall that $\iota: \msf E(\TSi) \xrightarrow{\sim} \msf E({\TSi}^{\! \vee})$. Specifically, in the notation of sec.~\ref{sec:abelian_lines}, one has
\begin{equation}
    (\iota^{-1})^* a(h_1,h_2)_\msf e 
    =
    \left\{
    \begin{array}{ll}
        h_i  & \text{if $\msf e \subset \gamma_i$}
        \\
        h_i^{-1}  & \text{if $-\msf e \subset \gamma_i$}
        \\
        \mathbb 1 & \text{otherwise}
    \end{array}
    \right. \, ,
\end{equation}
for any $\msf e \in \msf E({\TSi}^{\! \vee})$, which allows to simplify the factor $\prod_{\msf e \in \msf E({\TSi}^{\! \vee})} \chi_\msf e^\vee\big( (\iota^{-1})^* a(h_1,h_2)_\msf e\big)$. 

Let us suppose for now that $(h_1,h_2) = (\mathbb 1,\mathbb 1)$. Comparing with eq.~\eqref{eq:ZIsingA_gaugedBis}, we recognise the partition function of a model on the oriented lattice ${\TSi}^{\! \vee}$ with Boltzmann weights $\theta^\vee \in A^{\msf E({\TSi}^{\! \vee})}$ whose symmetry $\what A$ has been gauged. Specifically, in the notation of eq.~\eqref{eq:ZIsingA_fact}, we have the following equality:\footnote{On a simply connected lattice $\Sigma_\Upsilon$, one would have instead $\la \Sigma_\Upsilon, A \, | \, \Sigma_\Upsilon ; \theta \ra = |\what A|^{|\msf P(\Sigma_\Upsilon^\vee)|-1} \, \widetilde{\la \Sigma_\Upsilon^\vee ,\phantom{ A|}} \hspace{-1em} \widehat{A}  \,  | \, \Sigma_\Upsilon^\vee ; \theta^\vee \ra$.}
\begin{equation}
    \mc Z^A(\TSi;\theta)  =
    \la \TSi, A \, | \, \TSi ; \theta \ra 
    = |\what A|^{|\msf P({\TSi}^{\! \vee})|+1} \, 
    \widetilde{\la {\mathbb T^2_\Upsilon}^{\! \vee} \! ,\phantom{ A|}} \hspace{-1em} \widehat{A}  \,  | \, {\TSi}^{\! \vee} ; \theta^\vee \ra \, .
\end{equation}
This is consistent with the fact that gauging a symmetry $\what A$ results in a theory with a symmetry $A$. 
In particular, the topological state $\widetilde{| {\mathbb T^2_\Upsilon}^{\! \vee} \! ,\phantom{ A\ra}} \hspace{-1em} \widehat{A} \ra$, which belongs to a subspace of $\bigotimes_{\msf e \in \msf E({\TSi}^{\! \vee})} \mathbb C[\what A]$, can be conveniently parametrised as a tensor network of the form \eqref{eq:PEPSA_2}, where black and white tensors have been swapped. Indeed, as a function $\chi : A \to \mathbb C^\times$, we have $| \chi \ra = 1/|A|\sum_{a \in A}\chi(a) |a \ra \in \mathbb C[A]$, and conversely $|a \ra = \sum_{\chi \in \what A}\chi^\vee(a) | \chi \ra$, which one can interpret as a vector in $\mathbb C[\what A] \cong {\rm Span}_\mathbb C\{| \chi \ra \, | \, \chi \in \what A\}$. It follows that the white tensors
\begin{equation*}
\begin{split}
    \!\!\! \deltaT{2}{}{}{}{}{}{} \!\!\!\!\!
    &= \!\!
    \sum_{\{a_i\}_{i=1}^{p+q}} 
    \sum_{\{\chi_j\}_{j=1}^{p+q}}
    \delta_{\prod_{j=1}^p a_j \prod_{i=p+1}^{p+q-1}a_i^{-1},a_{p+q}}
    \bigg(\prod_{i=p+1}^{p+q}\chi_i(a_i) \bigg)
    \bigg(\prod_{j=1}^p \chi^\vee_j(a_j) \bigg)
    \bigotimes_{i=p+1}^{p+q} \! |\chi_i \ra \, \bigotimes_{j=1}^p \, \la \chi_j | 
    \\[-.7em]
    &= \!\! 
    \sum_{\{a_i\}_{i=1}^{p+q-1}} 
    \sum_{\{\chi_j\}_{j=1}^{p+q}}
    \bigg(\prod_{i=p+1}^{p+q-1}\chi_i(a_i)\chi_i^\vee(a_i) \bigg)
    \bigg(\prod_{j=1}^p \chi^\vee_j(a_j)\chi_j(a_j) \bigg)
    \bigotimes_{i=p+1}^{p+q} \! |\chi_i \ra \, \bigotimes_{j=1}^p \, \la \chi_j | 
    \\
    &=  
    |A|^{p+q-1}
    \sum_{\{\chi_i\}_{i=1}^{p+q}}
    \bigg(\prod_{i=1}^{p+q-1}\delta_{\chi_i,\chi_{i+1}} \bigg) 
    \bigotimes_{i=p+1}^{p+q} \! |\chi_i \ra \, \bigotimes_{j=1}^p \, \la \chi_j | 
\end{split}
\end{equation*}
behave as black tensors when treated as maps $\mathbb C[\what A]^{\otimes p} \to \mathbb C[\what A]^{\otimes q}$, and vice versa.
Moreover, it follows from previous derivations that 
\begin{equation}
    \widetilde{| {\mathbb T^2_\Upsilon}^{\! \vee} \! ,\phantom{ A\ra}} \hspace{-1em} \widehat{A} \ra = \frac{1}{|\widehat A|^2}\sum_{\chi_1,\chi_2 \in \what A} | {\TSi}^{\! \vee},\what A, (\chi_1,\chi_2) \ra \, ,
\end{equation}
where $| {\TSi}^{\! \vee},\what A, (\chi_1,\chi_2) \ra $ can be parametrised as a tensor network of the form \eqref{eq:PEPSA_1}, where black and white tensors have also been swapped, with topological lines $(\chi_1,\chi_2) \in \what A \times \what A$ inserted along their respective non-contractible cycles. Let us now go back to the case of an arbitrary topological sector $(h_1,h_2) \in A \times A$. We have the following equality:
\begin{equation}
    \mc Z^A(\TSi;\theta)(h_1,h_2)  =
    \la \TSi, A, (h_1,h_2) \, | \, \TSi ; \theta \ra 
    = |\what A|^{|\msf P({\TSi}^{\! \vee})|+1} \, 
    \widetilde{\la {\mathbb T^2_\Upsilon}^{\! \vee} \! ,\phantom{ A|}} \hspace{-1em} \widehat{A}, (h_1,h_2)  \,  | \, {\TSi}^{\! \vee} ; \theta^\vee \ra \, .
\end{equation}
where
\begin{equation}
    \widetilde{| {\mathbb T^2_\Upsilon}^{\! \vee} \! ,\phantom{ A\ra}} \hspace{-1em} \widehat{A} , (h_1,h_2)\ra = \frac{1}{|\widehat A|^2}\sum_{\chi_1,\chi_2 \in \what A} \chi_1(h_1) \chi_2(h_2)
    \, | {\TSi}^{\! \vee},\what A, (\chi_1,\chi_2) \ra \, .
\end{equation}
In words, a $A$-symmetric theory on $\TSi$ with Boltzmann weights $\theta$ in the topological sector $(h_1,h_2)$ is completely equivalent to the theory obtained after gauging the symmetry of an $\what A$-symmetric theory on ${\TSi}^{\! \vee}$ with Boltzmann $\theta^\vee$, where the gauging is performed via Fourier transform over the moduli space $H^1({\TSi}^{\! \vee},\what A)$ of flat $\what A$-connections with respect to the measure provided by $(h_1,h_2): \widehat A \times \what A \to \mathbb C^\times$.

\subsection{Abelian Kramers--Wannier duality}

Starting from a theory with configuration variables valued in $A$ and symmetry $A$, we have considered two distinct operations: gauging the symmetry $A$ and performing a Fourier transform on $A$ of the Boltzmann weights. Combining both operations relates theories with Pontrjagin dual symmetries, placed on Poincar\'e dual lattices, and with Boltzmann weights that are related by Fourier transform. We refer to this combination as a \emph{Kramers--Wannier} duality. As commented above, distinct moduli spaces organise the topological sectors before and after gauging the symmetry, so that equating partition functions of Kramers--Wannier dual theories requires matching the topological sectors. To do so, it is convenient to find a common parametrisation for topological sectors before and after gauging. The group $A \times \what A$ naturally provides this common parametrisation. Given a pair $(h_1,\chi_1) \in A \times \what A$, define
\begin{equation}
\begin{split}
    |\TSi , A , (h_1,\chi_1) \ra 
    &:=
    \frac{1}{|A|} \sum_{h_2 \in A} \chi^\vee_1(h_2) |\TSi , A , (h_1, h_2) \ra
    \\
    &=
    |A|^{-\frac{|\msf V(\TSi)|+|\msf E(\TSi)|+3}{2}}
    \!\!\!
    \sum_{h_2 \in A}
    \sum_{\si \in A^{\msf V(\TSi)}} 
    \chi^\vee_1(h_2)
    \bigotimes_{\msf e \in \msf E(\TSi)} | a(h_1,h_2)_\msf e \, \si^{-1}_{\pme} \si_{\ppe}^{\phantom{-1}} \ra 
\end{split}
\end{equation}
and
\begin{equation}
\begin{split}
    \widetilde{| \TSi,A,} \, (h_1,\chi_1)\ra 
    & := \sum_{\chi_2 \in \what A} \chi_2(h_1) \widetilde{| \TSi,A,} \, (\chi_2,\chi_1)\ra 
    \\
    &= |A|^{-\frac{|\msf V(\TSi)|+|\msf E(\TSi)|+3}{2}}
    \!\!\! 
    \sum_{\chi_2 \in \what A}
    \sum_{a \in \ker \rd^{(1)}} \chi_2^\vee \bigg( h_1^{-1} \prod_{\msf e \subset \gamma_1^\vee}a_\msf e \bigg) \chi_1^\vee \bigg( \prod_{\msf e \subset \gamma_2^\vee} a_\msf e\bigg)
    \bigotimes_{\msf e \in \msf E(\TSi)} |a_\msf e \ra \, .
\end{split}
\end{equation}
One can confirm that these topological states are indeed equal, i.e., $|\TSi , A , (h,\chi) \ra  =     \widetilde{| \TSi,A,} (h,\chi)\ra $, for any $(h,\chi) \in A \times \what A$, yielding the following equality of partition functions before and after gauging the symmetry $A$:
\begin{equation}
    \la \TSi,A, (h,\chi) \, | \, \TSi;\theta \ra
    =
    \widetilde{\la \TSi,A,} \, (h,\chi) \, | \, \TSi;\theta \ra \, .
\end{equation}
Subsequently performing a Fourier transform of the Boltzmann weights finally yields\footnote{On a simply connected lattice $\Sigma_\Upsilon$, this equality becomes $\mc Z^A(\Sigma_\Upsilon;\theta) = |\what A|^{|\msf P(\Sigma_\Upsilon^\vee)|-1} 
   \mc Z^{\what A}(\Sigma_\Upsilon^\vee;\theta^\vee)$.}
\begin{equation}
    \label{eq:ZIsingA_KW}
\begin{split}
    \mc Z^A(\TSi;\theta)(h,\chi) 
    &=
    \la \TSi,A, (h,\chi) \, | \, \TSi;\theta \ra 
    \\
    & =
   |\what A|^{|\msf P({\TSi}^{\! \vee})|+1} \la {\TSi}^{\! \vee}, \what A, (h,\chi) \, | \, {\TSi}^{\! \vee} ; \theta^\vee \ra 
   = |\what A|^{|\msf P({\TSi}^{\! \vee})|+1} 
   \mc Z^{\what A}({\TSi}^{\! \vee};\theta^\vee)(h,\chi) \, .
\end{split}
\end{equation}

\medskip \noindent
A particularly interesting scenario is whenever the theory is self-dual under Kramers--Wannier. Let us briefly review the case of the usual Ising model. Let $\Sigma$ be a two-dimensional \emph{connected} and \emph{simply connected} surface, $\Sigma_\Upsilon$ a lattice embedded in $\Sigma$, $A = \mathbb Z / 2 \mathbb Z \ni \{+1,-1\}$ and $\theta_{\msf e} \equiv \theta^\beta : \pm 1 \mapsto e^{\pm  \beta}$, for every $\msf e \in \msf E(\Sigma_\Upsilon)$, where $\beta \in \mathbb R$ is the \emph{inverse temperature}. The Fourier transform of the Boltzmann weights explicitly reads
\begin{equation}
    ({\theta^\beta})^\vee(\pm 1) = \theta^\beta(+1) \pm \theta^\beta(-1)
    =
    \sqrt{\frac{\sinh 2\beta}{2}} e^{\mp \frac{1}{2} \log \tanh \beta} \equiv \sqrt{\frac{\sinh 2 \beta}{2}} \theta^{\beta^\vee}(\pm 1) \, ,
\end{equation}
where $\beta^\vee := -\frac{1}{2} \log \tanh \beta$ satisfies $\sinh (2\beta) \sinh (2\beta^\vee) = 1$. It follows that eq.~\eqref{eq:ZIsingA_KW} specialises to 
\begin{equation}
    \frac{2^{|\msf P(\Sigma_\Upsilon)|/2}}{(\sinh 2\be)^{|\msf E(\Sigma_\Upsilon)|/4}} \mc Z^{\bZ/2\bZ}(\Sigma_\Upsilon;\theta^\be)
    =
    \frac{2^{|\msf P(\Sigma_\Upsilon^{\vee})|/2}}{(\sinh 2\be^\vee)^{|\msf E(\Sigma_\Upsilon^{\vee})|/4}} \mc Z^{\bZ/2\bZ}(\Sigma_\Upsilon^{\vee};\theta^{\be^\vee}) \, .
\end{equation}
Whenever the lattice $\Sigma_\Upsilon$ is itself self-dual, one can exploit this identity to extract the critical temperature of the model. Specifically, under the assumption that there is a single critical point separating high and low temperature phases, the inverse critical temperature $\beta_{\rm c}$ is such that $\beta_{\rm c}^\vee = \beta_{\rm c}$  \cite{Kramers:1941kn}. The simple connectedness of the lattice as a prerequisite for the self-duality was already pointed out in ref.~\cite{Kuroki:1996ck}, where gauging abelian symmetry was interpreted as a T-duality.

As we move from symmetry structures encoded into finite abelian groups to finite non-abelian groups, or even higher mathematical structures, it becomes exceedingly difficult to compute and compare partition functions of theories related by generalised gauging procedures and Fourier transform operations. In the remainder of this manuscript, we discuss a framework facilitating these tasks, which generalises that sketched in this section

\bigskip    
\section{Topological states\label{sec:topo}}

\noindent
\emph{We begin our demonstration by introducing families of topological states on the lattice that are parameterised by topological boundary conditions for various input spherical fusion categories. We also introduce corresponding families of topological lines.}

\subsection{Gluing boundaries\label{sec:gluing}}

In \cite{Atiyah:1989vu}, Atiyah axiomatised a $d$-dimensional TQFT as a symmetric monoidal functor $\mc Z : \msf{Bord}_d \to \Vect$, where $\msf{Bord}_d$ is the symmetric monoidal category whose (1-)morphisms are equivalence classes of oriented cobordisms between closed 1-codimensional manifolds and $\Vect$ is the symmetric monoidal category of (complex) vector spaces and linear maps. Concretely, this means that a TQFT $\mc Z$ is specified by a choice of vector space $\mc Z(\Sigma)$ to every oriented closed 1-codimensional manifold $\Sigma$, a choice of linear map $\mc Z(\Sigma) \to \mc Z(\Sigma')$ to every cobordism $\Sigma \to \Sigma'$, as well as isomorphisms $\mc Z(\varnothing) \cong \mathbb C$ and $\mc Z(\Sigma \sqcup \Sigma') \cong \mc Z(\Sigma) \otimes_\mathbb C \mc Z(\Sigma)'$, which are compatible with the associativity of the monoidal structures as well as the braidings. In particular, $\mc Z$ assigns a vector in $\mc Z(\Sigma)$ to every $d$-dimensional manifold bounded by $\Sigma$ by treating it as a cobordism $\varnothing \to \Sigma$. Furthermore, functoriality of $\mc Z$ implies that $\mc Z(\Sigma \times [0,1]) = 1_{\mc Z(\Sigma)}$, where $\Sigma \times [0,1]$ is here interpreted as a cobordism $\overline \Sigma \to \Sigma$, and $\mc Z(\Sigma \to \Sigma' \cup_{\Sigma'} \Sigma' \to \Sigma'') = \mc Z(\Sigma \to \Sigma') \circ \mc Z(\Sigma' \to \Sigma'')$. Henceforth, we refer to such boundaries along which manifolds can be glued as \emph{gluing boundaries}.

In three dimensions, the Turaev--Viro--Barrett--Westbury construction produces a state-sum TQFT $\mc Z_\mc C$ given the data of a spherical fusion category $\mc C$ \cite{Turaev:1992hq,Barrett:1993ab}. Given a choice of $\mc C$---in our exposition, it will typically either be the category $\Vect_G$ of $G$-graded vector spaces or the category $\Rep(G)$ of finite-dimensional representations of $G$, where $G$ is a (possibly non-abelian) finite group---we are interested in particular collections of basis states in the vector space $\mc Z_\mc C(\Sigma)$ assigned by $\mc Z_\mc C$ to a two-dimensional surface $\Sigma$. We shall obtain these basis states as the vectors assigned by $\mc Z_\mc C$ to specific cobordisms of the form $\mathbb C \to \Sigma$. Physically, these correspond to ground states of the Hamiltonian realisation of $\mc Z_\mc C$ on the gluing boundary $\Sigma$. More concretely, to every cell decomposition $\Sigma_\Upsilon$ of $\Sigma$, one can associate a microscopic Hilbert space $\mc H_\mc C(\Sigma_\Upsilon)$. Denoting by $\Sigma^\mathbb I_\Upsilon$ a cell decomposition of the cobordism $\Sigma^\mathbb I \equiv \Sigma \times [0,1]$ such that $\partial \Sigma^\mathbb I_\Upsilon = \overline \Sigma_\Upsilon \sqcup \Sigma_\Upsilon$, we have $\mc Z_\mc C(\Sigma^\mathbb I_\Upsilon) : \mc H_\mc C(\Sigma_\Upsilon) \to \mc H_\mc C(\Sigma_\Upsilon)$. It follows from $\Sigma^\mathbb I_\Upsilon \cup_{\Sigma_\Upsilon} \Sigma^\mathbb I_\Upsilon$ being diffeomorphic to $\Sigma^\mathbb I_\Upsilon$ that $\mc Z_\mc C(\Sigma_\Upsilon^\mathbb I)$ is a projector and we denote by $\mc Z_\mc C(\Sigma_\Upsilon)$ its image, i.e. $\mc Z_\mc C(\Sigma_\Upsilon) := {\rm Im} \, \mc Z_\mc C(\Sigma_\Upsilon^\mathbb I) \subseteq \mc H_\mc C(\Sigma_\Upsilon)$. Moreover, given two cellular decompositions $\Sigma_\Upsilon$ and $\Sigma_{\Upsilon'}$ of $\Sigma$, let ${}_\Upsilon \Xi_{\Upsilon'}$ be the cellular decomposition of a \emph{pinched interval} cobordism such that $\partial {}_\Upsilon \Xi_{\Upsilon'} = \overline \Xi_\Upsilon \cup_{\partial \Xi_\Upsilon} \Xi_{\Upsilon'}$ with $\partial \Xi_\Upsilon = \partial \Xi_{\Upsilon'}$, where $\Xi_\Upsilon \subseteq \Sigma_\Upsilon$ and $\Xi_{\Upsilon'} \subseteq \Sigma_{\Upsilon'}$. Naturally, such a cobordism defines a map $\mc Z_\mc C({}_\Upsilon \Xi_{\Upsilon'}): \mc H_\mc C(\Sigma_\Upsilon) \to \mc H_\mc C(\Sigma_{\Upsilon'})$, which in turn produces an isomorphism $\mc Z_\mc C( {}_\Upsilon \Xi_{\Upsilon'}): \mc Z_\mc C(\Sigma_\Upsilon) \xrightarrow{\sim} \mc Z_\mc C(\Sigma_{\Upsilon'})$. It follows that $\mc Z$ assigns the same vector space, up to isomorphisms, to every cell decomposition $\Sigma_\Upsilon$ of $\Sigma$, and we call $\mc Z_\mc C(\Sigma)$ this vector space.

\bigskip \noindent
Ultimately, we are interested in deriving basis states in $\mc Z_\mc C(\Sigma_\Upsilon) \cong \mc Z_\mc C(\Sigma)$ for various cell decompositions $\Sigma_\Upsilon$ of the gluing boundary $\Sigma$. Instead of strictly following the steps outlined above, we shall immediately evaluate the partition function on specific cobordisms of the form $\mathbb C \to \Sigma$. But first, we need to review the explicit definition of the microscopic Hilbert space $\mc H_\mc C(\Sigma_\Upsilon) \supseteq \mc Z_\mc C(\Sigma_\Upsilon)$.
Let $\mc C$ be a spherical fusion category over the ground field $\mathbb C$.  Representatives of the finitely many isomorphism classes of simple objects in $\mc C$ are denoted by $X_1, X_2 , \ldots \in \mc I_\mc C$ and the corresponding quantum dimensions by $d_{X_1}, d_{X_2}, \ldots \in \mathbb C$. In practice, we shall typically conflate isomorphism classes of simple objects and the corresponding representatives. The monoidal structure of $\mc C$ is notated as $(\otimes, \mathbb 1, F)$ where $\otimes : \mc C \times \mc C \to \mc C$ is the monoidal product, $\mathbb 1$ the unit object satisfying $\End(\mathbb 1) \cong \mathbb C$, and $F : (- \otimes -) \otimes - \xrightarrow{\sim} - \otimes (- \otimes -)$ the monoidal associator. Introducing the notation $\mc C^{X_1 X_2}_{X_3} := \Hom_\mc C(X_1 \otimes X_2, X_3) \ni | X_1X_2X_3,i\ra$, we have $X_1 \otimes X_2 \cong \bigoplus_{X_3}d^{X_1X_2}_{X_3} X_3$ with $d^{X_1 X_2}_{X_3} := \dim_{\mathbb C} \mc C^{X_1X_2}_{X_3}$ and components $F^{X_1 X_2 X_3}$ of the monoidal associator boil down to matrices
\begin{equation}
    F^{X_1X_2X_3}_{X_4} : 
    \bigoplus_{X_5} \mc C^{X_1X_2}_{X_5} \otimes \mc C^{X_5 X_3}_{X_4}
    \xrightarrow{\sim}
    \bigoplus_{X_6} \mc C^{X_1X_6}_{X_4} \otimes \mc C^{X_2 X_3}_{X_6} \, .
\end{equation}
These matrices can be depicted in terms of string diagrams as\footnote{By convention, we choose strings in string diagrams to be always implicitly oriented from top to bottom.} 
\begin{equation}
    \label{eq:Fmove}
    \monoidalAssociator{1} \hspace{-10pt}
    = \sum_{X_6} \sum_{k,l} \big(F^{X_1 X_2 X_3}_{X_4}\big)_{X_5,ij}^{X_6,kl}
    \monoidalAssociator{2} \, ,
\end{equation}
where $i$, $j$, $k$ and $l$ label basis vectors $|X_1X_2X_5,i\ra \in \mc C^{X_1X_2}_{X_5}$, $|X_5X_3X_4,j\ra \in \mc C^{X_5X_3}_{X_4}$, $|X_2X_3X_6,k\ra \in \mc C^{X_2X_3}_{X_6}$ and $|X_1X_6X_4,l\ra \in \mc C^{X_1X_6}_{X_4}$, respectively. We shall refer to the entries of these matrices as $F$-symbols. Whenever $\mc C$ is chosen to be $\Rep(G)$, these coincide with the ordinary $6j$-symbols. Unless otherwise specified, we work with basis vectors obeying the following diagrammatic property: 
\begin{equation}
    \label{eq:innerProduct}
    \innerProduct{1} = \delta_{i,j}\delta_{X_3,X_3'}\sqrt{\frac{d_{X_1}d_{X_2}}{d_{X_3}}}\innerProduct{2} \, ,
\end{equation}
which implies in particular the following normalisation conditions:
\begin{equation}
    \label{eq:normalisation}
    \la X_1 X_2X_3 ,j | X_1X_2X_3 ,i \ra \equiv \innerProduct{3} =  \delta_{i,j}\sqrt{d_{X_1}d_{X_2}d_{X_3}} \, .
\end{equation}
Following ref.~\cite{Kirillov:2011mk}, given simple objects $X_1, X_2, \ldots, X_N$ in $\mc C$, where $N \in \mathbb N$ is arbitrary, consider the following vector space
\begin{equation} 
    \label{eq:vecSpaceObj}
    \Hom_\mc C(\mathbb 1, X_1 \otimes X_2 \otimes \cdots \otimes X_N) \, .
\end{equation}
The pivotal structure of $\mc C$ produces isomorphisms
\begin{equation}
    \Hom_\mc C(\mathbb 1, X_1 \otimes \cdots \otimes X_N) 
    \cong
    \Hom_\mc C(X_1^\vee , X_2 \otimes \cdots X_N)
    \cong
    \Hom_\mc C(\mathbb 1,X_2 \otimes \cdots \otimes X_N \otimes X_1) \, ,
\end{equation}
where $X^\vee$ is the object dual to $X$.
This implies that the vector space \eqref{eq:vecSpaceObj} only depends on the cyclic ordering of the simple objects $X_1,\ldots, X_N$ up to canonical isomorphisms. As a rigid category, $\mc C$ is equipped with evaluation and coevaluation morphisms
\begin{align}
    {\rm ev}_X : X^\vee \otimes X \to \mathbb 1 \, , \q 
    {\rm coev}_X : \mathbb 1 \to X^\vee \otimes X \, ,
\end{align}
which gives rise to tensor contraction maps of the form
\begin{align}
    &\Hom_{\mc C}(\mathbb 1,X_1 \otimes \cdots \otimes X_N \otimes Z^\vee) 
    \otimes
    \Hom_{\mc C}(\mathbb 1,Z \otimes Y_1 \otimes \cdots \otimes Y_{N'})
    \\
    & \q
    \xrightarrow{{\rm ev}_Z}
    \Hom_{\mc C}(\mathbb 1,X_1 \otimes \cdots \otimes X_N \otimes Y_1 \otimes \cdots \otimes Y_{N'})
\end{align}
inducing in particular a non-degenerate pairing
\begin{equation}
    \label{eq:pairing}
    \Hom_\mc C(\mathbb 1,X_N^\vee \otimes \cdots \otimes X_1^\vee) \otimes 
    \Hom_\mc C(\mathbb 1,X_1 \otimes \cdots \otimes X_N) \to \End_\mc C(\mathbb 1) \cong \mathbb C 
\end{equation}
for any simple objects $X_1, \ldots, X_N$ in $\mc C$, and thus isomorphisms of the form
\begin{equation}
    \Hom_\mc C(\mathbb 1,X_N^\vee \otimes \cdots \otimes X_1^\vee)
    \cong
    \Hom_\mc C(\mathbb 1,X_1 \otimes \cdots \otimes X_N)^\vee \, .
\end{equation}
Given the cell decomposition $\Sigma_\Upsilon$ of an oriented two-dimensional surface $\Sigma$, we denote by $\msf V(\Sigma_\Upsilon) \ni \msf v$ and $\msf E(\Sigma_\Upsilon) \ni \msf e$ the set of oriented vertices and edges of $\Sigma_\Upsilon$ such that $-\msf e$ denotes the edge $\msf e$ with opposite orientation. We define a $\mc C$-labeling of the edges of $\Sigma_\Upsilon$ as a map $\fr X : \msf E(\Sigma_\Upsilon) \to \mc I_\mc C$ such that $\fr X(-\msf e)= \fr X(\msf e)^{\vee}$. Moreover, two labelings $\fr X$ and $\fr X'$ are defined to be equivalent whenever $\fr X(\msf e) \cong \fr X'(\msf e)$ for every $\msf e \in \msf E(\Sigma_\Upsilon)$. The microscopic space $\mc H_\mc C(\Sigma_\Upsilon)$ associated with $\Sigma_\Upsilon$ is then taken to be
\begin{equation}
    \label{eq:microHilb}
    \mc H_\mc C(\Sigma_\Upsilon) \equiv \bigoplus_{[\fr X]} \mc H(\Sigma_\Upsilon,\fr X) \cong \bigoplus_{[\fr X]} \bigotimes_{\msf v \in \msf V(\Sigma_\Upsilon)} \!
    \Hom_\mc C \big(\mathbb 1, \bigotimes_{\msf e \supset \msf v}\fr X(\msf e) \big) \, ,
\end{equation}
where the direct sum $\bigoplus_{[\fr X]}$ is over equivalences classes of $\mc C$-labelings, and the tensor product $\bigotimes_{\msf e \supset \msf v}$ is over edges incident to $\msf v$ ordered \emph{counterclockwise} and assumed to be oriented in the \emph{outward} direction. Of course, given a generic vertex $\msf v \in \msf V[\Sigma_\Upsilon]$, some edges $\msf e \supset \msf v$ incident to it would be oriented in the inward direction and others in the outward direction, in which case the local vector space $\Hom_{\mc C}\big(\mathbb 1, \bigotimes_{\msf e \supset \msf v} \fr X(\msf e)\big)$ would involve dual simple objects.

Given a state in $\mc H_\mc C(\Sigma_\Upsilon)$ that has a non-zero overlap with the subspace $\mc Z_\mc C(\Sigma_\Upsilon) \subseteq \mc H_\mc C(\Sigma_\Upsilon)$, one can obtain a state in $\mc Z_\mc C(\Sigma_\Upsilon)$ by applying the ground state projector $\mc Z_\mc C(\Sigma_\Upsilon^\mathbb I)$ to it. Instead, one will directly compute the linear map $\mc Z_\mc C(\varnothing \to \Sigma_\Upsilon) : \mathbb C \to \mc Z_\mc C(\Sigma_\Upsilon^\mathbb I)$ assigned by $\mc Z_\mc C$ to cell decompositions $\varnothing \to \Sigma_\Upsilon$ of specific cobordisms of the form $\varnothing \to \Sigma$ so that $\mc Z_\mc C(\varnothing \to \Sigma_\Upsilon)(1) \in \mc Z_\mc C(\Sigma_\Upsilon)$. But in order to introduce the family of cobordisms we are interested in, one requires another type of boundaries.

\subsection{Brane boundaries \label{sec:brane}}

We described above the concept of gluing boundary. This is the type of boundary that appears when cutting manifolds into pieces; or conversely, the type of (parametrised) boundary along which cobordisms are glued via choices of gluing maps. Furthermore, we explained how to compute the vector space a Turaev--Viro--Barrett--Westbury theory assigns to such a gluing boundary via the introduction of a cell decomposition. We shall now introduce the concept of \emph{brane boundary}.\footnote{Depending on the context, such boundaries are also referred to in the literature as `physical', `end-of-the-world', `coloured' or `free' boundaries.} The question is whether the theory can be extended to such a brane boundary while remaining topological. Crucially, this means that given a manifold with a non-empty brane boundary but an empty gluing boundary, a theory $\mc Z_\mc C$ would still assign a complex number to it interpreted as $\mc Z_\mc C(\varnothing \to \varnothing)(1)$. In the same vein, we interpret a manifold with a disjoint union of a non-empty brane boundary and a non-empty gluing boundary $\Sigma$ as a cobordism $\varnothing \to \Sigma$. These are the cobordisms we want to compute the topological invariants of.

From now on, we shall focus on the cobordism $\SIbg$ diffeomorphic to $\Sigma \times [0,1]$ where one treats $\Sigma \times \{0\}$ as a brane boundary and $\Sigma \times \{1\}$ as a gluing boundary so that it is a cobordism of the form $\varnothing \to \Sigma$ as far as gluing boundary components are concerned. A condition must be imposed on the brane boundary. It was established in ref.~\cite{kongBdries,Fuchs:2012dt} that given a spherical fusion category $\mc C$, elementary brane boundary conditions are labeled by indecomposable (finite semisimple $\mathbb C$-linear) module categories over $\mc C$. Let $\mc M$ be a right module category over $\mc C$. Representatives of the finitely many isomorphism classes of simple objects in $\mc M$ are denoted by $M_1, M_2, \ldots \in \mc I_\mc M$. Even though $\mc M$ is not necessarily monoidal, and a fortiori not spherical fusion, one can define a notion of quantum dimension for objects in $\mc M$, which we denote by $d_{M_1}, d_{M_2}, \ldots \in \mathbb C$ \cite{etingof2016tensor}. The (right) module structure of $\mc M$ is notated as $(\cat, \F{\cat})$ where $\cat: \mc M \times \mc C \to \mc M$ is the module action and $\F{\cat}: (- \cat -) \cat - \xrightarrow{\sim} - \cat (- \otimes -)$ is the module associator, which is required to fulfil a `pentagon axiom' involving  the monoidal associator in $\mc C$. Introducing the notation $\mc M^{M_1 X}_{M_2} := \Hom_{\mc M}(M_1 \cat X , M_2) \ni |M_1 X M_2,i\ra$, we have $M_1 \cat X \cong \bigoplus_{M_2} d^{M_1 X}_{M_2} M_2$ with $d^{M_1 X}_{M_2} := \dim_\mathbb C \mc M^{M_1 X}_{M_2}$ and components $\F{\cat}^{M_1 X_1 X_2}$ of the module associator boil down to matrices
\begin{equation}
    \F{\cat}^{M_1 X_1X_2}_{M_2} : 
    \bigoplus_{M_3} \mc M^{M_1X_1}_{M_3} \otimes \mc M^{M_3 X_2}_{M_2}
    \xrightarrow{\sim}
    \bigoplus_{X_3}  \mc M^{M_1 X_3}_{M_2} \otimes \mc C^{X_1X_2}_{X_3} \, .
\end{equation}
These matrices can be depicted in terms of string diagrams as
\begin{equation}
    \label{eq:FcatMove}
    \rightModuleAssociator{1} \hspace{-10pt}
    = \sum_{X_3} \sum_{k,l} \big(\F{\cat}^{M_1 X_1 X_2}_{M_2}\big)_{M_3,ij}^{X_3,kl}
    \rightModuleAssociator{2} \, ,
\end{equation}
where $i$, $j$, $k$ and $l$ label basis vectors $|M_1X_1M_3,i\ra \in \mc M^{M_1X_1}_{M_3}$, $|M_3X_2M_2,j\ra \in \mc M^{M_3X_2}_{M_2}$, $|X_1X_2X_3,k\ra \in \mc C^{X_1X_2}_{X_3}$ and $|M_1X_3M_2,l\ra \in \mc M^{M_1X_3}_{M_2}$, respectively. We shall refer to the entries of these matrices as $\F{\cat}$-symbols. Unless otherwise specified, we work with basis vectors obeying the analogue of the diagrammatic property \eqref{eq:innerProduct}. By convention, we set to zero $\F{\cat}$-symbols for which all the fusion rules are not satisfied. Given any spherical fusion category $\mc C$, one can always choose $\mc M$ to be $\mc C$ itself---a choice referred to as the \emph{regular} $\mc C$-module category---in which case the $\F{\cat}$-symbols coincide with the $F$-symbols. Henceforth, we refer to this choice as the \emph{Dirichlet} boundary condition. Another type of (brane) boundary condition, which does not always exist, is provided by forgetful monoidal functors $\mc C \to \Vect$, i.e. $\mc C$-module categories which are equivalent to $\Vect$ as categories. Monoidal functors of this form are referred to as \emph{fiber} functors and the corresponding boundary conditions as \emph{Neumann} boundary conditions.\footnote{We justify this terminology by the results of sec.~\ref{sec:lines}, where we show that when choosing $\mc C=\Vect_G$, Wilson lines condense on the Dirichlet boundary, while 't Hooft lines condense on Neumann boundaries. Indeed, given a gauge theory, the value of the gauge field is fixed on a Dirichlet boundary, while it remains dynamical on a Neumann boundary. As a consequence, Wilson lines condense and pick up fixed values at the Dirichlet boundary, while 't Hooft lines survive. The situation is reversed in the Neumann case. Thus, we classify a boundary condition as being either Dirichlet or Neumann depending on whether Wilson or 't Hooft lines condense. In the general case, whenever $\mc C$ admits a fiber functor, one can identify analogues of Wilson and 't Hooft lines, and name the corresponding boundary conditions by analogy with gauge theory.} 

\bigskip \noindent
Let us consider a couple of examples. Recall that $\Vect_G$ is the category whose objects are $G$-graded vector spaces of the form $V = \bigoplus_{g \in G} V_g$ equipped with the fusion structure $(V \otimes W)_g = \bigoplus_{x \in G} V_x \otimes W_{x^{-1}g}$ with unit $\mathbb 1$ such that $\mathbb 1_g = \delta_{g,\mathbb 1} \mathbb C$. Simple objects are provided by the one-dimensional vector spaces $\mathbb C_g$ with $(\mathbb C_g)_h = \de_{g,h} \mathbb C$ satisfying $\Hom_{\Vect_G}(\mathbb C_g,\mathbb C_h) \cong \delta_{g,h} \mathbb C$ and $\mathbb C_g \otimes \mathbb C_h \cong \mathbb C_{gh}$. Indecomposable finite semisimple module categories over $\Vect_G$ are given by pairs $(A, \psi)$ consisting of a subgroup $A \subseteq G$ and a normalised representative of a cohomology class $[\psi] \in H^2(G,\mathbb C^\times)$ \cite{2002math......2130O}. Let $\mc M(A,\psi)$ be the $\mathbb C$-linear finite semisimple category whose set of simple objects is a transitive $G$-set identified with the quotient $A \setminus G$ so that an object in $\mc M(A,\psi)$ is a graded vector space of the form $M= \bigoplus_{ Ar \in A \setminus G}M_{Ar}$. The (right) module structure over $\Vect_G$ is defined by 
\begin{equation}
    M \cat \mathbb C_g
    := \bigoplus_{ Ar \in A \setminus G} (M \otimes \mathbb C_g)_{Ar}
    = \bigoplus_{ Ar \in A \setminus G} M_{Aa \cat g^{-1}} \, ,
\end{equation}
for every $g \in G$ and $M \in \mc M(A,\psi)$. In particular, the action on simple objects reads $\mathbb C_{Ar} \cat  \mathbb C_g = \mathbb C_{Ar \cat g} \cong \mathbb C_{A(rg)}$ for every $g \in G$ and $Ar \in A \setminus G$. Assigning to every right coset in $A \setminus G$ a representative in $G$ via a map  ${\rm rep} : A \setminus G \to G$, we notice that given $g \in G$ and $Ar \in A \setminus G$, $(rg)$ may not be the representative in $G$ of $Ar \cat g$. For every $g \in G$ and $Ar \in A \setminus G$, we denote by $a_{Ar,g}$ the group element in $A$ such that $rg = a_{Ar,g} \cdot {\rm rep}(Ar \cat g)$. Associativity of the multiplication rule in $G$ imposes in particular that $a_{Ar,g_1g_2} =a_{Ar,g_1} a_{Ar \cat g_1,g_2}$, for all $g_1,g_2 \in G$ and $Ar \in A \setminus F$. Finally, the module associator is specified by the $\F{\cat}$-symbols 
\begin{equation}
    \big(\F{\cat}^{Ar \, g_1 \, g_2}_{Ar \cat (g_1g_2)}\big)^{g_1g_2,11}_{Ar \cat g_1,11} := \psi(a_{Ar,g_1},a_{Ar \cat g_1,g_2}) \, ,    
\end{equation}
for every $g_1,g_2 \in G$ and $Ar \in A \setminus G$. Interestingly, the same data also label indecomposable ($\mathbb C$-linear finite semisimple) module categories over the category $\Rep(G)$ of finite dimensional representations of $G$. As a matter of fact, we shall explain later that it is no mere coincidence. Given $(A,\psi)$, the corresponding $\Rep(G)$-module category is given by the category $\Rep^\psi(A)$ of projective representations of $A$ with Schur's multiplier $\psi$. Concretely, the module action is provided by the restriction functor ${\rm Res}^G_A : \Rep(G) \to \Rep(A)$ as 
\begin{equation}
    M \cat V := M \otimes {\rm Res}^G_A(V)  \, ,
\end{equation}
for every $V \in \Rep(G)$ and $M \in \Rep^\psi(A)$, while the module associator is provided by that in $\Vect$. More precisely, let $|V_1 V_2 V_3 ,i \ra$ and $|M_1 V M_2 ,j \ra$ be basis vectors in $\Hom_{\Rep(G)}(V_1 \otimes  V_2, V_3)$ and $\Hom_{\Rep^\psi(A)}(M_1 \cat V, M_2)$, respectively.
We choose the following basis vectors for the hom-spaces:
\begin{equation}
    \label{eq:basisVectors}
    \begin{split}
    | V_1 V_2 V_3,i \ra &\equiv
    \Big(\frac{d_{V_1} d_{V_2}}{d_{V_3}}\Big)^\frac{1}{4}
    \CC{V_1}{V_2}{V_3}{-}{-}{-}_i^*: V_1 \otimes V_2 \to V_3 
    \\
    \text{and} \q | V M_1 M_2,j \ra &\equiv
    \Big(\frac{d_{M_1}d_{V}}{d_{M_2}} \Big)^\frac{1}{4}
    \CC{M_1}{V}{M_2}{-}{-}{-}_j^* : M_1 \cat V \to M_2 
    \, ,
    \end{split}
\end{equation}
whose normalisation conditions \eqref{eq:normalisation} follow from that of the \emph{Clebsch--Gordan coefficients}:
\begin{equation}
\begin{split}
    \sum_{v_1, v_2} 
    \CC{V_1}{V_2}{V_3}{v_1}{v_2}{v_3}^*_i
    \CC{V_1}{V_2}{V_3}{v_1}{v_2}{v_3'}_j
    &= \delta_{i,j} \, \delta_{v_3,v_3'} \, ,
    \\
    \sum_{v, m_1} 
    \CC{M_1}{V}{M_2}{m_1}{v}{m_2}^*_i
    \CC{M_1}{V}{M_2}{m_1}{v}{m_2'}_j
    &= \delta_{i,j} \, \delta_{m_2,m_2'} \, .
\end{split}
\end{equation}
Finally, the $\F{\cat}$-symbols read 
\begin{equation}
    \label{eq:def6J}
    \big(\F{\cat}^{M_1 V_1 V_2}_{M_2} \big)^{V_3,kl}_{M_3,ij}
    = 
    \frac{1}{d_{M_2}} \! 
    \sum_{\substack{v_1,v_2,v_3 \\ m_1,m_2,m_3}} \!
    \CC{M_1}{V_1}{M_3}{m_1}{v_1}{m_3}_i^*
    \CC{M_3}{V_2}{M_2}{m_3}{v_2}{m_2}_j^*
    \CC{V_1}{V_2}{V_3}{v_1}{v_2}{v_3}_k
    \CC{M_1}{V_3}{M_2}{m_1}{v_3}{m_2}_l \, ,
\end{equation}
for every $V_1,V_2,V_3 \in \Rep(G)$, $M_1,M_2,M_3 \in \Rep^\psi(A)$ and basis vectors labelled by $i,j,k$ and $l$ in the relevant hom-spaces.

\subsection{State sum invariant\label{sec:state-sum}} 

We are now ready to compute topological states in $\mc Z_\mc C(\Sigma)$ closely following ref.~\cite{Lootens:2020mso}. Consider the manifold $\SIbg$, and impose the boundary condition labelled by the $\mc C$-module category $\mc M$ to its brane boundary $\Sigma \times \{0\}$. We consider the cell decomposition $\SIbg_\Upsilon$ of $\SIbg$ obtained by cutting transversely to the edges the cartesian product $\Sigma_\Upsilon \times [0,1]$ in such a way that there are no vertices in the interior of $\SIbg_\Upsilon$. In that spirit, we think of the gluing boundary $\Sigma \times \{1\}$ as containing neither edges nor plaquettes. Note that by virtue of $\mc Z_\mc C$ being topological, any other cell decomposition fitting $\Sigma_\Upsilon$ would do, but this minimal setting makes computations more amenable. We depict this procedure below:
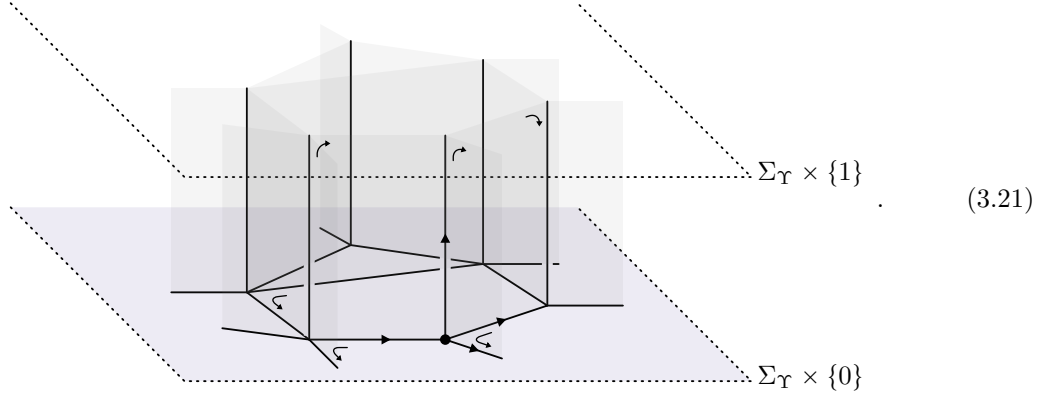
\begin{equation}
    \label{eq:cellDecomposition}
    \cellDecomposition \, .
\end{equation}
Recall that all the edges in $\Sigma_\Upsilon$ are oriented. We further choose an orientation for the plaquettes lying on the gluing boundary, which is the same for all of them, as well for the edges in the interior of $\SIbg_\Upsilon$, which is also the same for all of them. We illustrated such conventions for a few cells in eq.~\eqref{eq:cellDecomposition}. These choices induce an orientation for the plaquettes lying in the interior of $\SIbg_\Upsilon$ as well as a positive cyclic ordering for the set of plaquettes incident to any given edge. 

Borrowing the notations of sec.~\ref{sec:gluing}, consider a $\mc C$-labeling $\fr X: \msf P({\rm int}(\SIbg_\Upsilon)) \to \mc I_\mc C$ of the oriented plaquettes in the interior of $\SIbg_\Upsilon$ such that $\fr X(-\msf p) = \fr X(\msf p)^\vee$ and an $\mc M$-labelling $\fr M : \msf P(\Sigma_\Upsilon) \to \mc I_\mc M$ of the oriented plaquettes of the brane boundary $\Sigma_\Upsilon \times \{0\}$. The relevant plaquettes are shaded in eq.~\eqref{eq:cellDecomposition} in gray and purple, respectively. Two labellings $\fr M$ and $\fr M'$ are defined to be equivalent whenever $\fr M(\msf p) \cong \fr M'(\msf p)$ for every 2-cell $\msf p \in \msf P(\Sigma_\Upsilon)$. To every 0-cell $\msf v$ in the brane boundary, one can then associate a vector space $\mc H(\msf v,\fr X ,\fr M)$ as follows: First of all, we assign to the unique edge $\msf e^\text{int} \supset \msf v$ in the interior of $\SIbg_\Upsilon$ the hom-space $\mc H(\msf e^\text{int}, \fr X ,\fr M) := \Hom_{\mc C}(\mathbb 1, \bigotimes_{\msf p \supset \msf e^\text{int}} \fr X(\msf p))$, where the tensor product is over incident plaquettes ordered \emph{negatively} according to the cyclic ordering mentioned above and assumed to have the orientation inducing that of $\msf e^\text{int}$.\footnote{Notice that whenever $\msf e^\text{int}$ is in the interior of $\SIbg_\Upsilon$, the hom-space $\mc H(\msf e^\text{int},\fr X, \fr M)$ does not in fact depend on the $\mc C$-module category $\fr M$, but we still include it as we find it convenient to use the same notation for hom-spaces associated with all types of edges.} Then, to every edge $\msf e \supset \msf v$ in $\msf E(\Sigma_\Upsilon \times \{0\})$ oriented \emph{inwards}, we assign the hom-space $\mc H(\msf e,\fr X,\fr M) := \Hom_\mc M(\fr M(\dpe) \cat \fr X(\dinte), \fr M(\dme))$, where $\dinte \supset \msf e$ is in the interior of $\SIbg_\Upsilon$ and $\dpe,\dme \supset \msf e$ lay on the brane boundary such that the orientation of $\dpe$ induces that of $\msf e$ while that of $\dme$ does not. Similarly, to every edge $\msf e \supset \msf v$ oriented \emph{outwards} we assign the hom-space $\Hom_\mc M(\fr M(\dme), \fr M(\dpe) \cat \fr X(\dinte))$, which is isomorphic to $\mc H(\msf e,\fr X, \fr M)^\vee$. We then define $\mc H(\msf v,\fr X,\fr M)$ as
\beq
    \mc H(\msf v,\fr X,\fr M) := \bigotimes_{\msf e \supset \msf v}\mc H(\msf e, \fr X, \fr M) \label{eq:HVXM}
\eeq
where the tensor product is over all oriented edges incident to $\msf v$ and is unordered.

To every vertex $\msf v \in \msf V(\Sigma_\Upsilon)$, we then associate an evaluation map ${\rm ev_\msf v} : \mc H(\msf v, \fr X , \fr M) \to \mathbb C$ constructed as follows. Given a ball neighbourhood $\mathbb B_\msf v$ of $\msf v$, we consider the oriented graph $\Gamma_\msf v \subset \partial \mathbb B_\msf v$ constituted by the links formed by the intersection of $\partial \mathbb B_\msf v$ with the plaquettes in $\msf P(\SIbg_\Upsilon)$ such that the links of $\Gamma_\msf v$ inherit an orientation from the corresponding plaquettes. Moreover, the restrictions of the $\mc C$-labeling $\fr X$ and the $\mc M$-labeling $\fr M$ to the plaquettes sharing $\msf v$ induce a $(\mc C,\mc M)$-labeling of $\Gamma_\msf v$. Furthermore, we assign to every node of $\Gamma_\msf v$, which arises as the intersection of $\partial \mathbb B_\msf v$ with an oriented edge $\msf e \supset \msf v$, the corresponding hom-space $\mc H(\msf e,\fr X,\fr M)$. Removing any point from $\partial \mathbb B_\msf v$, it follows from $\mathbb S_2 \setminus {\rm pt} \simeq \mathbb R^2$ that one can treat $\Gamma_\msf v$ as a planar graph. We exemplify this procedure and our choice of conventions below:
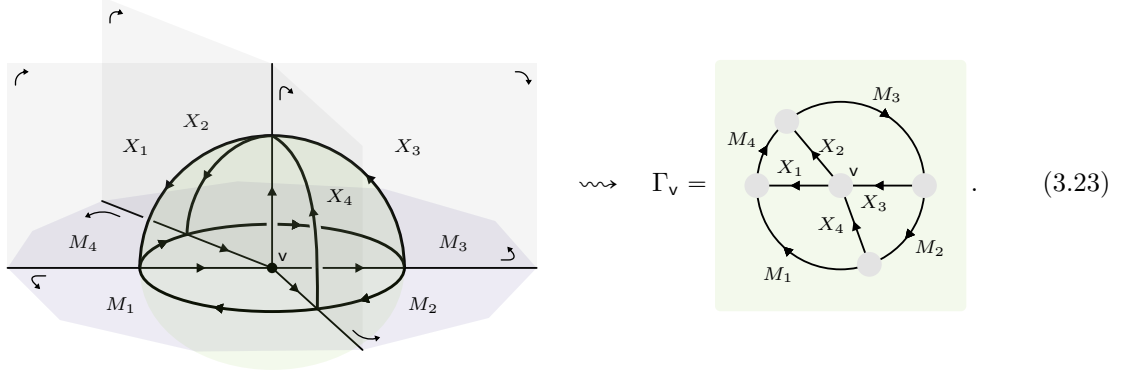
\begin{equation}
    \label{eq:evaluation}
    \evaluation 
    \q \rightsquigarrow \q
    \Gamma_\msf v = \planarDiagram{1} \, .
\end{equation}
For any labeling of the nodes of $\Gamma_\msf v$ by basis vectors $|i_\msf e \ra$ in the relevant hom-spaces, the graphical calculus of string diagrams assigns a quantum invariant ${\rm ev}_\msf v(\otimes_{\msf e \supset \msf v} | i_\msf e \ra )\in \mathbb C$.\footnote{Concretely, the complex number ${\rm ev}_\msf v(\otimes_{\msf e \supset \msf v} | i_\msf e \ra )$ is obtained by resolving using the coevaluation map the hom-space in $\mc C$ in terms of hom-spaces associated with 3-valent vertices before using combinations of diagrammatic identities of the form \eqref{eq:Fmove}, \eqref{eq:innerProduct} and \eqref{eq:normalisation}, as well as adaptations thereof in $\mc M$.} Whenever the vertex $\msf v$ is 3-valent, the result of this evaluation is proportional to an $\F{\cat}$-symbol of $\mc M$ (see below for an explicit example). We can then consider the tensor product ${\rm ev}_{\Sigma_\Upsilon}$ over all the vertices in $\Sigma_\Upsilon$ of these evaluations maps:
\begin{equation}
    \label{eq:globalEvaluation}
    {\rm ev}_{\Sigma_\Upsilon} := 
    \!\! \bigotimes_{\msf v \in \msf V(\Sigma_\Upsilon)} \!\! {\rm ev}_\msf v : \bigotimes_{\msf e \in \msf E(\SIbg_\Upsilon)} 
    \!\!\! \mc H(\msf e,\fr X,\fr M) \otimes \mc H(\msf e,\fr X,\fr M)^\vee
    \to \!\! \bigotimes_{\msf e \in \msf E({\rm int}(\SIbg_\Upsilon))} 
    \!\!\!\!\! \mc H(\msf e,\fr X,\fr M)^\vee \cong \mc H(\Sigma_\Upsilon, \fr X) \, , 
\end{equation}
where $\mc H(\Sigma_\Upsilon, \fr X)$ was defined in eq.~\eqref{eq:microHilb}. Indeed, it follows from our definitions that the hom-space $\mc H(\msf e^{\rm int},\fr X,\fr M)$ associated with the edge $\msf e^{\rm int} \supset \msf v$ in the interior of $\SIbg_\Upsilon$ is dual to that associated with $\msf v$ when defining $\mc H(\Sigma_\Upsilon, \fr X)$ in eq.~\eqref{eq:microHilb}. For instance in \eqref{eq:evaluation}, the hom-space associated with the vertex $\msf v$ when defining $\mc H(\Sigma_\Upsilon,\fr X)$ is found to be $\Hom_\mc C(\mathbb 1, X_4 \otimes X_3 \otimes X_2^\vee \otimes X_1^\vee)$, which is isomorphic to $\Hom_\mc C(X_1 \otimes X_2, X_4 \otimes X_3)$.
For every $\msf e \in \msf E(\SIbg_\Upsilon)$, fixing a choice of basis $\{|i_\msf e \ra\}_{i_\msf e}$ in $\mc H(\msf e, \fr X, \fr M)$ and denoting by $\{|i_\msf e\ra^\vee\}_{i_\msf e}$ the dual basis in $\mc H(\msf e,\fr X, \fr M)^\vee$ with respect to the non-degenerate pairing \eqref{eq:pairing}, we consider the following canonical vectors
\begin{equation}
    \label{eq:canonicalVector}
    \sum_{i_\msf e}| i_\msf e \ra \otimes |i_\msf e \ra^\vee = \sum_{i_\msf e} \frac{|i_\msf e \ra \la i_\msf e | }{\la  i_\msf e | i_\msf e \ra } \in \mc H(\msf e, \fr X ,\fr M) \otimes \mc H(\msf e, \fr X , \fr M)^\vee \, ,
\end{equation}
for every $\msf e \in \msf E(\SIbg_\Upsilon)$. Finally, we define
\begin{equation}
    \mc Z_\mc C(\SIbg_\Upsilon,\fr X , \fr M) := {\rm ev}_{\Sigma_\Upsilon}
    \bigg( \bigotimes_{\msf e \in \msf E(\SIbg_\Upsilon)} \sum_{i_\msf e}| i_\msf e \ra \otimes |i_\msf e \ra^\vee \bigg)
    \in \mc H(\Sigma_\Upsilon,\fr X) \, . \label{eq:ev-canon}
\end{equation}
Putting everything together, we define the state 
\begin{equation}
    \label{eq:TV}
    |\Sigma_\Upsilon, \mc C,\mc M \ra := \mc Z_\mc C(\SIbg_\Upsilon)(1) \, \propto
    \sum_{[\fr X], [\fr M]} \mc Z_\mc C(\SIbg_\Upsilon, \fr X , \fr M) 
    \prod_{\msf p \in \msf P(\Sigma_\Upsilon \times \{0\})} \hspace{-5pt} d_{\fr M(\msf p)} \prod_{\msf p \in \msf P({\rm int}(\SIbg_\Upsilon))} \hspace{-5pt} d_{\fr X(\msf p)}^\frac{1}{2}
    \, ,
\end{equation} 
where the sum is over equivalence classes of $(\mc C,\mc M)$-labellings.
By construction, the vector $| \Sigma_\Upsilon, \mc C, \mc M \ra$, in an element of the topologically invariant subspace $\mc Z_\mc C(\Sigma_\Upsilon) \subset \mc H_\mc C(\Sigma_\Upsilon)$. We defined the state $|\Sigma_\Upsilon, \mc C, \mc M \ra$ up to a numerical factor depending on $\mc C$ and $\mc M$, which is fixed by requiring the state to be normalised. Note that in general, different indecomposable $\mc C$-module categories yield different states in $\mc Z_\mc C(\Sigma_\Upsilon)$. In sec.~\ref{sec:lines}, we shall explain how to obtain complete bases of states in $\mc Z_\mc C(\Sigma_\Upsilon)$ for every simple object indecomposable $\mc C$-module category $\mc M$ in $\Mod(\mc C)$.

\bigskip \noindent
Let us now consider a couple of specific cases. Let $\Sigma$ be an oriented two-dimensional surface and $\Sigma_\triangle$ a choice of triangulation. We equip $\Sigma_\triangle$ with a total ordering of its 0-simplices. This total ordering in turn induces a relative orientation for the 1- and 2-simplices. Let $\Std$ be the Poincar\'e dual of $\Sigma_\triangle$ whose constitutive cells inherit an orientation from their respective dual simplices. We are interested in two families of topological states, namely states in $\mc Z_\mc C(\Sigma_\triangle)$ for the input $\mc C = \Vect_G$ and states in $\mc Z_\mc C(\Std)$ for the input $\mc C =\Rep(G)$. Let us begin with the latter. We shall focus on brane boundary conditions given by the $\Rep(G)$-module categories $\Rep(A)$ of \emph{linear representations} of a subgroup $A$ of $G$. By definition, $\Std$ only admits 3-valent vertices, and, according to our chosen orientation conventions, these vertices come in only two types. Given a vertex $\msf v$ of one such type and a $(\Rep(G),\Rep(A))$-labelling, applying the recipe described above yields an oriented graph $\Gamma_\msf v$ of the form  
\begin{equation}
    \label{eq:evaluationRepG}
    \Gamma_\msf v = \planarDiagram{2} \, .
\end{equation}
Assigning basis vectors of the form \eqref{eq:basisVectors} labelled by $i,j,k$ and $l$ to the four hom-spaces appearing in eq.~\eqref{eq:evaluationRepG}, respectively, it follows from eq.~\eqref{eq:FcatMove} and the chosen normalisation conditions that the corresponding string diagram evaluates to $\sqrt{d_{M_1}d_{V_1}d_{V_2}d_{M_2}} 
    \big( \F{\cat}^{M_1 V_1 V_2}_{M_2}\big)^{V_3,kl}_{M_3,ij} \in \mathbb C$.
But, when defining the state $| \Std, \Rep(G),\Rep(A) \ra$, we assign to every edge of the cell-complex canonical basis vectors of the form \eqref{eq:canonicalVector}, which introduces additional multiplicative factors that depend on the quantum dimension of the various objects. The multiplicative factors associated with the edges incident to a vertex can be `absorbed' by multiplying the result of the evaluation map at this vertex by the square root of these factors. In the case of eq.~\eqref{eq:evaluationRepG}, these amount to $(d_{V_1}d_{V_2}d_{V_3}d_{M_1}d_{M_2}d_{M_3})^{-\frac{1}{2}}$. Furthermore, the definition \eqref{eq:TV} includes multiplicative factors of the square root of the quantum dimension of the simple object labelling the plaquettes in the interior on the cell decomposition. Absorbing these factors in a similar manner finally produces the following amplitude associated with the vertex $\msf v$ in \eqref{eq:evaluationRepG}: $(d_{V_1}d_{V_2}d_{V_3}^{-1}d_{M_3}^{-2})^\frac{1}{4}  \big( \F{\cat}^{M_1 V_1 V_2}_{M_2}\big)^{V_3,kl}_{M_3,ij} \in \mathbb C$.
This invites us to consider the following object:
\begin{equation}
    \label{eq:PEPS}
    \PEPS{mod}{i}{k}{l}{j}{M_1}{M_2}{M_3}{V_1}{V_2}{V_3}{1} :=
    \Big(\frac{d_{V_1} d_{V_2}}{d_{V_3}} \Big)^{\frac{1}{4}}\frac{\big( \F{\cat}^{M_1 V_1 V_2}_{M_2}\big)^{V_3,kl}_{M_3,ij}}{\sqrt{d_{M_3}}} \, .
\end{equation}
Proceeding similarly for the other type of vertex appearing in $\Std$, one defines
\begin{equation}
    \PEPS{mod}{i}{l}{j}{k}{M_1}{M_2}{M_3}{V_1}{V_2}{V_3}{2} :=
    \Big(\frac{d_{V_1}d_{V_2}}{d_{V_3}} \Big)^\frac{1}{4}
    \frac{\big( \Fbar{\cat}^{M_1 V_1 V_2}_{M_2}\big)^{V_3,lj}_{M_3,ik}}{\sqrt{d_{M_3}}}
    \, ,
\end{equation}
where $\Fbar{\cat}$ here refers to the inverse of the module associator $\F{\cat}$.
We then use these graphical depictions to define 4-valent tensors of the form
\begin{equation}
    \sum_{\{V\},\{M\}}\sum_{i,j,k,l} 
    \PEPS{mod}{i}{k}{l}{j}{M_1}{M_2}{M_3}{V_1}{V_2}{V_3}{1} 
    | M_1V_1M_3,i \ra^\vee \otimes | M_3V_2M_2,j \ra^\vee \otimes | V_1V_2V_3,k\ra \otimes | M_1V_3M_2,l\ra \, , \label{eq:vertex_tensor}
\end{equation}
and similarly for the other type of vertex in $\Std$.
We assign these two types of tensors to the two types of vertices in $\Std$, and contract neighbouring tensors via the non-degenerate pairing \eqref{eq:pairing}. Adopting the convention established in ref.~\cite{Williamson:2017uzx} that every closed loop of blue strand labelled by a simple object in the module category---which appears in our case around every plaquette of $\Std$---is accompanied by a multiplicative factor of the quantum dimension of the corresponding object, this contraction scheme results in a state proportional to $| \Std, \Rep(G), \Rep(A) \ra$, as defined via eq.~\eqref{eq:TV}. 
Bringing everything together, one recovers the type of topological tensor network states that have been extensively studied in the context (2+1)d topological phases of matter, as was originally demonstrated in ref.~\cite{ Lootens:2020mso}.

Choosing instead $\mc C = \Vect_G$, let us now construct states in $\mc Z_{\mc C}(\Sigma_\triangle)$. In the same vein as the previous case, we shall focus on brane boundary conditions given by the $\Vect_G$-module categories $\mc M(A,1) \equiv \mc M(A)$ defined in sec.~\ref{sec:brane}. 
Unlike the previous case, since we are now working with $\Sigma_\triangle$, vertices can have arbitrary valence. But, provided that all the hom-spaces are non-zero, the result of the evaluation map at any vertex is always one. Moreover, quantum dimensions of simple objects in $\Vect_G$ and $\mc M(A)$ are also equal to one. To every vertex $\msf v$ in $\Sigma_\triangle$, we thus assign a tensor whose entries are of the form\footnote{When specialising to $G$ abelian, and choosing the module category to be $\Vect_G$ or $\Vect$, one recovers tensor networks of the form \eqref{eq:PEPSA_1} and \eqref{eq:PEPSA_2}, respectively. First of all, the microscopic Hilbert space $\mc H_{\Vect_G}(\Sigma_\triangle)$ is isomorphic to $\bigotimes_{\msf e \subset \Sss} \mathbb C[G]$, in which the tensor networks \eqref{eq:PEPSA_1} and \eqref{eq:PEPSA_2} live. For $\mc M= \Vect$, the white tensors in eq.~\eqref{eq:PEPSA_2} implement the fact that hom-spaces in eq.~\eqref{eq:GAHex} associated with vertices $\msf v \in \msf V(\Sigma_\triangle)$ are non-zero if and only if the fusion rules in $\Vect_G$ are satisfied. For $\mc M = \Vect_G$, the tensor depicted in eq.~\eqref{eq:GAHex} does not live in a tensor product of hom-spaces since they depend on common choices of degrees of freedom in $\mc M$. The black tensors in eq.~\eqref{eq:PEPSA_1} play the role of the loops of blue strands encoding such a structure, while the white tensors impose the conditions expressed by the Kronecker deltas in eq.~\eqref{eq:GAHex}.}
\beq
\begin{aligned}
    \PEPSHex{mod}{\mathbb C_{A r_4}}{\mathbb C_{A r_6}}{\mathbb C_{A r_5}}{\mathbb C_{A r_3}}{\mathbb C_{Ar_1}}{\mathbb C_{Ar_2}}{\mathbb C_{g_3}}{\mathbb C_{g_4} }{\mathbb C_{g_1}}{\mathbb C_{g_6}}{\mathbb C_{g_2}}{\mathbb C_{g_5}} 
    = \delta_{g_1g_2g_3,g_4g_5g_6} 
    &\cdot \delta_{Ar_1 \cat g_1,A r_2}
    \cdot \delta_{Ar_2 \cat g_2,A r_4}
    \cdot \delta_{Ar_4  \cat g_3,A r_6}
    \\[-5em] 
    &\cdot \delta_{Ar_1 \cat g_4,A r_3}
    \cdot \delta_{Ar_3 \cat g_5,A r_5}
    \cdot \delta_{Ar_5 \cat g_6,A r_6} \, ,
    \\[3em] 
\end{aligned} \label{eq:GAHex}
\eeq
where the orientation of the black lines is compatible with that of the edges incident to $\msf v$. Contracting these tensors along the edges of $\Std$ yields a state proportional to $|\Sigma_\triangle, \Vect_G,\mc M(A)\ra$, as defined via eq.~\eqref{eq:TV}. In the following, we shall explain how to construct complete bases of states in $\mc Z_{\Vect_G}(\Sigma_\triangle)$ and $\mc Z_{\Rep(G)}(\Std)$ by acting with topological operators on the states $|\Std,\Rep(G),\Rep(A) \ra$ and $| \Sigma_\triangle,\Vect_G,\mc M(A)\ra$, respectively.

\subsection{Topological lines\label{sec:lines}}

Given a spherical fusion category $\mc C$, bulk topological lines in $\mc Z_\mc C$ are organised into the Drinfel'd centre $\ms Z(\mc C)$ of $\mc C$ \cite{Majid1991,JOYAL199143}. Let us recall that $\ms Z(\mc C)$ is the (modular tensor) category whose objects are pairs $(Z,R_{Z,-})$ consisting of objects $Z \in \mc C$ and natural isomorphisms $R_{Z,-} : Z \otimes -\xrightarrow{\sim}- \otimes Z$ fulfilling a `hexagon axiom' involving the monoidal associator in $\mc C$. This category is the quantum invariant $\mc Z_\mc C$ assigns to the circle. 

Given a brane boundary labelled by $\mc M \in \Mod(\mc C)$, certain bulk topological lines condense on it. Those that survive are organised into the so-called Morita dual $\mc C^\vee_\mc M$ of $\mc C$ with respect to $\mc M$ defined as the category $\FunC_\mc C(\mc M,\mc M)$ of $\mc C$-module endofunctors of $\mc M$. An object in $\FunC_\mc C(\mc M,\mc M)$ is a pair $(\fr F,\omF{\fr F})$ consisting of a functor $\fr F: \mc M \to \mc M$ and a natural isomorphism $\omF{\fr F} : \fr F(-) \cat - \xrightarrow{\sim} \fr F(- \cat -)$ fulfilling a `pentagon axiom' involving the module associator in $\mc M$. By virtue of $\mc M$ being indecomposable, $\mc C^\vee_\mc M$ has the structure of a fusion category where the monoidal product is given by the composition of $\mc C$-module functors \cite{etingof2016tensor}. Crucially, the Drinfel'd centre is an invariant of Morita equivalence, namely there is a canonical tensor equivalence $\ms Z(\mc C) \simeq \ms Z(\mc C^\vee_\mc M)$ \cite{MUGER200381,etingof2016tensor}. It will often be convenient to refer to an object in the fusion category $\mc C^\vee_\mc M$ via a single label $Y$, in which case $(\FF{Y}, \omF{Y})$ will denote the corresponding $\mc C$-module functor.  Given a simple object $Y \in \mc C^\vee_\mc M$, components $\omF{Y}^{XM}$ of the module structure of the corresponding functor boil down to matrices
\begin{equation}
    \omF{Y}^{M_1 X}_{M_2} : \bigoplus_{M_3} \mc M^{Y M_1}_{M_3}  \otimes \mc M^{ M_3 X}_{M_2}  \xrightarrow{\sim} \bigoplus_{M_4} \mc M^{M_1 X}_{M_4} \otimes \mc M^{Y M_4}_{M_2} \, ,
\end{equation}
where $\mc M^{Y M_1}_{M_2} := \Hom_\mc M(\FF{Y}(M_1),M_2)$.
Drawing from the fact that the composition of module functors endows $\mc M \simeq \FunC_\mc C(\mc C,\mc M)$ with the structure of an (invertible) $(\mc C,\mc C^\vee_\mc M)$-bimodule category, we depict the above matrices in terms of string diagrams as follows:
\begin{equation}
    \label{eq:moduleFunctor}
    \moduleFunctor{1}
    \!\!\!\!\!\! 
    = \sum_{M_4} \sum_{k,l}
    \big( \omF{Y}^{M_1X}_{M_2}\big)^{M_4,kl}_{M_3,ij}
    \moduleFunctor{2} ,
\end{equation}
where $i,j,k$ and $l$ label basis vectors $|YM_1M_3,i\ra \in \mc M^{YM_1 }_{M_3}$, $|M_3 XM_2,j\ra \in \mc M^{M_3 X}_{M_2}$, $|M_1X M_4,k\ra \in \mc M^{M_1 X}_{M_4}$ and $|Y M_4 M_2,l\ra \in \mc M^{Y M_4 }_{M_2}$, respectively. 

\bigskip \noindent
Let us suppose that the two-dimensional surface $\Sigma$ is diffeomorphic to the two-torus $\mathbb T^2$ and let us denote as before by $\mathbb T^2_\Upsilon$ a choice of cell decomposition. It is well-known that basis states of $\mc Z_\mc C(\mathbb T^2)$ are in one-to-one correspondence with simple objects in $\ms Z(\mc C)$ \cite{Levin:2004mi,Kirillov:2011mk}. Given a simple object in $\ms Z(\mc C)$, we could define a non-local \emph{ribbon operator} explicitly acting on $\mc H_\mc C(\mathbb T^2_\Upsilon)$, as is usually done when dealing with lattice Hamiltonian realisations of $\mc Z_\mc C$ \cite{KITAEV20032,Levin:2004mi}, which upon acting on say a state $|\mathbb T^2_\Upsilon,\mc C,\mc M\ra$ would yield another basis state. But these non-local operators are usually quite intricate. In the spirit of our derivation of $|\Sigma_\Upsilon,\mc C,\mc M\ra$, and the study carried out in sec.~\ref{sec:QD}, we shall rather proceed as follows. 

Given a brane boundary labelled by $\mc M \in \Mod(\mc C)$, we mentioned above that topological lines living on it are labelled by $\mc C^\vee_\mc M$. We think of such a closed oriented line labelled by $Y \in \mc C^\vee_\mc M$ as cutting edges of $\Sigma_\Upsilon$ transversely in such a way that it divides plaquettes in $\msf P(\Sigma_\Upsilon)$ into parts, to which an $\mc M$-labelling $\fr M$ assigns their own simple object in $\mc M$. Any $\mc C$-labelling $\fr X$ is left unaffected. As described in ref.~\cite{Lootens:2020mso}, one can then adapt the evaluation map ${\rm ev}_{\Sigma_\Upsilon}$ to accommodate the presence of such lines. In addition to associating an evaluation map ${\rm ev}_\msf v$ to every vertex $\msf v \in \Sigma_\Upsilon$, we associate a new evaluation map to every crossing $\msf c$ of a topological line with an edge in the brane boundary. As before, this map is constructed by considering a ball neighbourhood $\mathbb B_\msf c$ of $\msf v$ and the oriented graph $\Gamma_\msf c$ constituted by the links formed by the intersection of $\partial \mathbb B_\msf c$ with the plaquettes of $\msf P(\SIbg)$, as well as the restriction of the topological line to the interior of $\mathbb B_\msf c$. The restrictions of the $\mc C$-labeling $\fr X$ and the $\mc M$-labeling $\fr M$ to the plaquettes adjacent to the crossing $\msf c$ induce a $(\mc C,\mc M)$-labelling of $\Gamma_\msf c$. The oriented and labelled graph $\Gamma_\msf c$ always contains four nodes, two of which are of the same type as in eq.~\eqref{eq:evaluation}, and we assign to them the same hom-space $\mc H(\msf e,\fr X,\fr M)$ as before. To the other two, we assign hom-spaces $\Hom_\mc M(\FF{Y}(\fr M(\dpe)), \fr M(\dme))$ or $\Hom_\mc M(\FF{Y}(\fr M(\dme)),\fr M(\dpe))$ depending on whether the corresponding piece of the topological line is oriented inwards or outwards with respect to $\msf c$, respectively, where $\dpe$ and $\dme$ are the plaquettes such that orientation of $\dpe$ induces that of the line while that of $\dme$ does not. We illustrate this procedure below:
\begin{equation}
    \label{eq:evaluationLine}
    \evaluationLine     \q \rightsquigarrow \q
    \Gamma_\msf c = \planarDiagram{3} \, .
\end{equation}
We finally bring everything together and proceed as in eq.~\eqref{eq:TV} in order to obtain a state in $\mc Z_\mc C(\Sigma_\Upsilon)$. The fact that the resulting state is indeed in $\mc Z_\mc C(\Sigma_\Upsilon)$ follows from the topological invariance of the line, which can be demonstrated as follows: Assigning basis vectors labelled by $i,j,k$ and $l$ to the four hom-spaces appearing on the r.h.s. of eq.~\eqref{eq:evaluationLine}, respectively, it follows from eq.~\eqref{eq:moduleFunctor} and our chosen normalisation conditions that the corresponding string diagram evaluates to $\sqrt{d_{Y}d_Xd_{M_1}d_{M_2}} \big( \omF{Y}^{M_1 X}_{M_2}\big)^{M_4,kl}_{M_3,ij}$. Absorbing the additional multiplicative factors as we did before invites us to consider the following objects \cite{Lootens:2020mso}:\footnote{Note that in ref.~\cite{Lootens:2020mso}, monoidal associators $F$, module associators $\F{\cat}$, and module structures $\omF{-}$, are notated via $\F{4}$, $\F{3}$ and $\F{2}$, respectively.}
\begin{equation}
    \label{eq:MPO}
    \MPO{mod}{mod}{i}{l}{j}{k}{M_3}{M_2}{M_4}{M_1}{X}{Y}{1} :=
    \frac{\big( \omF{Y}^{M_1 X}_{M_2}\big)^{M_4,kl}_{M_3,ij}}{\sqrt{d_{M_3}d_{M_4}}} 
\end{equation}
and
\begin{equation}
    \MPO{mod}{mod}{k}{j}{i}{l}{M_3}{M_2}{M_4}{M_1}{X}{Y}{2}
    :=
    \frac{\big( \omFbar{Y}^{M_3 X}_{M_4}\big)^{M_2,kl}_{M_1,ij}}{\sqrt{d_{M_3}d_{M_4}}} 
     \, ,
\end{equation}
where $\omFbar{Y}$ refers to the inverse of the module structure $\omF{Y}$. We then employ these graphical depictions to define 4-valent tensors of the form
\begin{align}
    \sum_{X,Y,\{M\}}\sum_{i,j,k,l} \; 
    \MPO{mod}{mod}{i}{l}{j}{k}{M_3}{M_2}{M_4}{M_1}{X}{Y}{1} \; 
     | YM_1M_3,i \ra \otimes | M_3XM_2,j \ra^\vee \otimes | M_1XM_4,k\ra \otimes | YM_4M_2,l\ra^\vee \, ,
\end{align}
and similarly for the other orientation. The whole topological line is obtained by contracting `horizontally' the above tensors via the non-degenerate pairing \eqref{eq:pairing}. The state that $\mc Z_C$ assigns to $\SIbg_\Upsilon$ in the presence of such a topological line is simply obtained by inserting the corresponding tensor network in between tensors resulting from the procedure depicted in eq.~\eqref{eq:evaluation}. Topological invariance of the line 
is then ensured by the pentagon axiom satisfied by the module structure $\omF{Y}$, which involves the module associator of $\mc M$. In the case of a three-valent vertex in $\Sigma_\Upsilon$, this pentagon axiom translates into tensor network identities of the form\footnote{Recall that we work under the convention that every closed loop formed by blue strands such as the one on the l.h.s. is implicitly accompanied by a multiplicative factor of the quantum dimension of the corresponding simple object in $\mc M$.}
\begin{equation}
    \label{eq:pulling}
    \sum_{M_3} \;  \pulling{1} = \pulling{2} \, ,
\end{equation}
which is true for every combination of simple objects in $\mc C$, simple objects in $\mc M$, and basis vectors in the relevant hom-spaces. Analogous identities hold for higher-valent vertices. Later, when specializing to $\mc C = \Vect_G$ and $\mc C=\Rep(G)$, we shall more explicitly compute tensors \eqref{eq:MPO} and verify equations of the form \eqref{eq:pulling}. In the tensor network literature, these are referred to as the `pulling-through' conditions of Matrix Product Operators (MPOs) \cite{PhysRevB.79.085119,Bultinck:2015bot, Sahinoglu:2014upb, Williamson:2017uzx, Lootens:2020mso}. Topological invariance confirms that the resulting state is indeed in $\mc Z_\mc C(\Sigma_\Upsilon)$. Whenever the support of the closed topological line is contractible, we recover $|\Sigma_\Upsilon,\mc C,\mc M\ra$, otherwise it yields a different state in $\mc Z_{\mc C}(\Sigma_\Upsilon)$. It is quick to convince oneself that this way of computing the action of topological lines is significantly simpler than explicitly working out the action of the corresponding ribbon operators on $\mc H_\mc C(\Sigma_\Upsilon)$ \cite{KITAEV20032,Levin:2004mi}.

We are left to explain how to use these topological lines in order to construct a complete basis of states in $\mc Z_\mc C(\Sigma_\Upsilon)$. Specialising to the two-torus, it follows from $\ms Z(\mc C) \simeq \ms Z(\mc C^\vee_\mc M)$ that basis states in $\mc Z_\mc C(\mathbb T^2_\Upsilon)$ are also in one-to-one correspondence with simple objects in $\ms Z(\mc C^\vee_\mc M)$. Given a simple object $(Z,R_{Z,-})$ in $\ms Z(\mc C^\vee_\mc M)$, the so-called `half-braiding' isomorphisms $R_{Z,X_1} : Z \otimes X_1 \xrightarrow{\sim} X_1 \otimes Z$ can be depicted as
\begin{equation}
    R_{Z,X_1} = \halfBraiding{Z}{X_1} = \sum_{X_2}\sum_{i,j} \big(R_{Z,X_1}^{X_2} \big)_{ij} \!\!\! \resolution{X_1}{\, X_2}{X_1}{Z}{i}{j} \, ,
\end{equation}
which we expressed on the r.h.s. in the basis of $\Hom_{\mc C^\vee_\mc M}(Z \otimes X_1, X_1 \otimes Z)$ given by $|Z X_1 X_2,i\ra \la X_1 Z X_2,j |$ where $1 \leq i,j \leq \dim_\mathbb C \Hom_\mc C(Z \otimes X_1, X_2)$. At this point, it is important to note that even though $(Z,R_{Z,-})$ is a simple object in $\ms Z(\mc C^\vee_\mc M)$, $Z$ typically is not a simple object in $\mc C^\vee_\mc M$. Ground states in $\mc Z_\mc C(\mathbb T^2_\Upsilon)$ labelled by simple objects $(Z,R_{Z,-})$ in $\ms Z(\mc C^\vee_\mc M)$ are finally obtained by computing the state assigned by $\mc Z_\mc C$ in the presence of the following insertion of topological lines
\begin{equation}
    \label{eq:linesC}
    \frac{1}{\dim \mc C^\vee_\mc M} \sum_{X \in \mc I_{\mc C^\vee_{\! \mc M}}} d_X \; \;
    \networkLines{Z}{X} \, ,
\end{equation}
where $\dim \mc C^\vee_\mc M = \sum_{X }d_X^2$ is the global dimension of the spherical fusion category $\mc C^\vee_\mc M$. More concretely, this boils down to inserting superpositions of tensors of the form \eqref{eq:MPO} contracted to a pair of tensors of the form \eqref{eq:PEPS} that evaluate to the module associator of $\mc M$ over $\mc C^\vee_\mc M$, where the \emph{left} module structure of $\mc M$ is given by the composition of $\mc C$-module functors \cite{Lootens:2020mso}. The resulting basis state will be denoted by $| \mathbb T^2_\Upsilon,\mc C,\mc M,(Z,R_{Z,-}) \ra$. We will not seek to express these basis states too explicitly in general since it is enough for our purpose to know that one can associate a basis of $\mc Z_\mc C(\mathbb T^2)$ to any indecomposble $\mc C$-module category $\mc M$ in $\Mod(\mc C)$.

Finally, recall that in the context of the Hamiltonian realisation of $\mc Z_\mc C$, simple objects in the Drinfel'd centre $\ms Z(\mc C)$ also label the topological anyon-like excitations. As a matter of fact, the operators defined above mapping topological basis states on $\mathbb T^2$ onto each other amounts to nucleating a pair of anyon-like excitations, move one around one of the non-contractible cycles of the torus, and annihilating them. Generally, these point-like excitations are created at the endpoints of open bulk topological lines. For instance, the excited state associated with a simple object in $\mc C^\vee_\mc M \subset \ms Z(\mc C^\vee_\mc M)$ can be created by acting on $| \Sigma_\Upsilon, \mc C , \mc M \ra$ with an open version of the corresponding topological line, as constructed above. 
\bigskip
\section{Boundary states and symmetric theories\label{sec:symTh}}

\emph{Exploiting the previous formalism, we explain in this section how to construct a family of so-called `boundary states' that are typically not topological. By considering the inner product between topological and boundary states, we construct partition functions of theories with arbitrary (non-)invertible symmetries encoded into spherical fusion categories. We work out two classes examples: Finite group generalisations of the Ising model, and renormalisation group fixed points of gapped symmetric phases.}

\subsection{Boundary states\label{sec:non_topo}}

Given a cell decomposition $\Si_\Up$ of $\Sigma$ and a spherical fusion category $\cC$, we defined in eq.~\eqref{eq:microHilb} the so-called microscopic Hilbert space $\cH_{\cC}(\Si_\Up)$. With the additional data of an indecomposable $\mc C$-module category $\cM$, the state sum procedure of \ref{sec:state-sum} produces a topological state $|\Si_\Up, \cC,\cM\ra$ belonging to the subspace $\mc Z_\mc C(\Sigma_\Up) \subset \cH_{\cC}(\Si_\Up)$. Let us now explain how to construct certain families of states in the same microscopic Hilbert space $\mc H_{\mc C}(\Sigma_\Up)$, which are typically not in $\mc Z_\mc C(\Sigma_\Upsilon)$, extending the exposition in ref.~\cite{Aasen:2020jwb}. We proceed in close analogy with the construction of topological states $|\Sigma_\Upsilon, \mc C, \mc M \ra$, as was described in sec.~\ref{sec:state-sum} (see also ref.~\cite{Freed:2018cec} for a closely related construction). We consider an inverted version $\Sigma_{\Upsilon}^{\mathbb I,{\rm g}|{\rm b}}$ of the cell decomposition $\Sigma_{\Upsilon}^{\mathbb I,{\rm b}|{\rm g}}$ considered previously, which is obtained by cutting transversely the Cartesian product $\Sigma_\Upsilon \times [1,2]$ so that $\Sigma_\Up \times \{1\}$ is a gluing boundary, whereas $\Sigma_\Up \times \{2\}$ plays a role analogous to that of the brane boundary $\Sigma_\Up \times \{0\}$ in eq.~\eqref{eq:cellDecomposition}. We depict this configuration below:
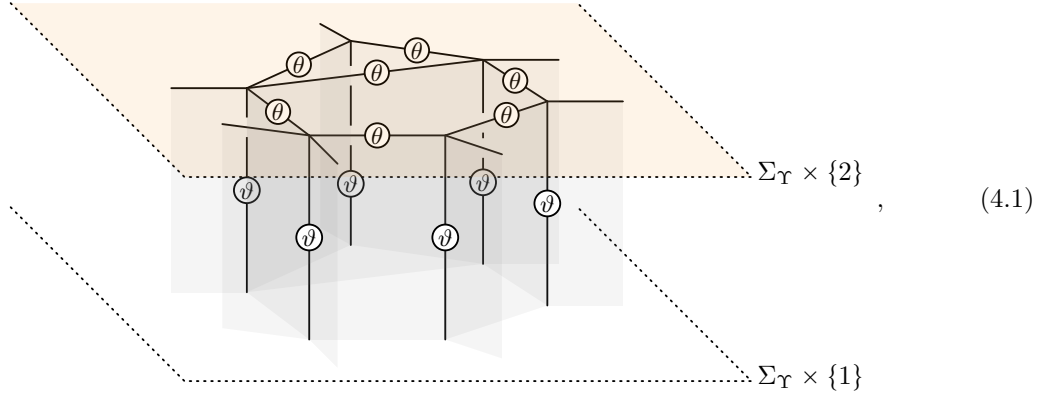
\beq
    \label{eq:cellDecompositionW}
    \cellDecompositionW \, ,
\eeq
where the plaquettes shaded in orange constitute the brane boundary $\Sigma_\Upsilon \times \{2\}$. The blobs labelled by $\theta$ and $\vartheta$
along the edges of $\Sigma_\Upsilon \times \{2\}$ and those in the interior of $\SIgb_\Upsilon$, respectively, embody modifications to the state sum construction of sec.~\ref{sec:state-sum} to be described momentarily. 
Following sec.~\ref{sec:state-sum}, we impose the brane boundary condition labelled by the indecomposable $\mc C$-module category $\mc N$ to $\Sigma_\Up \times \{2\}$. Moreover, given a (typically not simple) object $\mc A = \bigoplus_{X \in \mc I_\mc C}\la X , \mc A \ra X \in \mc C$, where $\la X , \mc A \ra \in \mathbb Z_{\geq 0}$, we consider a $(\mc C,\mc N)$-labelling $(\fr A, \fr N)$ of the plaquettes in $\msf P(\text{int}(\Sigma_{\Upsilon}^{\mathbb I,{\rm g}|{\rm b}}))$ and $\msf P(\Sigma_\Upsilon \times \{2\})$ by simple objects appearing in $\mc A$ and representatives in $\mc I_\mc N$, respectively.\footnote{Note that when decomposing $\mc A$ into simple objects, the same simple object may appear several times.} To every 0-cell $\msf v$ in $\Sigma_\Up \times \{2\}$, one associates a vector space $\mc H(\msf v, \fr A ,\fr N)$, as defined previously (see eq.~\eqref{eq:HVXM}), together with the corresponding evaluation map $\text{ev}_\msf v : \mc H(\msf v, \fr A ,\fr N) \to \mathbb C$. As in eq.~\eqref{eq:globalEvaluation}, we consider the tensor product $\text{ev}_{\Sigma_\Up} := \bigotimes_{\msf v \in \msf V(\Sigma_\Up)} \text{ev}_\msf v$ over all vertices of these evaluation maps. Furthermore, let $\theta = \{\theta(\msf e)\}_{\msf e \in \msf E(\Sigma_\Up \times \{2\})}$ be a collection of elements
\begin{equation}
    \theta_\msf e \equiv \theta(\msf e) \in 
    \bigoplus_{X \in \mc A} \bigoplus_{N_1,N_2 \in \mc I_\mc N} \Hom_\mc N(N_1 \cat X,N_2) \otimes \Hom_\mc N(N_2, N_1 \cat X)
\end{equation}
that are left invariant under orientation reversal of $\Si_\Up$, and let $\vartheta = \{\vartheta(\msf e)\}_{\msf e \in \msf E(\text{int}(\SIgb_\Upsilon))}$ be a collection of elements\footnote{Notice that we are not making any positivity assumptions regarding $\theta$ and $\vartheta$ so the resulting theory may not necessarily be interpreted as a statistical mechanical model.}
\begin{equation}
    \vartheta_\msf e \equiv \vartheta(\msf e) \in \bigoplus_{\{X_\msf p \in \mc A\}_{\msf p \supset \msf e}} \Hom_\mc C(\mathbb 1, \bigotimes_{\msf p \supset \msf e}X_\msf p ) \otimes \Hom_\mc C( \bigotimes_{\msf p \supset \msf e}X_\msf p, \mathbb 1 )  \, .
\end{equation}
Henceforth, we refer to such a pair $(\theta, \vartheta)$ as `Boltzmann weights'.
Previously, we evaluated the tensor product over all the edges of canonical vectors \eqref{eq:canonicalVector}. Given the $(\mc C,\mc N)$-labelling $(\fr A,\fr N)$, to every $\msf e \in \msf E(\text{int}(\SIgb_\Upsilon))$, we now attach the vector
\begin{equation}
    \vartheta_{\msf e}(\fr A, \fr N) := \sum_{i_\msf e} (\vartheta_\msf e)^{i_\msf e}_{i_{\msf e}} \, | i_\msf e \ra \otimes | i_\msf e \ra^\vee \in \mc H(\msf e, \fr A , \fr N) \otimes \mc H(\msf e, \fr A , \fr N)^\vee \, ,
\end{equation}
whereas to every $\msf e \in \msf E(\Sigma_\Up \times \{2\})$, we now attach the vector
\begin{equation}
    \theta_{\msf e}(\fr A,\fr N) := \sum_{i_\msf e, j_\msf e} (\theta_\msf e)_{i_\msf e}^{j_\msf e} \,  |i_\msf e \ra \otimes | j_\msf e \ra^\vee \in \mc H(\msf e, \fr A , \fr N) \otimes \mc H(\msf e, \fr A , \fr N)^\vee \, ,
\end{equation}
where the sums are over basis vectors in $\mc H(\msf e, \fr A, \fr N)$. Proceeding as before, we consider
\begin{equation}
    \mc Z_\mc C(\SIgb_\Upsilon,\mc A, \fr A,\fr N;\theta,\vartheta)
    = {\rm ev}_{\Sigma_\Up}
    \left(
    \bigotimes_{\msf e \in \msf E(\Sigma_\Up \times \{2\})} \!\!
    \tht_{\msf e}(\fr A,\fr N)
     \!\! \bigotimes_{\msf e' \in \msf E({\rm int}(\SIgb_\Upsilon))} \!\! \vartheta_{\msf e'}(\fr A ,\fr N) 
    \right) \, .
\end{equation}
Putting everything together, one finally defines
\begin{equation}
\begin{split}
    \label{eq:nonTV}
    |\Sigma_\Up, \mc C, \mc A,\mc N ;\theta, \vartheta \ra 
    \, \propto \,
    \sum_{[\fr A],[\fr N]} 
    \mc Z_{\mc C}(\SIgb,\mc A, \fr A, \fr N ; \theta, \vartheta) \!\!
    \prod_{\msf p \in \msf P(\Sigma_\Upsilon \times \{2\})} \hspace{-6pt} d_{\fr N(\msf p)} \prod_{\msf p \in \msf P({\rm int}(\SIgb_\Upsilon))} \hspace{-6pt} d_{\fr A(\msf p)}^\frac{1}{2}
    \, \in \mc H_{\mc C}(\Sigma_\Up) \, ,
\end{split}
\end{equation}
where the sum is over equivalence classes of $(\mc C,\mc N)$-labellings. The state is defined up to a numerical factor depending on $\mc C$ and $\mc N$, which will implicitly be fixed later when used in practice. 
Given this data, we claim that the following inner product in $\mc H_\mc C(\Sigma_\Upsilon)$ defines the partition function of a $\mc C^\vee_\mc M$-symmetric model:
\begin{equation}
    \mc Z^{\mc C^\vee_\mc M}(\Sigma_\Upsilon;\theta , \vartheta)
    := \la \Sigma_\Upsilon, \mc C, \mc M \, | \, \Sigma_\Upsilon, \mc C ,\mc A, \mc N; \theta, \vartheta \ra \, .
\end{equation}
First of all, let us make a couple of remarks about the notation: (i) The partition function $\mc Z^{\mc C^\vee_\mc M}(\Sigma_\Upsilon; \theta, \vartheta)$ of a two-dimensional $\mc C^\vee_\mc M$-symmetric theory is not to be confused with the three-dimensional topological theory $\mc Z_{\mc C}$ with input spherical fusion category $\mc C$. (ii) The choice of $\mc C$-module categories $\mc N$ and object $\mc A$ are implicitly encoded into the Boltzmann weights $(\theta, \vartheta)$, and are thus omitted for conciseness. This choice is further justified by the following remarks. (iii) Whenever we choose the object $\mc A$ to be $\bigoplus_{X \in \mc I_\mc C}X$---which will typically be the case---we omit writing $\mc A$ when defining the boundary state. (iv) Moreover, whenever discussing examples associated with fusion categories that admit fiber functors, we typically choose $\mc N$ to be $\Vect$ so that $\fr N(\msf p) \cong \mathbb C$ for every 2-cell $\msf p \in \msf P(\Sigma_\Up \times \{2\})$. In these cases, since different choices of fiber functors can be absorbed into a choice of $\vartheta$, we also omit writing $\mc N$ when writing the boundary state.

The $\mc C^\vee_\mc M$-symmetry of the partition function $\mc Z^{\mc C^\vee_\mc M}(\Sigma_\Upsilon; \theta , \vartheta)$ follows from our ability to couple the theory to (closed) topological lines in $\mc C^\vee_\mc M$, which is in turn guaranteed by the possibility of defining topological states in $\mc Z_\mc C(\Sigma_\Upsilon)$ in the presence of topological lines in $\mc C^\vee_\mc M$, as demonstrated in sec.~\ref{sec:lines}. Whenever the support of a closed topological line is contractible, it leaves the topological state, and a fortiori the partition function, invariant, hence the symmetry. Whenever the support is not contractible, this amounts to considering a different state in $\mc Z_{\mc C}(\Sigma_\Upsilon)$. When $\Si = \mathbb T^2$ is the 2-torus, a basis of $\mc Z_{\mc C}(\Sigma_\Upsilon)$ is labelled by simple objects in the Drinfel'd centre $\ms Z(\mc C^\vee_\mc M)$. The inner product between this new topological state $| \Sigma_\Upsilon,\mc C,\mc M,(Z,R_{Z,-}) \ra$ associated with the simple object $(Z,R_{Z,-})$ in $\ms Z(\mc C^\vee_\mc M)$ and $|\Sigma_\Upsilon, \mc C, \mc A, \mc N; \theta,\vartheta \ra$ results in the partition function of the theory in a non-trivial (topological) sector:\footnote{Despite the notation, it is important to note that the partition function in the trivial sector is not necessarily that associated with the identity object in $\ms Z(\mc C^\vee_\mc M)$ as it depends on choices of conventions. More generally, there is no canonical way of assigning a topological sector to a given simple object in $\ms Z(\mc C^\vee_\mc M)$.}
\begin{equation}
    \label{eq:twistedZ}
    \mc Z^{\mc C^\vee_\mc M} \big(\Sigma_\Upsilon; \theta, \vartheta) (Z,R_{Z,-}) = \la \Sigma_\Upsilon, \mc C, \mc M,(Z,R_{Z,-}) \, | \, \Sigma_\Upsilon, \mc C, \mc A, \mc N; \theta, \vartheta \ra \, .  
\end{equation}
Within this context, open bulk topological lines associated with simple objects in $\ms Z(\mc C^\vee_\mc M)$ correspond to \emph{generalised} charges \cite{Freed:2018cec,Chang:2018iay, Bhardwaj:2023wzd, Bartsch:2023wvv}. In particular, simple objects in $\mc C^\vee_\mc M$ play the role of \emph{disorder operators}. Conversely, open bulk topological lines condensing on the brane boundary labelled by $\mc M$ give rise to pairs of local \emph{order operators} that are charged under the symmetry $\mc C^\vee_\mc M$ symmetry. 

Notice that choosing the object $\mc A$ to be the direct sum $\bigoplus_{X \in \mc I_\mc C} X$ of simple objects in $\mc C$ and the `trivial' Boltzmann weights $(\theta^{\rm triv.},\vartheta^{\rm triv.})$, such that $(\theta^{\rm triv.}_\msf e)_{i_\msf e}^{j_\msf e}=1$ and $(\vartheta^{\rm triv.}_\msf e)_{i_\msf e}^{i_\msf e}=1$ for all $i_\msf e,j_\msf e$ in the relevant hom-spaces $\mc H(\msf e,\fr A,\fr N)$, yields the identification $|\Sigma_\Up, \mc C,\mc A,\mc N;\theta^{\rm triv}, \vartheta^{\rm triv.}\ra  \equiv |\Sigma_\Up,\mc C, \mc N \ra$. Whenever we choose $\vartheta$ to be trivial---which is often the case---we omit including it in the definition of the partition function for convenience. In sec.~\ref{sec:RG}, we discuss how  to choose the Boltzmann weights so as to obtain partition functions of renormalisation group fixed points for $\mc C^\vee_\mc M$-symmetric gapped phases. Finally, note that the boundary states are overparametrised, as multiple combinations of triples $(\mc N,\theta, \vartheta)$ yield the same states; we exploit this overparametrisation in sec.~\ref{sec:FT_fixed}, for instance.

\subsection{Generalised Ising models\label{sec:GIsing}}

Let us illustrate the above construction with a concrete example. 
Let $\Sigma$ be a two-dimensional oriented surface endowed with a triangulation $\Sigma_\triangle$. We import notations and conventions from sec.~\ref{sec:topo}. In particular, relative orientations of the 1- and 2-simplices are implied from a total ordering of the 0-simplices. Given a finite (possibly non-abelian) group $G$, let $\tht : \msf E(\Si_\triangle^\vee) \to \mathbb C^G$ be a collection of \emph{even} functions, one for each edge of the Poincar\'e dual $\Si_\triangle^\vee$. We define a $\Vect_G$-symmetric theory on $\Si_\triangle^\vee$ with Boltzmann weights $\tht$ via its partition function 
\begin{equation}
     \label{eq:ZIsingG0}
    \mc Z^{\Vect_G}(\Std;\theta) := 
    \!\!\!
    \sum_{\sigma \in G^{\msf V(\Stdss)}} 
    \prod_{\msf e \in \msf E(\Std)} \tht_\msf e \big(\sigma_{\pme}^{-1} \sigma_{\ppe}^{\phantom{-1}}  \big) \, ,
\end{equation}
where we are employing the shorthand notations $\theta_\msf e \equiv \theta(\msf e)$ and $\sigma_\msf v \equiv \sigma(\msf v)$. Henceforth, we refer to such a theory as a $G$-Ising model. Let us now lift this theory to a boundary theory of the state-sum TQFT with input datum $\Vect_G$. To begin with, we can proceed as in the abelian case (c.f. sec.~\ref{sec:QD}). Given the Hilbert space $\bigotimes_{\msf e \in \msf E(\Std)} \mathbb C[G]$, we define two of its states as
\beq
    \label{eq:ground_tht_G}
    | \Std, G \ra
    \; \propto
    \!
    \sum_{\sigma \in G^{\msf V(\Stdss)}} 
    \bigotimes_{\msf e \in \msf E(\Std)} 
    \big| \sigma_{\pme}^{-1} \sigma_{\ppe}^{\phantom{-1}}  \big\ra
    \q \text{and} \q
    | \Std;\theta \ra \; \propto \! \bigotimes_{\msf e \in \msf E(\Si_\triangle^\vee)} \! \ket{\tht_{\msf e}} \, ,
\eeq
respectively, where $|\theta_\msf e \ra := \frac{1}{|G|}\sum_{g \in G} \theta_\msf e(g)|g \ra$.
It follows that the partition function \rf{eq:ZIsingG0} can be rewritten as the inner product
\beq
    \mc Z^{\Vect_G}(\Std;\theta) = \la  \Si_\triangle^\vee, G \, | \, \Std; \tht \ra \, ,
\eeq
as desired. 
Let us rather apply our previous construction of boundary state to show that, upon choosing $\Si_\Up = \Si_\triangle$, $\cC = \Vect_G$, and $\cM = \Vect_G$, the partition function \rf{eq:ZIsingG0} can equivalently be written as an inner product between $|\Si_\triangle,\Vect_G,\Vect_G\ra $ and a boundary state from $\cH_{\Vect_G}(\Si_\triangle)$ encoding the Boltzmann weights. Consider the topological state $|\Si_\triangle,\Vect_G,\Vect_G\ra$ defined in eq.~\eqref{eq:TV}. Recall that this state can be obtained by assigning to every vertex $\msf v$ in $\Sigma_\triangle$ a tensor whose entries are of the form \eqref{eq:GAHex} for the $\Vect_G$-module category $\mc M(\{\mathbb 1\}) \simeq \Vect_G$, take the tensor product of all the tensors, and contract indices along the edges of $\Sigma_\triangle$ via the non-degenerate pairing \eqref{eq:pairing}. Remembering that the module structure of $\Vect_G$ over itself is simply given by the group multiplication in $G$, we find that $|\Sigma_\triangle,\Vect_G,\Vect_G\ra$ is a normalised sum over group labels associated with plaquettes of $\Sigma_\triangle$---each such labeling fixing a labeling for the plaquettes in the interior of $\Sigma_\triangle^{\mathbb I, {\rm b}| {\rm g}}$ by definition of the $\Vect_G$-module structure---of configurations of the unique basis vectors in the hom-spaces of $\Vect_G$ associated with the vertices of $\Sigma_\triangle$, which are always non-zero by construction. Overall, one finds
\begin{align}
    \label{eq:state_VecG_D}
    |\Si_\triangle,\Vect_G,\Vect_G \ra 
    \; \propto \sum_{\sigma \in G^{\msf P(\Sss)}} \bigotimes_{\msf v \in \msf V(\Si_\triangle)} 
    \big| \{\sigma_{\dpe}^{-1}\sigma^{\phantom{-1}}_{\dme}\}_{\msf e \supset \msf v},1\big\ra \, ,
\end{align}
where the notations $\dpe$ and $\dme$ were introduced in sec.~\ref{sec:state-sum}.

Next, we need to construct the state in $\cH_{\Vect_G}(\Si_\triangle)$ encoding the Boltzmann weights. Consider the state $| \Sigma_\Upsilon, \mc C ; \theta, \vartheta \ra$ defined in eq.~\eqref{eq:nonTV} specialised to $\mc C= \Vect_G$ and $\Sigma_\Upsilon = \Sigma_\triangle$.\footnote{Recall that by convention $| \Sigma_\triangle , \Vect_G ; \theta , \vartheta \ra \equiv | \Sigma_\triangle, \Vect_G,\bigoplus_{g \in G}\mathbb C_g,\Vect ; \theta , \vartheta \ra$.} Using the notations of sec.~\ref{sec:state-sum}, for an edge $\msf e_\text{int}$ in the interior of $\SIgb_\triangle$, the hom-space $\cH(\msf e_\text{int}, \fr A, \fr N) = \Hom_{\Vect_G}(\mathbb 1, \bigotimes_{\msf p \supset \msf e} \fr A(\msf p))$, where $\fr A(\msf p) \equiv g_\msf p$, is one dimensional if $\bigotimes_{\msf p \supset \msf e} \fr A(\msf p) = \mathbb 1$, and empty otherwise. Moreover, since $\mathbb C \cat \mathbb C_g \simeq \mathbb C$ for every simple object $\mathbb C_g$ in $\Vect_G$, hom-spaces $\mc H(\msf e, \fr A ,\fr N)$ associated with edges $\msf e$ on the brane boundary are all one-dimensional. It follows that Boltzmann weights $\theta$ boil down to functions $\theta_\msf e : G \to \mathbb C$, as expected. We choose these functions to be those entering the definition of \eqref{eq:ZIsingG0}. Moreover, we choose $\vartheta$ to be trivial. Putting everything together, we find the following (boundary) state:
\begin{equation}
    \label{eq:state_VecG_nonTopo}
    |\Sigma_\triangle,\Vect_G;\theta, \vartheta^{\rm triv.}\ra 
    \; \propto
    \sum_{g \in G^{\msf E(\Sss)}} \!\!
    \bigg(\prod_{\msf v \in \msf V(\Sigma_\triangle)}\delta_{\mathbb 1, g_\msf v} \bigg)
    \bigg(\prod_{\msf e \in \msf E(\Sigma_\triangle)} \!\! \theta_\msf e(g_\msf e) \bigg)
    \bigotimes_{\msf v \in \msf V(\Sigma_\triangle)} \!\!
    | \{g_\msf e\}_{\msf e \supset \msf v},1 \ra \, ,
\end{equation}
where we identified $g_\msf e \equiv g_{\dinte}$ so that $g_\msf v := \prod_{\msf e \supset \msf v} g_{\msf e}$ is a product of group elements over edges incident to $\msf v$ ordered counterclockwise and assumed to be oriented in the outward direction. By inspection, we immediately find that\footnote{Normalisation of the state \eqref{eq:state_VecG_nonTopo} is chosen so the inner product with \eqref{eq:state_VecG_D} exactly equates the partition function $\mc Z^{\Vect_G}(\Std;\theta)$.}
\begin{equation}
    \mc Z^{\Vect_G}(\Std;\theta) = \la \Sigma_\triangle,\Vect_G,\Vect_G \, | \, \Sigma_\triangle,\Vect_G;\theta, \vartheta^{\rm triv.} \ra \, ,
\end{equation}
as requested. The $\Vect_G$-symmetry of the theory follows from the results of the general construction together with the fact that $(\Vect_G)^\vee_{\Vect_G} \simeq \Vect_G$. More specifically, insertions of topological lines in $\Vect_G$ are performed via tensors whose non-vanishing entries are of the form
\begin{equation}
    \MPO{mod}{mod}{1}{1}{1}{1}{x \sigma}{x \sigma g}{\sigma g}{\sigma}{g}{x}{1} =1 \, ,    
\end{equation}
for any $\sigma, g ,x \in G$. It is an interesting exercise to confirm that the action of such topological lines on $\mc Z_{\Vect_G}(\Sigma_\triangle)$ precisely coincide with the action of closed \emph{magnetic operators} in the context of the Hamiltonian realisation of $\mc Z_{\Vect_G}$ \cite{KITAEV20032}.

\bigskip \noindent
Expression \eqref{eq:ZIsingG0} provides the partition function of the $G$-Ising model in the trivial (or singlet) topological sector. By definition of being symmetric, it can be coupled to background flat gauge fields so that different equivalence classes of background fields correspond to different sectors of the theory. Generally, coupling the theory to a background gauge field amounts to inserting a fine enough network of topological lines in $\Vect_G$. Specialising to $\Sigma_\triangle = \mathbb T^2_\triangle$, one can always invoke topological invariance to reduce any fine enough network of lines to a pair $(\mathbb C_{h_1}, \mathbb C_{h_2})$ of lines in $\Vect_G$ such that $h_1h_2=h_2h_1$, each wrapping either a meridional or a longitudinal non-contractible cycle of $\mathbb T^2_\triangle$, respectively. Group elements $h_1,h_2 \in G$ should here be interpreted as the \emph{holonomies} along the corresponding transverse non-contractible cycles, and the commutation relation $h_1h_2h_1^{-1}h_2^{-1} = \mathbb 1$ merely encodes that the holonomy along a contractible cycle around the intersection of the two lines must be trivial. But, due to the $\Vect_G$-symmetry, which acts on $(h_1,h_2)$ by conjugation, topological sectors of the $G$-Ising model are not simply labelled by pairs of commuting group elements. More precisely, we define an equivalence relation on $\{(h_1,h_2) \in G \times G\ \, | \, h_1h_2=h_2h_1 \}$ via $(h_1,h_2) \sim (h_1',h_2')$, if there exists an $x \in G$ such that $(h_1',h_2') = (xh_1x^{-1},xh_2x^{-1})$. The moduli space of flat connections is then found to be 
\begin{equation}
    \{(h_1,h_2) \in G \times G \, | \, h_1h_2 = h_2h_1\} \, / \! \sim \; 
    \cong \; \Hom(\pi_1(\mathbb T^2),G) \, / \! \sim \, ,
\end{equation}
representatives of which label the topological sectors of the $G$-Ising model. Coupling such an equivalence class $[h_1,h_2]$ of background flat gauge fields to the $G$-Ising model \eqref{eq:ZIsingG0} results in the partition function
\begin{equation}
    \label{eq:ZIsingG}
    \mc Z^{\Vect_G}(\Sigma_\triangle^\vee;\theta)([h_1,h_2])\; \propto
    \sum_{\sigma \in G^{\msf V(\Stdss)}} 
    \prod_{\msf e \in \msf E(\Std)} \tht_\msf e \big(\sigma_{\pme}^{-1} \,  g(h_1,h_2)_\msf e \, \sigma_{\ppe}^{\phantom{-1}} \big) \, ,
\end{equation}
where $g : \Hom(\pi_1(\mathbb T^2),G)\to G^{\msf E(\Std)}$ is a background gauge field with holonomies $h_1,h_2 \in G$ around their respective non-contractible cycles.\footnote{The background gauge field can defined in analogy to the abelian connection \eqref{eq:gaugeFieldA}.} Crucially, $\mc Z^{\Vect_G}(\Std; \theta)( [h_1,h_2])$ does not depend on a choice of representative in $[h_1,h_2]$.

Let us now confirm that this characterisation is compatible with the general statement that topological sectors are in one-to-one correspondence with simple objects in the Drinfel'd center of the symmetry fusion category. It is well known that isomorphism classes of simple objects in $\ms Z(\Vect_G)$ are labelled by pairs $([c_1],V_{c_1})$ consisting of a conjugacy class $[c_1]$ in $G$ represented by $c_1$ and an irreducible representation $V_{c_1}$ of the centraliser $Z_G(c_1)$ of $c_1$ in $G$. Given $([c_1],V_{c_1})$, the corresponding simple object in $\ms Z(\Vect_G)$ is the $G$-graded vector space $V = \bigotimes_{g \in G} V_g$ such that $V_g \cong V_{c_1}$ whenever $g \in [c_1]$, and zero otherwise. The half-braiding isomorphisms can be explicitly constructed from the knowledge of the action of $Z_G(c_1)$ on the vector space $V_{c_1}$. We explained in sec.~\ref{sec:lines} how to construct a topological state in $\mc Z_{\Vect_G}(\mathbb T^2_\triangle)$ associated with a given simple object in $\ms Z(\Vect_G)$. It is the state assigned by the topological theory to $\mathbb T^2_\triangle$ in the presence of the network of topological lines depicted in eq.~\eqref{eq:linesC}. We denote the resulting topological state by $|\Sigma_\triangle,\Vect_G,\Vect_G,([c_1],V_{c_1}) \ra \in \mc Z_{\Vect_G}(\Sigma_\triangle)$. As per eq.~\eqref{eq:twistedZ}, the partition function of the $G$-Ising model in the  topological sector $([c_1],V_{c_1})$ reads
\begin{align}
    \label{eq:ZIsingG_Z}
    \mc Z^{\Vect_G}(\Std;\theta)([c_1],V_{c_1}) 
    &=
    \la \Sigma_\triangle,\Vect_G, \Vect_G,([c_1],V_{c_1}) \, | \, \Sigma_\triangle, \Vect_G ; \theta, \vartheta^{\rm triv. }\ra 
    \\ \nn & \; \propto \!  
    \sum_{\substack{h_1 \in [c_1] \\ h_2 \in Z_{G}(h_1)}} \!\!
    {\rm tr}\big(\eta^\vee(q_{h_1}^{-1}h_2q_{h_1})\big) \!\!
    \sum_{\sigma \in G^{\msf V(\Stdss)}} 
    \prod_{\msf e \in \msf E(\Std)} \tht_\msf e \big(\sigma_{\pme}^{-1} \,  g(h_1,h_2)_\msf e \, \sigma_{\ppe}^{\phantom{-1}} \big) \, ,
\end{align}
where $\eta : Z_G(c_1) \to \End_\mathbb C(V_{c_1})$ and $q_{h_1} \in G$ is such that $c_1 = q_{h_1}^{-1}h_1q_{h_1}$ for all $h_1 \in [c_1]$. Of course, we can consider arbitrary linear combinations of states in $\mc Z_{\Vect_G}(\Sigma_\triangle)$. Given $h_1,h_2 \in G$ such that $h_1h_2=h_2h_1$, consider in particular the state
\begin{equation}
    \frac{1}{|Z_G(h_1)|}\sum_{V_{h_1} \in \widehat{Z_G(h_1)}}
    \!\!\!
    \tr\big(\eta(h_2) \big)\, | \Sigma_\triangle, \Vect_G,\Vect_G,([h_1],V_{h_1}) \ra
\end{equation}
where $\eta : Z_G(h_1) \to \End_\mathbb C(V_{h_1})$. By inspection, one finds that the inner product between this topological state and $|\Sigma_\triangle, \Vect_G ; \theta, \vartheta^{\rm triv.} \ra$ coincides with $\mc Z^{\Vect_G}(\Std;\theta)([h_1,h_2])$.

\subsection{Renormalisation group fixed points of gapped phases\label{sec:RG}}

Given a fusion category $\mc C$, it was argued in ref.~\cite{Thorngren:2019iar,Komargodski:2020mxz,Bhardwaj:2023fca} that one-dimensional $\mc C$-symmetric gapped phases of matter are in one-to-one correspondence with (finite semisimple) $\mc C$-module categories, such that vacua are labelled by simple objects in the corresponding module categories.  Given such a finite semisimple $\mc C$-module category $\mc M$, there exists an \emph{associative algebra object} $\mc A$ in $\mc C$ such that $\mc M$ is equivalent to the category $\Mod_\mc C(\mc A)$ of $\mc A$-modules in $\mc C$ \cite{etingof2016tensor}. Endowing the algebra $A$ with a $\Delta$-separable symmetric Frobenius structure \cite{Fuchs:2002cm}, one can build a renormalisation group fixed point as follows:\footnote{An analogous strategy was employed in ref.~\cite{Inamura:2021szw} in order to construct fixed point Hamiltonians in (1+1)d.} Let us denote the multiplication and the section of the algebra by $\mu : \mc A \otimes \mc A \to \mc A$ and $\Delta: \mc A \to \mc A \otimes \mc A$, respectively. It follows from the semisimplicity of $\mc C$ that one can decompose $\mc A$ into simple objects. Given a simple object $X$ in $\mc C$, we notate as $|X \mc A,i \ra$ a choice of basis in the hom-space $\Hom_\mc C(X,\mc A) \equiv \mc C^X_\mc A$. The multiplication $\mu$ boils down to the \emph{structure constants} of the algebra defined as follows \cite{Fuchs:2002cm}:
\begin{equation}
    \label{eq:mSymbols}
    \algebraObj{1} \hspace{-10pt} = \sum_{i}  \big(\mu^{X_1X_2}_{X_3 } \big)_{jkl}^i
    \!\!\!\!\!\!\!\! \algebraObj{2} \!\! ,
\end{equation}
where $j,k,l$ and $i$ label basis vectors $|X_1 \mc A,j\ra \in \mc C^{X_1}_\mc A$, $|X_2 \mc A,k\ra \in \mc C^{X_2}_\mc A$, $\la X_3 \mc A,l| \in \big(\mc C^{X_3}_\mc A \big)^{\! \vee}$ and $|X_1X_2X_3,i\ra \in \mc C^{X_1X_2}_{X_3}$, respectively. The algebra object $\mc A$ labels a gapped $\mc C$-symmetric phase that preserves the topological lines labelled by simple objects in $\mc A$. The partition function of the corresponding fixed point can be realised as the inner product between the topological state $| \Si_\Up, \mc C, \mc C \ra$ and a state in the kinematical Hilbert space $\mc H_\mc C(\Si_\Up)$ loosely defined as follows: Assign to every oriented edge in $\Si_\Up$ the object $\mc A$ and choose at every junction the vector in the hom-space provided by the multiplication.\footnote{Strictly speaking, not all junctions in $\mc H_\mc C(\Si_\Up)$ are of the type depicted in eq.~\eqref{eq:mSymbols}, in which case the morphism is not quite the multiplication $\mu$. For instance, the junction obtained by turning these string diagrams upside down is rather provided by the section $\Delta$. Additional junctions would require the symmetric Frobenius structure of $\mc A$.}

Importantly, the construction sketched above is valid for any fusion category $\mc C$, but the resulting state only falls within the class of boundary states considered in sec.~\ref{sec:non_topo} whenever $\mc C$ admits a fiber functor. Supposing $\mc C$ does admit a fiber functor and choosing the module structure over $\Vect$ to be trivial---which does not necessarily mean that the $\F{\cat}$-symbols are trivial---the desired state is in fact of the form $| \Sigma_\Upsilon,\mc C,\mc A;\theta^{\rm triv.}, \vartheta^\mc A \ra$. Indeed, given $\msf e \in \msf E({\rm int}(\SIgb_\Upsilon))$, suppose that $\mc H(\msf e, \fr A, \fr N)$ is of the form $\Hom_\mc C(X_1^j \otimes X_2^k , X_3^l)$, where $j,k$ and $l$ are multiplicity labels for the simple objects $X_1, X_2$ and $X_3$ in $\mc A$, respectively. Given basis vectors $| i_\msf e \ra \equiv | X_1^j X_2^k X_3^l,i \ra$, one defines $ \vartheta^\mc A_\msf e(\fr A,\fr N) : \mc H(\msf e, \fr A, \fr N) \to \mc H(\msf e, \fr A, \fr N)$ via
\begin{equation}
   (\vartheta_{\msf e}^{\mc A})^{i_{\msf e}}_{i_{\msf e}} = \langle i_{\msf e} | \vartheta_{\msf e}^{\mc A} | i_{\msf e} \rangle
   := \big(\mu^{X_1X_2}_{X_3} \big)_{jkl}^i \, . 
\end{equation}
We then obtain the partition function of the renormalisation group fixed point associated with the algebra $\mc A$ as $\la \Sigma_\Up,\mc C,\mc C\, | \, \Sigma_\Upsilon,\mc C, \mc A; \theta^{\rm triv.},\vartheta^\mc A \ra$. Whenever $\mc C$ does not admit a fiber functor, it remains possible to realise the partition function of the renormalisation group fixed point within the framework of sec.~\ref{sec:non_topo}, but the boundary state needs to be defined differently, choosing for instance $\mc N$ to be $\Mod_\mc C(\mc A)$ itself and specific Boltzmann weights $\theta$, while $\vartheta$ can be trivial. We discuss such a case in sec.~\ref{sec:FT_fixed}. Finally, note that considering instead the topological state $| \Sigma_\Upsilon, \mc C, \mc M \ra$ would still yield a renormalisation group fixed point but for a different gapped symmetric phase with respect to the symmetry $\mc C^\vee_\mc M$. We delve deeper into this aspect in sec.~\ref{sec:gauged_fixed}.

Let us examine a couple of explicit examples. Every indecomposable $\Vect_G$-module category being of the form $\Mod_\mc C(\mc A)$, Morita classes of separable algebra objects in $\Vect_G$ are also labelled by pairs $(A,\psi)$ consisting of a subgroup $A \subseteq G$
and a normalised representative of a cohomology class $[\psi] \in H^2(G,\mathbb C^\times)$. The algebra object in $\Vect_G$ associated with $(A,1)$ is simply given by the group algebra $\mathbb C[A] \cong \bigoplus_{a \in A}\mathbb C_a$ so that $\big(\mu^{a_1 \, a_2}_{(a_1 a_2)}\big)_{111}^1 =1$, i.e. $\vartheta^{\mathbb C[A]} = \vartheta^{\rm triv.}$. The corresponding gapped phase is that preserving the subsymmetry $A$. It follows from comments in sec.~\ref{sec:brane} that separable algebra objects in $\Rep(G)$ are also labelled by pairs $(A,\psi)$. The algebra object in $\Rep(G)$ associated with $(A,1)$ is the $G$-algebra of functions in $\mathbb C^{A \setminus G}$ with pointwise multiplication such that $\Mod_{\Rep(G)}(\mathbb C^{A \setminus G}) \simeq \Rep(A)$ \cite{DAVYDOV2010319}. As an object in $\Rep(G)$, $\mathbb C^{A \setminus G}$ defines the permutation representation, which is isomorphic to the induced representation ${\rm Ind}_A^G(\ub 0_A)$ in $G$ of the trivial representation $\ub 0_A$ of $A$. Working out the structure constants in this case is more delicate. It requires decomposing the permutation representation in terms of irreducible representations and expressing the pointwise multiplication in a chosen basis for the hom-spaces in $\Rep(G)$. But the final result can be conveniently expressed in terms of the module associator of $\Rep(A)$ over $\Rep(G)$ as \cite{Fuchs:2002cm}
\begin{equation}
    \label{eq:muSym}
    \big( \mu^{V_1 V_2}_{V_3} \big)_{jkl}^i =
    \big(\F{\cat}^{\ub 0_A V_1 V_2}_{\ub 0_A} \big)_{V_3,il}^{\ub 0_A, kj} \, ,
\end{equation}
for every $V_1,V_2,V_3 \in \Rep(G)$ and basis vectors in the relevant hom-spaces.
\bigskip
\section{Fourier transform of $\Vect_G$-symmetric theories \label{sec:VecG-FT}}

\emph{In this section, we consider the $\Vect_G$-symmetric generalisations of the Ising model introduced previously, realised as boundary theories of topological theories, and perform in this setting the Fourier transform of the corresponding Boltzmann weights in arbitrary (topological) sectors.}

\subsection{Fourier transform}

Previously, we realised the (generalised) $G$-Ising model with partition function \eqref{eq:ZIsingG0} as a boundary theory of the Turaev--Viro--Barrett--Westbury theory $\mc Z_{\Vect_G}$. We were able to do so because we showed in sec.~\ref{sec:lines} that we could construct topological states in the presence of topological lines labelled by $(\Vect_G)^\vee_{\Vect_G} \simeq \Vect_G$, guaranteeing that the $G$-Ising model can be coupled with topological lines in $\Vect_G$. But the same $\Vect_G$ symmetry can be realised in different ways, stemming from the fact that there exist multiple combinations of fusion category $\mc C$ and $\mc C$-module category $\mc M$ such that $\mc C^\vee_\mc M \simeq \Vect_G$. In particular, one has $(\Rep(G))^\vee_\Vect \simeq \Vect_G$. Therefore, we can expect a dual description of the $G$-Ising model where configuration variables are valued in $\Rep(G)$. This alternative description can be found starting from eq.~\eqref{eq:ZIsingG0} by performing a Fourier transform of all the Boltzmann weights. 

Let us carry out this derivation as explicitly as possible. Recall that for every function $\theta : G \to \mathbb C$, one can define its Fourier transform $\theta^\vee$ via
\begin{equation}
    \label{eq:IFT}
    \theta^\vee (\rho) = \frac{1}{|G|}\sum_{g \in G} 
    \theta(g) \rho(g) \in \End_\mathbb C(V) \, , 
\end{equation}
for any $\rho : G \to \End_\mathbb C(V)$. It follows from the orthogonality of the representation matrices that the inverse Fourier transform reads
\begin{equation}
\label{eq:FT}
    \theta(g) = \sum_{V \in \widehat G} d_V  \, {\rm tr}_{V} \big( \theta^\vee(\rho) \rho^\vee(g) \big) 
\end{equation}
such that $\rho^\vee(g)_v^w = \rho(g^{-1})_w^v$.
Applying eq.~\rf{eq:FT} to the weight $\tht_{\msf e}$ in the partition function \rf{eq:ZIsingG0}, and using the defining property of a group representation, we find
\beq
    \label{eq:FTG1}
    \mc Z^{\Vect_G}(\Std;\theta) 
    = 
    \sum_{V \in \widehat G^{\msf E(\Stdss)}} \! 
    \sum_{\sigma \in G^{\msf V(\Stdss)}}
    \prod_{\msf e \in \msf E(\Std)} \!
    d_{V_\msf e} \,  
    \theta^\vee_{\msf e}(\rho_{\msf e})_{v_{\msf e}}^{w_{\msf e}} \,
    \rho_{\msf e}^\vee(\si_{\pme}^{-1})_{w_\msf e}^{u_\msf e} \, \rho_{\msf e}^\vee(\si^{\phantom{-1}}_{\ppe})_{u_{\msf e}}^{v_{\msf e}} \, ,
\eeq
where $\widehat G \equiv \mc I_{\Rep(G)}$, $\rho_\msf e \equiv \rho(\msf e) : G \to \End_\mathbb C(V_\msf e)$.
Here, for a simple object $\rho:G \to \End_\mathbb C(V)$, we have written the trace over $V$ in terms of explicit matrix indices, using the convention that repeated indices are summed over. In particular, for each edge $\msf e \in \msf E(\Std)$, we have introduced the set of indices $\{u_{\msf e}, v_{\msf e}, w_{\msf e}\}$, the range of each index depending on the corresponding representation, namely $1,\ldots,\dim_\mathbb C V_\msf e$. The product over the edges can be reorganised so as to obtain\footnote{Notice that even though eq.~\eqref{eq:FTG2} does not naively display any repeated matrix indices, writing down explicitly the various products would reveal repeated indices associated with the same edges in $\Std$, which would then be summed over.}
\begin{align}
    \label{eq:FTG2}
    \mc Z^{\Vect_G}(\Std;\theta) = 
    \sum_{V \in \widehat G^{\msf E(\Stdss)}} \!\!
    \bigg( \prod_{\msf e \in \msf E(\Si^\vee_\triangle)} \!\! \theta^\vee_{\msf e}(\rho_{\msf e})_{v_{\msf e}}^{w_{\msf e}} \bigg) 
    \!\!
    \prod_{\msf v \in \msf V(\Si^\vee_\triangle)}
    \sum_{\si_{\msf v} \in G}
    \bigg(
    \prod_{\msf e' \leftarrow \msf v} d_{V_{\msf e'}}^\frac{1}{2} \, \rho_{\msf e'}(\si_{\msf v})_{u_{\msf e'}}^{w_{\msf e'}}
    \bigg)
    \bigg(
    \prod_{\msf e'' \rightarrow \msf v} d_{V_{\msf e''}}^\frac{1}{2} \,\rho^\vee_{\msf e''}(\si_{\msf v})_{u_{\msf e''}}^{v_{\msf e''}}  
    \bigg) .
\end{align}
This allows us to perform the sum over $\si \in G^{\msf V(\Std)}$, yielding invariant tensors at each vertex. Recall that as per the conventions of sec.~\ref{sec:state-sum}, $\Std$ only admits 3-valent vertices. Moreover, these vertices come in only two types: Either there are exactly two incoming edges, or there are exactly two outgoing edges. Given any vertex $\msf v \in \msf V(\Std)$, we notate via $\msf e_{\msf v, 1}$ and $\msf e_{\msf v, 2}$ the two incoming or outgoing edges and via $\msf e_{\msf v, 3}$ the remaining one. Whenever there are two incoming edges at $\msf v$, summing over the configuration variable $\sigma_\msf v\in G$ yields 
\begin{equation}
\begin{split}
    \label{eq:CC*1}
    &\frac{1}{|G|} \sum_{\si_{\msf v} \in G}
    (d_{V_{\msf e_{\msf v, 3}}})^\frac{1}{2} \, 
    \rho_{\msf e_{\msf v, 3}}(\si_{\msf v})_{u_{\msf e_{\msf v, 3}}}^{w_{\msf e_{\msf v, 3}}}
    (d_{V_{\msf e_{\msf v,1}}} d_{V_{\msf e_{\msf v,2}}})^\frac{1}{2} \, 
    \rho^\vee_{\msf e_{\msf v, 1}} (\si_{\msf v})_{u_{\msf e_{\msf v, 1}}}^{v_{\msf e_{\msf v, 1}}} \,
    \rho^\vee_{\msf e_{\msf v, 2}} (\si_{\msf v})_{u_{\msf e_{\msf v, 2}}}^{v_{\msf e_{\msf v, 2}}} \,
    \\
    & \q = 
    \sum_{i_\msf v}
    \Big(\frac{d_{V_{\msf e_{\msf v,1}}} d_{V_{\msf e_{\msf v,2}}}}{d_{V_{\msf e_{\msf v,3}}} }\Big)^{\frac{1}{4}} 
    \CC{V^\vee_{\msf e_{\msf v, 1}}}{V^\vee_{\msf e_{\msf v, 2}}}{V^\vee_{\msf e_{\msf v, 3}}}{u_{\msf e_{\msf v, 1}}}{u_{\msf e_{\msf v, 2}}}{u_{\msf e_{\msf v, 3}}}_{i_\msf v}
    \Big(\frac{d_{V_{\msf e_{\msf v,1}}} d_{V_{\msf e_{\msf v,2}}}}{d_{V_{\msf e_{\msf v,3}}} }\Big)^{\frac{1}{4}} 
    \CC{V^\vee_{\msf e_{\msf v, 1}}}{V^\vee_{\msf e_{\msf v, 2}}}{V^\vee_{\msf e_{\msf v, 3}}}{v_{\msf e_{\msf v, 1}}}{v_{\msf e_{\msf v, 2}}}{w_{\msf e_{\msf v, 3}}}_{i_\msf v}^*.
\end{split}
\end{equation}
where $i_\msf v = 1 , \ldots, \dim_\mathbb C \Hom_{\Rep(G)}(V_{\msf e_{\msf v,1}} \otimes V_{\msf e_{\msf v,2}}, V_{\msf e_{\msf v,3}})$. Similarly, whenever there are two outgoing edges at $\msf v$, one obtains
\begin{equation}
\begin{split}
    \label{eq:CC*2}
    &\frac{1}{|G|} \sum_{\si_{\msf v} \in G}
    (d_{V_{\msf e_{\msf v,1}}} d_{V_{\msf e_{\msf v,2}}})^\frac{1}{2} \, 
    \rho_{\msf e_{\msf v, 1}} (\si_{\msf v})_{u_{\msf e_{\msf v, 1}}}^{w_{\msf e_{\msf v, 1}}} \,
    \rho_{\msf e_{\msf v, 2}} (\si_{\msf v})_{u_{\msf e_{\msf v, 2}}}^{w_{\msf e_{\msf v, 2}}} \,
    (d_{V_{\msf e_{\msf v, 3}}})^\frac{1}{2} \, 
    \rho^\vee_{\msf e_{\msf v, 3}}(\si_{\msf v})_{u_{\msf e_{\msf v, 3}}}^{v_{\msf e_{\msf v, 3}}}
    \\
    & \q = 
    \sum_{i_\msf v}
    \Big(\frac{d_{V_{\msf e_{\msf v,1}}} d_{V_{\msf e_{\msf v,2}}}}{d_{V_{\msf e_{\msf v,3}}} }\Big)^{\frac{1}{4}} 
    \CC{V_{\msf e_{\msf v, 1}}}{V_{\msf e_{\msf v, 2}}}{V_{\msf e_{\msf v, 3}}}{u_{\msf e_{\msf v, 1}}}{u_{\msf e_{\msf v, 2}}}{u_{\msf e_{\msf v, 3}}}_{i_\msf v}
    \Big(\frac{d_{V_{\msf e_{\msf v,1}}} d_{V_{\msf e_{\msf v,2}}}}{d_{V_{\msf e_{\msf v,3}}} }\Big)^{\frac{1}{4}} 
    \CC{V_{\msf e_{\msf v, 1}}}{V_{\msf e_{\msf v, 2}}}{V_{\msf e_{\msf v, 3}}}{w_{\msf e_{\msf v, 1}}}{w_{\msf e_{\msf v, 2}}}{v_{\msf e_{\msf v, 3}}}_{i_\msf v}^* ,
\end{split}
\end{equation}
At this point, a couple of observations are in order: (i) Both \rf{eq:CC*1} and \rf{eq:CC*2} provide the non-vanishing entries of the contraction of a pair of tensors, which evaluate to Clebsch-Gordan coefficients (or their conjugate) times some multiplicative factors that depend on the dimension of representations, along multiplicity indices in $\Rep(G)$. These tensors precisely coincide with eq.~\eqref{eq:PEPS} for $(\mc C,\mc M) = (\Rep(G),\Vect)$. For instance, whenever there are two incoming edges, these tensors are of the form\footnote{Recall from eq.~\eqref{eq:evaluation} that the strings supporting the objects in $\mc C$ in the definition of the tensors are oriented backwards compared to the corresponding edges in $\Sigma_\Upsilon$.}
\begin{equation}
    \PEPS{dotmod}{v_1}{i}{v_3}{v_2}{}{}{}{V_1}{V_2}{V_3}{1} =
    \Big(\frac{d_{V_1} d_{V_2}}{d_{V_3}} \Big)^{\frac{1}{4}}
    \CC{V_1}{V_2}{V_3}{v_1}{v_2}{v_3}_i  \, , 
\end{equation}
where dotted lines depict the unique simple object in $\Vect$.
(ii) Expression  \rf{eq:FTG2} for the partition function further involves the Fourier transform $\theta^\vee_\msf e$ of the Boltzmann weights. Importantly, the matrices $\theta^\vee_\msf e(\rho_\msf e)$ appear with indices of the type $v_\msf e$ and $w_\msf e$, but not $u_\msf e$. Moreover, notice that in eq.~\eqref{eq:CC*1} and \eqref{eq:CC*2}, there are no Clebsch-Gordan coefficients mixing indices of the type $u_\msf e$ with others. This means that the tensors of the form \eqref{eq:PEPS} are naturally divided into two disjoint sets such that one set carries indices of the types $(v_\msf e, w_\msf e)$, whereas the other set only carries indices of the $u_\msf e$. The former set of tensors are contracted to each other via the matrices $\theta^\vee_\msf e(\rho_\msf e)$, whereas the latter one are directly contracted to each other, while multiplicity indices in $\Rep(G)$ remain free. The result of these contractions are the tensor network states $|\Std,\Rep(G), \theta^\vee \ra$ and $| \Std, \Rep(G), \Vect \ra$, respectively. Finally, contracting these tensor network states along multiplicity indices as per eq.~\eqref{eq:CC*1} and \eqref{eq:CC*2} precisely recovers the partition function \eqref{eq:ZIsingG0}. In symbols,
\begin{equation}
    \label{eq:FTGf}
    \mc Z^{\Vect_G}(\Std;\theta) = \la \Std, \Rep(G), \Vect \, | \, \Std, \Rep(G), \theta^\vee, \vartheta^{\rm triv.} \ra \, ,
\end{equation}
as requested. Practically, it means that, instead of explicitly carrying out the Fourier transform of the partition function of a $\Vect_G$ symmetric theory, we can immediately invoke eq.~\eqref{eq:FTGf}, and generalisations thereof. It is worth emphasising that eq.~\eqref{eq:FTGf} is a very natural way of organising the result of the Fourier transform, and can be used to motivate a posteriori the realisation of the $G$-Ising model as a boundary theory of a topological theory. 

It is interesting to note that it is possible to define a generalised Fourier transform. More specifically, for every representative of the Morita class of $\Vect_G$, one can define a Fourier transform, the same way the ordinary Fourier transform is associated with the $\Vect_G$-module category $\Vect$. In this scenario, the configuration variables are valued in $(\Vect_G)^\vee_{\mc M(A)}$ rather than $(\Vect_G)^\vee_{\Vect} \simeq \Rep(G)$. Similarly, we expect that similar transformations should exist for any spherical fusion category $\mc C$ admitting distinct indecomposable module categories. We postpone a study of this more general scenario to another manuscript.

\subsection{Fourier transform in the presence of topological lines\label{sec:FT_lines}}

We have established above the following equality of partition functions for the $G$-Ising model in the trivial topological sector:
\begin{equation}
     \la \Sigma_\triangle,\Vect_G,\Vect_G \, | \, \Sigma_\triangle,\Vect_G;\theta, \vartheta^{\rm triv.} \ra 
    =
     \la \Std, \Rep(G), \Vect \, | \, \Std, \Rep(G), \theta^\vee, \vartheta^{\rm triv.} \ra \, .
\end{equation}
Let us now consider the partition function $\mc Z_{\Vect_G}(\Std;\theta)([h_1,h_2])$ obtained by inserting topological lines in $\Vect_G$. Even though the $\Vect_G$ symmetry is obviously preserved by the Fourier transform, its realisation differs, as embodied by tensors \eqref{eq:MPO}. We wish to confirm this statement by explicitly computing the Fourier transform in the presence of topological lines. The derivation largely follows the same steps as previously with slight modifications. Consider the Fourier transform of the Boltzmann weights in \eqref{eq:ZIsingG}:
\begin{equation}
\begin{split}
    \tht_\msf e \big(\sigma_{\pme}^{-1} \,  g(h_1,h_2)_\msf e \, \sigma_{\ppe}^{\phantom{-1}} \big)
    =
    \sum_{V_\msf e}d_{V_\msf e} \, 
    \theta^\vee_{\msf e}(\rho_{\msf e})_{v_{\msf e}}^{w_{\msf e}} \,
    \rho_{\msf e}^\vee \! (\si^{-1}_{\pme})_{w_{\msf e}}^{u_{\msf e}} \,
    \rho^\vee_\msf e \! \big(g(h_1,h_2)_\msf e\big)_{u_{\msf e}}^{t_{\msf e}}\,
    \rho_{\msf e}^\vee(\si^{\phantom{-1}}_{\ppe})_{t_{\msf e}}^{v_{\msf e}} \, .
\end{split}
\end{equation}
Since factors of the form $\rho^\vee (g(h_1,h_2)_\msf e)_{u_{\msf e}}^{t_{\msf e}}$ do not depend on the configuration $\si$ that is summed over, these can be placed beside Boltzmann weights $\theta^\vee$ in eq.~\rf{eq:FTG2}. The subsequent steps remain unchanged. We must simply include these additional factors in the tensor network producing the topological state following the contraction pattern. But these factors precisely coincide with the non-vanishing entries of tensors \eqref{eq:MPO} encoding the coupling of topological lines in $\Vect_G$ for input spherical fusion category $\Rep(G)$ and $\Rep(G)$-module category $\Vect$, namely
\beq
    \MPO{dotmod}{dotmod}{1}{1}{u}{t}{}{}{}{}{V}{\mathbb C_g}{1} = \rho (g)_u^t \, ,
\eeq
where dotted lines depict the unique simple object in $\Vect$ and $\rho : G \to \End_\mathbb C(V)$. Naturally, the same statement holds for any linear combination of topological sectors so that
\begin{equation}
    \label{eq:ZIsingGFT}
    \mc Z^{\Vect_G}(\Std;\theta)([c_1],V_{c_1}) = \la \Std, \Rep(G), \Vect,([c_1],V_{c_1}) \, | \, \Std, \Rep(G), \theta^\vee, \vartheta^{\rm triv.} \ra \, ,
\end{equation}
for every simple object $([c_1],V_{c_1})$ in the Drinfel'd center of $\Vect_G$.

\subsection{Fourier transform of renormalisation group fixed points\label{sec:FT_fixed}}

Given a subgroup $A \subseteq G$, we explained in sec.~\ref{sec:RG} how to construct the partition function of a renormalisation group fixed point for the gapped phase preserving the symmetry $A$. Spelling out the definition, one obtains
\begin{equation}
\begin{split}
    \label{eq:ZFixedA}
    \mc Z^{\Vect_G}(\Std;\theta^{\rm triv.}, \vartheta^{\mathbb C[A]}) :&= 
    \la \Sigma_\triangle, \Vect_G, \Vect_G \, | \, \Sigma_\triangle, \Vect_G, \mathbb C[A] ; \theta^{\rm triv.}, \vartheta^{\mathbb C[A]} \ra
    \\
    &=     
    \!\!
    \sum_{\sigma \in G^{\msf V(\Stdss)}} 
    \prod_{\msf e \in \msf E(\Std)} 
    1_A \big(\sigma_{\pme}^{-1} \sigma_{\ppe}^{\phantom{-1}} \big) 
    \\
    &=
    \la \Sigma_\triangle,\Vect_G,\Vect_G \, | \, \Sigma_\triangle, \Vect_G;1_A,\vartheta^{\rm triv.} \ra \, ,
\end{split}
\end{equation}
where $1_A : G \to \{0,1\}$ is the characteristic function of the subgroup $A$. Despite the different setting, the alternative parametrisation in the last line of eq.~\eqref{eq:ZFixedA} makes it possible to employ formula \eqref{eq:FTGf}:
\begin{equation}
    \mc Z^{\Vect_G}(\Std; \theta^{\rm triv.}, \vartheta^{\mathbb C[A]}) = \la \Std, \Rep(G),\Vect \, | \, \Std, \Rep(G); 1_A^\vee, \vartheta^{\rm triv.} \ra \, .
\end{equation}
By definition of the inverse Fourier transform, one has
\begin{equation}
    1^\vee_A(\rho) = \frac{1}{|G|} \sum_{g \in G}1_A(g) \, \rho(g)
    = \frac{1}{|G|}\sum_{a \in A} \rho(a) \, ,
\end{equation}
for all $\rho : G \to \End_\mathbb C(V)$ in $\widehat G$. 
Schur's orthogonality relation together with Frobenius reciprocity stipulate that $1^\vee_A(\rho)$ is zero unless $\rho$ appears in the decomposition into irreducible representations of the $G$-representation ${\rm Ind}_A^G(\ub 0_A)$ induced from the trivial representation of $A$. Moreover, recall that as a $G$-representation, ${\rm Ind}_A^G(\ub 0_A)$ is isomorphic to $\mathbb C^{A \setminus G}$. But, invoking the trivial representation $\eta : A \to \End_\mathbb C(\ub 0_A)$, one can write the matrix entries of $1^\vee_A(\rho)$ in terms of Clebsch-Gordan coefficients in $\Rep(A)$ as follows: 
\begin{equation}
    \label{eq:characDual}
    1^\vee_A(\rho)_{v}^w
    = 
    \frac{1}{|G|}\sum_{a \in A} \eta(a) \, \rho(a)_v^w \, \eta(a)
    = |A \setminus G|
    \sum_{i} \CC{\ub 0_A}{V}{\ub 0_A}{1}{v}{1}_i
    \CC{\ub 0_A}{V}{\ub 0_A}{1}{w}{1}_i^* \, ,
\end{equation}
where the range of $i$ is equal to the multiplicity of $\ub 0_A$ in the restriction ${\rm Res}^G_A(V)$ of $V$. Inspecting the definition of the state $|\Std, \Rep(G);1^\vee_A,\vartheta^{\rm triv.}\ra$, one can invoke eq.~\eqref{eq:characDual} to rewrite it as a contraction of tensors of the form \eqref{eq:MPO} for $(\mc C, \mc M) = (\Rep(G),\Rep(A))$ but where the simple objects in $\Rep(A)$ are restricted to be $\ub 0_A$, which can be imposed via an appropriate choice of Boltzmann weights $\theta$. The non-vanishing entries of these tensors are of the form $(d_{V_1} d_{V_2} /d_{V_3})^\frac{1}{4} \big(\F{\cat}^{\ub 0_A V_1 V_2}_{\ub 0_A} \big)_{V_3,il}^{\ub 0_A,kj}$. Finally, it follows from eq.~\eqref{eq:muSym} that we can equivalently rewrite the partition function of the fixed point model as
\begin{equation}
    \mc Z^{\Vect_G}(\Std; \theta^{\rm triv.}, \vartheta^{\mathbb C[A]}) 
    =
    \la \Std, \Rep(G), \Vect \, | \, \Std, \Rep(G), \mathbb C^{A \setminus G}; \theta^{\rm triv.}, \vartheta^{\mathbb C^{A \setminus G}} \ra \, .
\end{equation}
In the following, we shall contemplate the interplay between algebra objects and gapped symmetric phases as one changes the basis of topological states entering the definition of the partition function.
\bigskip
\section{Gauging the $\Vect_G$ symmetry\label{sec:gauging}}

\emph{In this section, we perform the partial or total gauging of the $\Vect_G$ symmetry of the $G$-Ising model, both in terms of the original formulation and its Fourier transform. The actions of the resulting topological lines are explicitly computed in terms of the tensor network operators.}

\subsection{Gauging via a choice of brane boundary condition\label{sec:gauging_brane}}

We explained in sec.~\ref{sec:GIsing} that by virtue of its symmetry, the $G$-Ising model can be coupled to background flat gauge fields, which amounts to inserting a fine enough network of topological lines in $\Vect_G$. Given a subgroup $A \subseteq G$, gauging the corresponding subsymmetry is performed by summing over all possible such insertions of topological lines in $\Vect_A$. 
The partition function of the resulting theory reads
\begin{equation}
    \label{eq:gauged_ZIsing0_1}
    \mc Z^{(\Vect_G)^\vee_{\mc M(A)}}(\Std;\theta)
    \; \propto
    \sum_{\sigma \in G^{\msf V(\Stdss)}}
    \sum_{a \in A^{\msf E(\Stdss)}}
    \bigg(\prod_{\msf p \in \msf P(\Std)} \delta_{\mathbb 1,a_\msf p} \bigg)
    \prod_{\msf e \in \msf E(\Std)} \tht_\msf e \big(\sigma_{\pme}^{-1} \,  a_\msf e \, \sigma_{\ppe}^{\phantom{-1}} \big) \, ,
\end{equation}
where we are using the shorthand notation $a_\msf e \equiv a(\msf e)$, and $a_\msf p := \prod_{\msf e \subset \partial \msf p}a_\msf e$ is over edges in the boundary of $\msf p$ ordered counterclockwise and assumed to have the orientation opposite to that induced by the orientation of $\msf p$.
Our choice of notation for the partition function, namely $\mc Z^{(\Vect_G)^\vee_{\mc M(A)}}(\Std;\theta)$, will be justified in sec.~\ref{sec:lines_gauging}. By construction, every `matter' degree of freedom $\sigma_\msf v \in G$ can always be decomposed as $\sigma_\msf v = a_\msf v \tilde \sigma_\msf v$, where $a_\msf v \in A$ and $\tilde \sigma_\msf v$ is the representative of a coset in $A \setminus G$. Redefining the gauge field $a_\msf e \mapsto a_{\pme}^{\phantom{-1}}a_\msf e a_{\ppe}^{-1}$ yields
\begin{equation}
    \label{eq:gauged_ZIsing0_3}
    \mc Z^{(\Vect_G)^\vee_{\mc M(A)}}(\Std;\theta)
    \; \propto
    \sum_{A\sigma \in (A \setminus G)^{\msf V(\Stdss)}}
    \sum_{a \in A^{\msf E(\Stdss)}}
    \bigg(\prod_{\msf p \in \msf P(\Std)} \delta_{\mathbb 1,a_\msf p} \bigg)
    \prod_{\msf e \in \msf E(\Std)} \tht_\msf e \big(\sigma_{\pme}^{-1} \,  a_\msf e \, \sigma_{\ppe}^{\phantom{-1}} \big) \, .
\end{equation}
In the spirit of ref.~\cite{Lootens:2021tet,Delcamp:2023kew}, and more generally of the SymTFT perspective \cite{Gaiotto:2020iye}, we rather perform this gauging via a choice of brane boundary condition. Recall from sec.~\ref{sec:brane} that brane boundary conditions are labelled by indecomposable module categories over $\Vect_G$. We reviewed in sec.~\ref{sec:brane} that indecomposable module categories over $\Vect_G$ or $\Rep(G)$ are labeled by the same data, namely pairs $(A, \psi)$ consisting of a subgroup $A \subseteq G$ and normalised representative $\psi$ of a cohomology class $[\psi] \in H^2(G,\mathbb C^\times)$.
Invoking our construction, we claim that gauging the subsymmetry can be equivalently performed by replacing the brane boundary condition $\Vect_G$ by $\mc M(A)$ in the definition of the topological state so that 
\begin{equation}
    \label{eq:gauged_ZIsing0_2}
    \mc Z^{(\Vect_G)^\vee_{\mc M(A)}}(\Sigma_\triangle^\vee ; \theta) = \la \Sigma_\triangle, \Vect_G, \mc M(A) \,|\, \Sigma_\triangle, \Vect_G ; \theta, \vartheta^{\rm triv.} \ra \, .
\end{equation}
Let us write down the right-hand side of this equality more explicitly. On the one hand, spelling out the definition of the tensors \eqref{eq:GAHex} in terms of the notations of sec.~\ref{sec:state-sum}, the topological state explicitly reads
\begin{equation}
    |\Sigma_\triangle,\Vect_G, \mc M(A)\ra 
    \; \propto \!\!\!\!\! 
    \sum_{\substack{g \in G^{\msf E(\Sss)} \\ A\sigma \in (A \setminus G)^{\msf P(\Sss)}}} 
    \!\!\!\!\!
    \bigg(\prod_{\msf v \in \msf V(\Sigma_\triangle)} \!\! \delta_{\mathbb 1, g_\msf v} \bigg) \! 
    \bigg(\prod_{\msf e \in \msf E(\Sigma_\triangle)} \!\! \delta_{A\sigma_{\dpe} \!\! \cat g_\msf e, A\sigma_{\dme}} \bigg)
    \bigotimes_{\msf v \in \msf V(\Sigma_\triangle)} \!\!
    | \{g_\msf e\}_{\msf e \supset \msf v},1 \ra \, ,
\end{equation}
where $g_\msf v := \prod_{\msf e \supset \msf v} g_{\msf e}$ is a product of group elements over edges incident to $\msf v$ ordered counterclockwise and assumed to be oriented in the outward direction. On the other hand, the non-topological state was provided in eq.~\eqref{eq:state_VecG_nonTopo}. By inspection, one obtains
\begin{equation}
\begin{split}
    &\la \Sigma_\triangle, \Vect_G, \mc M(A) \, |\, \Sigma_\triangle, \Vect_G ; \theta, \vartheta^{\rm triv.} \ra
    \\
    & \q \propto \!\!\!
    \sum_{\substack{g \in G^{\msf E(\Stdss)} \\ A\sigma \in (A \setminus G)^{\msf V(\Stdss)}}} 
    \!\!\!\!\!
    \bigg(\prod_{\msf p \in \msf P(\Std)} \!\! \delta_{\mathbb 1, g_\msf p} \bigg)  
    \!
    \bigg(\prod_{\msf e \in \msf E(\Std)} \!\! \delta_{A\sigma_{\pme} \! \cat g_\msf e, A\sigma_{\ppe}} \bigg)
    \prod_{\msf e \in \msf E(\Std)} \theta_\msf e(g_\msf e)  \, ,
\end{split}
\end{equation}
where $g_\msf p := \prod_{\msf e \subset \partial \msf p}g_\msf e$. Since for every $\msf e \subset \msf E(\Std)$, there is a Kronecker delta function imposing $A\sigma_{\pme} \cat g_\msf e =  A\sigma_{\ppe}$, one can always construct a unique $a_{\sigma_{\pme},g_\msf e} \in A$ such that $g_\msf e = \sigma_{\pme}^{-1} \, a_{\sigma_{\pme},g_\msf e} \, \si_\ppe$. In particular, it implies that $g_\msf p = \prod_{\msf e \subset \partial \msf p}a_{\si_{\pme},g_\msf e}$. Moreover, given a pair $(\sigma_{\ppe}, \sigma_{\pme})$, there are exactly $|A|$-many variables $g_\msf e  \in G$ such that $A\si_{\pme} \cat g_\msf e = A \si_{\ppe}$. Summing $a_{\si_{\ppe,g_\msf e}}$ over such variables $g_\msf e$ in $G$ thus amounts to summing over a group variable $a_\msf e$ in $A$. Finally, the comparison with eq.~\eqref{eq:gauged_ZIsing0_3} establishes equality \eqref{eq:gauged_ZIsing0_2}. The main two advantages of employing expression \eqref{eq:gauged_ZIsing0_2} over \eqref{eq:gauged_ZIsing0_1} are: (i) It makes identifying the topological lines in the gauged theory much more straightforward (see sec.~\ref{sec:lines_gauging}). (ii) It teaches us how to gauge the subsymmetry $A$ using the Fourier transformed theory, which is less intuitive given that the configuration variables are valued in $\Rep(G)$ (see sec.~\ref{sec:equivariant}). 

\bigskip \noindent
Before moving on, let us comment on alternative ways of formalising this gauging procedure. In ref.~\cite{Tachikawa:2017gyf,Bhardwaj:2017xup}, the process of summing over background flat gauge fields was expressed in terms of algebra objects in $\Vect_G$. Specifically, consider the algebra object $\mathbb C[A]= \bigoplus_{a \in A}\mathbb C_a$. Inserting a network of one-codimensional defects in the dual lattice labelled by $\mathbb C[A]$, in such a way that every junction implements the multiplication of the algebra, precisely reproduces eq.~\eqref{eq:gauged_ZIsing0_1}. Within this context, the above demonstration that the partition function of the gauged theory can be equally expressed as \eqref{eq:gauged_ZIsing0_2} is the lattice realisation of the equivalence $\Mod_{\Vect_G}(\mathbb C[A]) \simeq \mc M(A)$ of $\Vect_G$-module categories.   

When summing over gauge fields in eq.~\eqref{eq:gauged_ZIsing0_1}, one made a specific choice of measure. Analogously to the abelian case discussed in sec.~\ref{sec:dyn_gauge}, alternative choices can be made. Specifically, the sum over the moduli space of flat gauge fields can be weighted by characters of the Drinfel'd double of $G$---the category of modules thereof being equivalent to $\ms Z(\Vect_G)$---in the same vein as eq.~\eqref{eq:ZIsingG_Z}. This amounts to performing the gauging as a Fourier transform over the moduli space of flat connections.

\subsection{Equivariantisation and de-equivariantisation\label{sec:equivariant}}

We commented in sec.~\ref{sec:brane} that indecomposable module categories over $\Vect_G$ and $\Rep(G)$ are labelled by the same data.
As a matter of fact, not only do they share the same data, but we have 2-functors
\begin{equation}
    \label{eq:Equivariant}
    \Mod(\Vect_G) \rightleftarrows \Mod(\Rep(G)) \, ,
\end{equation}
referred to as \emph{equivariantisation} and \emph{de-equivariantisation}, respectively, which are weak inverses of each other when restricting to semisimple module categories \cite{Drinfeld2010,Bruguieres2000}. The existence of such 2-functors is crucial to the definition of Kramers-Wannier dualities: It is employed to deduce which choice of $\Rep(G)$-module category amounts to the gauging of a given subsymmetry of the Fourier transformed theory, which is difficult to motivate from basic principles otherwise. 

Let us focus on the equivariantisation since the de-equivariantisation can be treated analogously. By definition, given a (finite semisimple $\mathbb C$-linear) category $\mc M$ with a (right) $G$-action---which is the same as saying that $\mc M$ is a module category over $\Vect_G$---we define a $G$-equivariant object in $\mc C$ as a pair $(X,\gamma)$ consisting of an object $X$ in $\mc M$ and a collection $\gamma$ of isomorphisms $\gamma_g : X \cat g \xrightarrow{\sim} X$ fulfilling a coherence axiom involving the module associator in $\mc M$. The collection of $G$-equivariant objects in $\mc M$ form a category $\mc M^G$ referred to as the equivariantisation of $\mc M$. Keeping in mind that a functor $\fr F:\Vect \to \mc M$ is specified by an object $\fr F(\mathbb C)$ in $\mc M$, one can readily check that the category $\mc M^G$ is equivalent to the category of $\Vect_G$-module functors $\FunC_{\Vect_G}(\Vect,\mc M)$. But $\FunC_{\Vect_G}(\Vect,\mc M)$ has the structure of a module category over $\Rep(G) \simeq \FunC_{\Vect_G}(\Vect,\Vect)$ via the composition of $\Vect_G$-module functors. It follows that $\mc M \mapsto \mc M^G$ is indeed a 2-functor $\Mod(\Vect_G) \to \Mod(\Rep(G))$. 

More concretely, consider the indecomposable  $\Vect_G$-module category $\mc M(A)$ defined in sec.~\ref{sec:brane}. Let us compute $\mc M(A)^G \simeq \FunC_{\Vect_G}(\Vect,\mc M(A))$. By definition, an object $(\fr F,\omF{\fr F})$ in $\FunC_{\Vect_G}(\Vect,\mc M(A))$ consists of a functor $\fr F: \Vect \to \mc M(A)$, which is fully specified by an object $\fr F(\mathbb C) := \bigoplus_{As \in A \setminus G}M_{As}$ in $\mc M(A)$, and a natural isomorphism $\omega$ prescribed by isomorphisms
\begin{equation}
    \label{eq:moduleFunCEquiv}
    \omF{\fr F}^{\mathbb C,\mathbb C_g} : 
     \Big(\bigoplus_{As \in A \setminus G}M_{As} \Big) \cat \mathbb C_g
    \cong \bigoplus_{As \in A \setminus G} M_{As \cat g^{-1}}
    \xrightarrow{\sim}
    \bigoplus_{As \in A \setminus G} M_{As}
    \, ,
\end{equation}
for every $g \in G$. The transitivity of the $G$-action on $A \setminus G$ requires $M_{As \cat g} \cong M_{As}$ for every $g \in G$. Writing  components of \eqref{eq:moduleFunCEquiv} as 
\begin{equation}
    \omF{\fr F}(As\xrightarrow{g})  : M_{As} \xrightarrow{\sim} M_{As \cat g} \cong M_{As} \, ,
\end{equation}
it follows from the pentagon axiom satisfied by $\omF{\fr F}$ that we have $\omF{\fr F}(As \xrightarrow{g_1}) \circ \omF{\fr F}(As \cat g_1 \xrightarrow{g_2}) = \omF{\fr F}(As \xrightarrow{g_1g_2})$, for every $g_1,g_2 \in G$ and $As \in G/A$. Given any $As \in A \setminus G$, there are $|A|$-many variables in $G$ stabilising it. Restricting to the stabiliser subgroup of any $As \in A \setminus G$, we find that $(\fr F, \omF{\fr F})$ provides an object in $\Rep(A) $. 
Conversely let $\eta: A \to \End_\mathbb C(M)$ be a simple object in $\Rep(A)$. The corresponding $\Vect_G$-module functor $M \equiv (\fr F, \omF{\fr F})$ in $\FunC_{\Vect_G}(\Vect, \mc M(A))$ is constructed as follows: Construct the functor $\fr F: \mathbb C \mapsto \bigoplus_{As \in A \setminus G} M_{As}$ such that $M_{As} \cong  M$ for every $As \in A \setminus G$. Then, define isomorphisms $\omF{\fr F}^{\mathbb C,\mathbb C_g} : \fr F(\mathbb C) \cat \mathbb C_g \xrightarrow{\sim} \fr F(\mathbb C)$ via 
\begin{equation}
    \label{eq:omegaEta}
    \omF{\fr F}(As \xrightarrow{g}) := \eta(a_{As,g}) :M \xrightarrow{\sim} M \, ,
\end{equation}
for every $g \in G$ and $As \in G/A$, where $a_{As,g} \in A$ was defined in sec.~\ref{sec:brane}. It follows from 
\begin{equation}
    \eta(a_{As,g_1g_2}) = \eta(a_{As,g_1}) \circ \eta(a_{As \cat g_1,g_2}) \, ,
\end{equation} 
which holds for every $As \in A \setminus G$ and $g_1,g_2 \in G$, that the natural isomorphism $\omF{\fr F}$ endows $\fr F$ with a module structure.  Therefore $(\fr F,\omF{\fr F})$ defines a $\Vect_G$-module functor from $\Vect$ to $\mc M(A)$. In terms of string diagrams, components of the module structure reads
\begin{equation}
    \label{eq:modFuncEquiv}
    \moduleFunctor{6} \!\!\!\!\!\!\! 
    = \sum_{n=1}^{{\rm dim}_\mathbb C M}
    \eta(a_{As,g})_{m}^{n} \!\!
    \moduleFunctor{5}  .
\end{equation}
Putting everything together, we obtained that $\mc M(A)^G \simeq \Rep(A)$ as categories. It remains to show that this equivalence holds as $\Rep(G)$-module categories. Since $\Rep(G) \simeq \FunC_{\Vect_G}(\Vect, \Vect)$, the composition of module functors  naturally endows $\mc M(A)^G$ with the structure of a $\Vect_G$-module category, which coincides with the action of the restriction functor ${\rm Res}^G_A$. Indeed, consider the composition 
\begin{equation}
    \circ: \FunC_{\Vect_G}(\Vect,\mc M(A)) \times \FunC_{\Vect_G}(\Vect,\Vect)
    \to \FunC_{\Vect_G}(\Vect,\mc M(A)) \, .
\end{equation}
Given simple objects $\eta_1 : A \to \End_\mathbb C(M_1)$ and $\rho : G \to \End_\mathbb C(V)$ in $\Rep(A)$ and $\Rep(G)$, respectively, the composition of the corresponding module functors yields an object in $\Rep(A) \simeq \mc M(A)^G$, which may or may not be simple. Therefore, there must exist symbols defined in terms of string diagrams as
\begin{equation}
    \label{eq:modCG}
    \leftModuleAssociator{2} = 
    \sum_{v=1}^{{\rm dim}_\mathbb C V}
    \sum_{m_1=1}^{{\rm dim}_\mathbb C M_1} \, 
    \CC{M_1}{V}{M_2}{m_1}{v}{m_2}_{i,As}^* \hspace{-12pt}
    \leftModuleAssociator{1} \, ,
\end{equation}
where $\eta_2 : A \to \End_\mathbb C(M_2)$ is a simple object in $\Rep(A)$ and $i$ labels a basis vector in the hom-space $\Hom_{\Rep(A)}(M_1 \cat V, M_2)$. By inspection, one finds
\begin{equation}
    \label{eq:ModCGeq}
    \CC{M_1}{V}{M_2}{m_1}{v}{m_2}_{i,As}^*
    = \sum_{w=1}^{{\rm dim}_\mathbb C V} \, 
    \CC{M_1}{V}{M_2}{m_1}{w}{m_2}_{i}^* \rho(s)_{w}^{v} \, ,
\end{equation}
where the symbols on the r.h.s. are the Clebsch-Gordan coefficients we introduced in eq.~\eqref{eq:basisVectors}.
These modified symbols satisfy a generalisation of a familiar relation involving Clebsch-Gordan coefficients. Given two cosets $Ar,As \in A \setminus G$, there are exactly $|A|$-many group variables in $G$ satisfying $Ar \cat g = As$. This constraint can be implemented by means of a Kronecker delta $\delta_{Ar \cat g, As}$. For any $g \in G$ such that $Ar \cat g = As$, we mentioned in sec.~\ref{sec:brane}, that there was a unique $a_{Ar,g} \in A$ such that $g = r^{-1}a_{Ar,g}s$. It follows that
\begin{equation}
    \label{eq:defModCG}
    \frac{1}{|A|} \sum_{g \in G} 
    \delta_{Ar \cat g , As} \, 
    \eta_1 (a_{Ar,g})_{m_1}^{n_1} \,
    \rho(g)_{v}^{w} \,
    \eta_2^\vee(a_{Ar,g})_{m_2}^{n_2}
    = \frac{1}{d_{M_2}}
    \sum_{i=1}^{d^{M_1 V}_{M_2}}
    \CC{M_1}{V}{M_2}{m_1}{v}{m_2}_{i,Ar}
    \CC{M_1}{V}{M_2}{n_1}{w}{n_2}_{i,As}^* \, ,
\end{equation}
where $d^{M_1 V}_{M_2} := {\rm dim}_\mathbb C \Hom_{\Rep(A)}(M_1 \cat V, M_2)$. In the same vein, by combining \eqref{eq:modFuncEquiv} and \eqref{eq:modCG} one obtains a coherence relation that translates into the following intertwining property:
\begin{equation}
    \label{eq:interModCG}
    \sum_{n_1,w}
    \eta_1 (a_{Ar,g})_{m_1}^{n_1} \,
    \rho_2(g)_{v}^{w} \, 
    \CC{M_1}{V}{M_2}{n_1}{w}{m_2}_{i,Ar \cat g} 
    =
    \sum_{n_2}
    \CC{M_1}{V}{M_2}{n_1}{w}{n_2}_{i,Ar} \,
    \eta_2(a_{Ar,g})_{n_2}^{m_2} \, ,
\end{equation}
which is true for any $g \in G$ and $Ar \in A \setminus G$.

In conclusion, the equivalence of $\Rep(G)$-module categories $\Rep(A) \simeq \mc M(A)^G$ stipulates that gauging the subsymmetry $A \subseteq G$ of the Fourier transformed theory is obtained by choosing the brane boundary condition $\Rep(A)$ in $\Mod(\Rep(G))$. Specifically, gauging the subsymmetry $A$ of a $G$-Ising model with partition function $\la \Sigma_\triangle^\vee, \Rep(G), \Vect \, | \, \Sigma_\triangle^\vee, \Rep(G) ; \theta^\vee, \vartheta^{\rm triv.} \ra
$ results in a gauged theory with partition function $\la \Sigma_\triangle^\vee, \Rep(G), \Rep(A) \, |\,  \Sigma_\triangle^\vee, \Rep(G) ; \theta^\vee, \vartheta^{\rm triv.} \ra
$. In sec.~\ref{sec:RepG-FT}, we shall confirm that the latter is indeed equal to $\mc Z^{(\Vect_G)^\vee_{\mc M(A)}}(\Std;\theta)$. Crucially, equivariantisation and de-equivariantisation further implies equivalence of the following categories of module endofunctors \cite{etingof2010fusion,GREENOUGH20101818}:
\begin{equation}
    \FunC_{\Vect_G}(\mc M(A), \mc M(A)) \simeq \FunC_{\Rep(G)}(\Rep(A),\Rep(A)) \, ,
\end{equation}
which will ultimately ensure that topological lines before and after Fourier transform are labelled by the same data, regardless of the amount of the initial symmetry being gauged, as expected. We define the explicit action of the corresponding topological lines in the following.

\subsection{Topological lines\label{sec:lines_gauging}}

It follows from the results of sec.~\ref{sec:lines} that the theory resulting from gauging the subsymmetry $A$, via the choice of indecomposable $\Vect_G$-module category $\mc M(A)$, can be coupled with topological lines labelled by objects in $(\Vect_G)^\vee_{\mc M(A)}$. In order to understand the action of such topological lines, i.e. to compute the tensors eq.~\eqref{eq:MPO}, it is useful to briefly review the explicit computation of $(\Vect_G)^\vee_{\mc M(A)}$. We proceed as for $\mc M(A)^G$. By definition, an object $(\fr F, \omF{\fr F})$ in $\FunC_{\Vect_G}(\mc M(A),\mc M(A))$ consists of an endofunctor $\fr F: \mc M(A) \to \mc M(A)$ specified by a collection of objects $\fr F(\mathbb C_{Ar}) := M^{Ar} = \bigoplus_{As \in A \setminus G}M^{Ar}_{As}$ in $\mc M(A)$, as well as a natural isomorphism $\omF{\fr F}$ prescribed by isomorphisms
\begin{equation}
    \label{eq:moduleFunCMA}
    \omF{\fr F}^{\mathbb C_{Ar},\mathbb C_g} : 
     \Big(\bigoplus_{As \in A \setminus G}M^{Ar}_{As} \Big) \cat \mathbb C_g
    \cong \bigoplus_{As \in A \setminus G} M_{As \cat g^{-1}}^{Ar}
    \xrightarrow{\sim}
    \bigoplus_{As \in A \setminus G} M^{Ar \cat g}_{As}
    \, ,
\end{equation}
for every $Ar \in A \setminus G$ and $g \in G$. Given $Ar,As \in A \setminus G$, the transitivity of the $G$-action requires that $M^{Ar}_{As} \cong  M^{Ar \cat g}_{As \cat g}\in \Vect$ for every $g \in G$. Writing components of \eqref{eq:moduleFunCMA} as 
\begin{equation}
    \omF{\fr F}\big((Ar,As)\xrightarrow{g}\big )  : M_{As}^{Ar} \xrightarrow{\sim} M^{Ar \cat g}_{As \cat g} \cong M^{Ar}_{As} \, ,
\end{equation}
it follows from the pentagon axiom satisfied by $\omF{\fr F}$ that we have
\begin{equation}
    \omF{\fr F}\big( (Ar,As) \xrightarrow{g_1} \big) \circ \omF{\fr F}\big( (Ar \cat g_1,As \cat g_1)) \xrightarrow{g_2}\big)
    = \omF{\fr F}\big( (Ar,As) \xrightarrow{g_1g_2} \big) \, ,
\end{equation}
for every $g_1,g_2 \in G$ and $Ar,As \in A \setminus G$.
Putting everything together, this is the statement that $(\fr F, \omF{\fr F})$ is a representation of the groupoid algebra $\mathbb C[{}_A\mathbb G_A]$, where ${}_A \mathbb G_A$ is the groupoid consisting of objects $(Ar,As) \in A \setminus G \times A \setminus G$ and morphisms of the form $(Ar,As) \xrightarrow{g} (Ar \cat g,As \cat g)$, for every $g \in G$. Groupoid representation theory then dictates that an irreducible representation of the groupoid algebra is labelled by a connected component of the groupoid ${}_A \mathbb G_A$ and an irreducible representation of the stabiliser subgroup in $G$ generated by morphisms stabilising the representative of this connected component. The set of connected components is isomorphic to $A \setminus G / A$ such that given $AxA \in A \setminus G / A$, a representative of the corresponding connected component is given by $(Ax,A)$. The subgroup stabilising the connected component represented by $(Ax,A)$ is then found to be $A \cap x^{-1} A x$. We infer from this analysis the following equivalence of categories \cite{2002math......2130O}
\begin{equation}
    \FunC_{\Vect_G}(\mc M(A),\mc M(A))
    \simeq 
    \Mod(\mathbb C[{}_A \mathbb G_A]) \simeq \!\! \bigboxplus_{AxA \in A \setminus G /A} \!\! \Rep(A \cap x^{-1}Ax) \, .
\end{equation}
This equivalence can be lifted to an equivalence of fusion categories, where the fusion structure is given by the composition of $\Vect_G$-module functors. 

It is useful to work out the converse explicitly. Consider a simple object in $\Mod(\mathbb C[{}_A \mathbb C_A])$, which amounts to a pair $V_x \equiv¯ (AxA \in A \setminus G / A,\; \eta : A \cap x^{-1} A x \to \text{End}_\mathbb C(V))$. The corresponding $\Vect_G$-module functor $V_x \equiv (\fr F, \omF{\fr F})$ in $\FunC_{\Vect_G}(\mc M(A),\mc M(A))$ is constructed as follows: Define the functor $\fr F : \mathbb C_{Ar} \mapsto \bigoplus_{As \in A \setminus G} M_{As}^{Ar}$ such that $M^{Ar}_{As} \cong \delta_{Ars^{-1}A,AxA} \, V$ for every $As, Ar \in A \setminus G$. Then, define isomorphisms $\omF{\fr F}^{\mathbb C_{Ar}, \mathbb C_g} : \fr F(\mathbb C_{Ar}) \cat \mathbb C_g \xrightarrow{\sim}
\fr F(\mathbb C_{Ar} \cat \mathbb C_g)$ via components
\begin{equation}
    \omF{\fr F}\big((Ar,As)\xrightarrow{g}\big ) := \eta(a_{As,g}) : V \xrightarrow{\sim} V \, ,
\end{equation}
for every $Ar,As \in A \setminus G$ such that $Ars^{-1}A = AxA$, where $a_{As,g} \in A \cap x^{-1} A x$ was defined in sec.~\ref{sec:brane}. It follows from 
\begin{equation}
    \label{eq:Shapiro}
    \eta(a_{As,g_1g_2}) = \eta(a_{As,g_1}) \circ \eta(a_{As \cat g_1,g_2}) \, ,
\end{equation} 
which holds for every $As \in A \setminus G$ and $g_1,g_2 \in G$, that the natural isomorphism $\omF{\fr F}$ endows $\fr F$ with a module structure. Therefore $(\fr F,\omF{\fr F})$ defines a $\Vect_G$-module endofunctor of $\mc M(A)$. In terms of string diagrams, components of the module structure read
\begin{equation}
    \moduleFunctor{4} \!\!\!\!\!\!\! 
    = \delta_{Ars^{-1}A,AxA}\sum_{w=1}^{{\rm dim}_\mathbb C V_x}
    \eta(a_{As,g})_{v}^{w} \!\!
    \moduleFunctor{3}  .
\end{equation}
We are now ready to define the action of the topological line labelled by $V_x \equiv (\fr F, \omF{\fr F})$ in $(\Vect_G)^\vee_{\mc M(A)}$ in terms of tensors whose components read:
\begin{equation}
    \MPO{mod}{mod}{v}{w}{1}{1}{\mathbb C_{As}\;\,}{\q\;\;\; \mathbb C_{As \cat g}}{\q\;\;\; \mathbb C_{Ar \cat g}}{\mathbb C_{Ar}\;\,}{\mathbb C_g}{V_x}{1} \!\!\!\!\! = \delta_{Ars^{-1}A,AxA} \, \eta(a_{As,g})_{v}^{w} \, ,
\end{equation}
for every $Ar,As \in A \setminus G$ and $g \in G$, with $\eta : A \cap x^{-1}Ax \to \text{End}_\mathbb C(V)$. Finally, topological invariance is guaranteed by eq.~\eqref{eq:Shapiro}. Let us examine a couple of explicit cases: On the one hand, we recover 't Hooft lines labelled by simple objects in $\Vect_G$ whenever $A = \{\mathbb 1\}$, i.e. whenever the brane boundary condition is Dirichlet. On the other hand, we recover Wilson lines labelled by simple objects in $\Rep(G) \simeq {(\Vect_G)}^\vee_\Vect$ whenever $A = G$, i.e. whenever the brane boundary condition is Neumann. 

\bigskip \noindent
The derivations above produce the topological lines of the theory resulting from gauging the subsymmetry $A \subseteq G$ in the original theory. Similarly, one can compute the action of topological lines resulting from gauging the subsymmetry $A \subseteq G$ of the Fourier transformed theory. It follows from ${(\Vect_G)}^\vee_{\mc M(A)} \simeq {(\Rep(G))}^\vee_{\Rep(A)}$ that topological lines are still labelled by a pair $V_x \equiv (AxA \in A \setminus G / A, \; \eta: A \cap x^{-1}Ax \to \End_\mathbb C(V))$. The corresponding $\Rep(G)$-module functor $V_x \equiv (\fr G, \omF{\fr G})$ in ${(\Rep(G))}^\vee_{\Rep(A)}$ can be constructed as follows: Realising $\Rep(A)$ as $\mc M(A)^G$, the functor $\fr G : \Rep(A) \to \Rep(A)$ takes a $\Vect_G$-module functor $\Vect \to \mc M(A)$ to the composite $\Vect \to \mc M(A) \xrightarrow{\fr F} \mc M(A)$, where the $\Vect_G$-module functor $\fr F : \mc M(A) \to \mc M(A)$ is that we defined above given the data underlying $V_x$. The $\Rep(G)$-module structure of $\fr G$, which follows from the composition of $\Vect_G$-module functors, produces symbols
\begin{equation}
    \moduleFunctor{8} \!\!\!\!\!\!\! 
    = 
    \sum_{M_4} \sum_{k,l}\big(\omF{V_x}^{M_1 W}_{M_2} \big)_{M_3,ij}^{M_4,kl}
    \!\!
    \moduleFunctor{7}  \, ,
\end{equation}
where $W$ is a simple object in $\Rep(G)$, while $M_1,M_2,M_3$ and $M_4$ are simple objects in $\Rep(A)$. One obtains that the action of the topological line labelled by $V_x \equiv (\fr G, \omF{\fr G})$ in ${(\Rep(G))}^\vee_{\Rep(A)}$ is realised by tensors whose components read:
\begin{equation}
    \MPO{mod}{mod}{i}{l}{j}{k}{M_3}{M_2}{M_4}{M_1}{W}{V_x}{1} =
    \big(\omF{V_x}^{M_1 W}_{M_2} \big)^{M_4,kl}_{M_3,ij} \, .
\end{equation}
As before, let us examine the two extreme scenarios: On the one hand, one obtains 't Hooft lines labelled by simple objects in $\Rep(G)$ whenever the brane boundary condition is Dirichlet. On the other hand, we recover Wilson lines labelled by simple objects in $\Vect_G \simeq {(\Rep(G))}^\vee_\Vect$ whenever the brane boundary condition is Neumann.

\subsection{Gauging renormalisation group fixed points\label{sec:gauged_fixed}}

Previously, we explained how to construct the partition function $\mc Z^{\Vect_G}(\Std;\theta^{\rm triv.},\vartheta^{\mathbb C[B]})$ of a renormalisation group fixed point for the $\Vect_G$-symmetric gapped phase with unbroken subgroup $B \subseteq G$, namely $ \la \Sigma_\triangle, \Vect_G, \Vect_G \, | \, \Sigma_\triangle, \Vect_G , \mathbb C[B]; \theta^{\rm triv.}, \vartheta^{\mathbb C[B]} \ra$. What happens to the gapped phase as we gauge the subgroup $A \subseteq G$? In other words, which symmetric gapped phase does the partition function $\la \Sigma_\triangle, \Vect_G, \mc M(A) \, | \, \Sigma_\triangle, \Vect_G, \mathbb C[B] ; \theta^{\rm triv.}, \vartheta^{\mathbb C[B]} \ra$ correspond to? The answer to this question requires a generalisation of the concept of equivariantisation, stipulating that, given a fusion category $\mc C$ and a finite semisimple $\mc C$-module category $\mc M$, the 2-functor
\begin{equation}
    \mc N \mapsto \FunC_\mc C(\mc M,\mc N): \Mod(\mc C) \to \Mod(\mc C^\vee_\mc M)
\end{equation}
is a 2-equivalence \cite{etingof2016tensor}. Recall that the category $\Mod_{\Vect_G}(\mathbb C[B])$ of module objects over $\mathbb C[B]$ in $\Vect_G$ is equivalent to the (indecomposable) $\Vect_G$-module category $\mc M(B)$. After gauging the symmetry $A$, we confirmed above that the symmetry of the resulting model is given by $(\Vect_G)^\vee_{\mc M(A)}$. It still encodes the renormalisation group fixed point of a gapped phase, but with respect to the dual symmetry $(\Vect_G)^\vee_{\mc M(A)}$. Therefore, it is the gapped phase encoded into the category $\FunC_{\Vect_G}(\mc M(A),\mc M(B))$, which is equipped with a $(\Vect_G)^\vee_{\mc M(A)}$-module category structure provided by the composition of $\Vect_G$-module functors \cite{Komargodski:2020mxz}. Consider the scenario where the whole $\Vect_G$ symmetry is gauged. We established in sec.~\ref{sec:FT_fixed} the equality
\begin{equation*}
    \la \Si_\triangle, \Vect_G, \Vect_G \, | \, \Si_\triangle, \Vect_G, \mathbb C[B]; \theta^{\rm triv.}, \vartheta^{\mathbb C[B]} \ra 
    = 
    \la \Std, \Rep(G), \Vect \, | \, \Std, \Rep(G), \mathbb C^{B \setminus G}; \theta^{\rm triv.}, \vartheta^{\mathbb C^{B \setminus G}} \ra \, .
\end{equation*}
After gauging the whole $\Vect_G$-symmetry, we thus obtain the Kramers--Wannier dual model with partition function $\la \Std, \Rep(G),\Rep(G) \, | \, \Std, \Rep(G), \mathbb C^{B \setminus G};\theta^{\rm triv.}, \vartheta^{\mathbb C^{B \setminus G}}\ra$, at which point one can identify the gapped phase as that breaking the $\Rep(B)$ symmetry.

Concretely, the reasoning above implies that the interpretation of a given phase of matter in terms of symmetry breaking pattern depends on the representative of the corresponding Morita class of symmetries. For instance, by changing representative via a gauging procedure, it is possible to change the `amount' of symmetry being broken. As a matter of fact, it is always possible to find a representative of a given class of models where its whole symmetry is broken. In the notations of the previous paragraph, it suffices to choose $A = B$, since the symmetry broken phase is always associated with choosing $(\Vect_G)^\vee_{\mc M(A)}$ as a module category over itself, in such a way that degenerate ground states are in one-to-one correspondence with simple objects in $(\Vect_G)^\vee_{\mc M(A)}$.
\bigskip
\section{Fourier transform of $(\Vect_G)^\vee_{\mc M(A)}$-symmetric models\label{sec:RepG-FT}}

\emph{In this section, we extend the derivations of sec.~\ref{sec:VecG-FT} to compute the Fourier transform of the partition function of the theory obtained after gauging some subsymmetry. We then bring everything together to equate the partition function of a $G$-Ising model and that of its Kramers--Wannier dual.}

\subsection{Fourier transform}

By performing a Fourier transform of the Boltzmann weights, we established in sec.~\ref{sec:VecG-FT} an equality between two expressions for the partition function $\mc Z^{\Vect_G}(\Std;\theta)$ of the $G$-Ising model. These two expressions can be obtained by realising the $G$-Ising model as a boundary theory of the topological theory $\mc Z_\mc C$ with input datum $\mc C=\Vect_G$ and $\mc C = \Rep(G)$, respectively. Gauging the subsymmetry $A \subseteq G$ was then performed in the previous section by choosing the brane boundary conditions provided by $\mc M(A)$ and $\Rep(A)$, respectively. We now wish to prove that the resulting partition functions are also related by a Fourier transform of the corresponding Boltzmann weights. 

Consider the partition function \eqref{eq:gauged_ZIsing0_2} we obtained in the previous section:
\begin{equation}
    \mc Z^{(\Vect_G)^\vee_{\mc M(A)}}(\Sigma_\triangle^\vee;
    \theta) = \la \Sigma_\triangle, \Vect_G, \mc M(A) \, | \, \Sigma_\triangle, \Vect_G ; \theta, \vartheta^{\rm triv.} \ra \, .
\end{equation}
For convenience, we reproduce below its explicit form:
\begin{equation*}
    \mc Z^{(\Vect_G)^\vee_{\mc M(A)}}(\Std;\theta)
    \; \propto \!\!\!
    \sum_{\substack{g \in G^{\msf E(\Stdss)} \\ A\sigma \in (A \setminus G)^{\msf V(\Stdss)}}} 
    \!\!\!\!\!
    \bigg(\prod_{\msf p \in \msf P(\Std)} \!\! \delta_{\mathbb 1, g_\msf p} \bigg)  
    \!
    \bigg(\prod_{\msf e \in \msf E(\Std)} \!\! \delta_{A\sigma_{\pme} \! \cat g_\msf e, A\sigma_{\ppe}} \bigg)
    \prod_{\msf e \in \msf E(\Std)} \theta_{\msf e}(g_{\msf e})  \, .
\end{equation*}
As in sec.~\ref{sec:gauging_brane}, one begins our computation by employing the fact that for every $\msf e \subset \partial \msf p$, there is a Kronecker delta imposing $A\sigma_{\pme} \cat g_\msf e =  A\sigma_{\ppe}$, so that there is a unique $a_{\sigma_{\pme},g_\msf e} \in A$ such that $g_\msf e = \sigma_{\pme}^{-1} \, a_{\sigma_{\pme},g_\msf e} \, \si_\ppe$. In particular, it implies that $g_\msf p = \prod_{\msf e \subset \partial \msf p}a_{\si_{\pme},g_\msf e}$. Performing an inverse Fourier transform of $\delta_{\mathbb 1,g_\msf p}$ then yields
\begin{equation}
    \label{eq:deltaFT}
    \delta_{\mathbb 1, g_\msf p} 
    = \sum_{M_\msf p \in \widehat A}
    d_{M_\msf p} \, \tr\bigg(\prod_{\msf e \subset \partial\msf p}\eta^\vee_\msf p(a_{\sigma_{\pme},g_\msf e}) \bigg)
    \equiv 
    \sum_{M_\msf p \in \what A} d_{M_\msf p}
    \prod_{\msf e \subset \partial \msf p} \eta^\vee_\msf p(a_{\si_{\pme},g_\msf e})_{m_{\msf p,\pme}}^{m_{\msf p,\ppe}}\, ,
\end{equation}
where $\eta^\vee_\msf p : A \to \End_\mathbb C(M_\msf p)$. As before, we use the convention that repeated indices are summed over. Applying eq.~\eqref{eq:deltaFT} to $\delta_{\mathbb 1,g_\msf p}$, for every $\msf p \in \msf P(\Std)$, and eq.~\eqref{eq:FT} to the Boltzmann weights $\theta_\msf e$, for every $\msf e \in \msf E(\Std)$, yields
\begin{align}
    \mc Z^{(\Vect_G)^\vee_{\mc M(A)}}(\Std;\theta)
    \; \propto \!\!
    \sum_{\substack{V \in \widehat G^{\msf E(\Stdss)} \\ M \in \widehat A^{\msf P(\Stdss)}}}
    \sum_{\substack{g \in G^{\msf E(\Stdss)} \\ A\sigma \in (A \setminus G)^{\msf V(\Stdss)}}} 
    \!\! &\bigg(\prod_{\msf p \in \msf P(\Std)} \!\! d_{M_\msf p} 
    \prod_{\msf e \subset \partial \msf p} \eta^\vee_\msf p(a_{\si_{\pme},g_{\msf e}})_{m_{\msf p,\pme}}^{m_{\msf p,\ppe}}
    \bigg) 
    \\[-.8em] \nn
    &\bigg( \prod_{\msf e \in \msf E(\Std)}
    \!\! \delta_{A\sigma_{\pme} \! \cat g_{\msf e}, A\sigma_{\ppe}} \,
    d_{V_{\msf e}} \, \theta^\vee_{\msf e}(\rho_{\msf e})_{v_{\msf e}}^{w_{\msf e}} \, \rho_{\msf e}^\vee(g_{\msf e})_{w_{\msf e}}^{v_{\msf e}} \bigg).
\end{align}
The product over the plaquettes can be reorganised so as to obtain
\begin{align}
    \mc Z^{(\Vect_G)^\vee_{\mc M(A)}}(\Std;\theta)
    \; \propto \!\!
    \sum_{\substack{V \in \widehat G^{\msf E(\Stdss)} \\ M \in \widehat A^{\msf P(\Stdss)}}}
    \sum_{A\sigma \in (A \setminus G)^{\msf V(\Stdss)}} \!
    &\bigg(\prod_{\msf p \in \msf P(\Std)} \!\! d_{M_\msf p}\bigg) \!
    \bigg(\prod_{\msf e \in \msf E(\Std)} \!\! d_{V_{\msf e}} \,  \theta^\vee_{\msf e}(\rho_{\msf e})_{v_{\msf e}}^{w_{\msf e}} \bigg)
    \\[-5pt] \nn
    &\bigg( \prod_{\msf e \in \msf E(\Std)}
    \sum_{g_\msf e \in G}
    \delta_{A\sigma_{\pme} \! \cat g_\msf e, A\sigma_{\ppe}} \;
    \eta^\vee_{\dpe}(a_{\sigma_{\pme},g_\msf e})_{m_{\dpe,\pme}}^{m_{\dpe,\ppe}} 
    \\[-7pt] \nn
    &\hspace{5.7em}\cdot \rho^\vee_\msf e(g_\msf e)_{w_\msf e}^{v_\msf e} \,
    \eta_{\dme}(a_{\sigma_{\pme},g_\msf e})_{m_{\dme,\pme}}^{m_{\dme,\ppe}} \bigg) \, .
\end{align}
We are precisely within the framework of sec.~\ref{sec:equivariant}. Invoking eq.~\eqref{eq:defModCG}, the sum over the gauge field $g_\msf e$ can be carried out so as to obtain
\begin{equation}
\begin{split}
    &\frac{1}{|A|} \sum_{g_\msf e \in G}
    \delta_{A\sigma_{\pme} \! \cat g_\msf e, A\sigma_{\ppe}} \, 
    \eta^\vee_{\dpe}(a_{\sigma_{\pme},g_\msf e})_{m_{\dpe,\pme}}^{m_{{\rm d}^+\msf e,\ppe}} \,
    \rho_\msf e^\vee(g_\msf e)_{w_\msf e}^{v_\msf e} \,
    \eta_{\dme}(a_{\sigma_{\pme},g_\msf e})_{m_{\dme,\pme}}^{m_{\dme,\ppe}} 
    \\
    & \q = 
    \sum_{i_\msf e}
    \frac{1}{(d_{M_{\dme}})^\frac{1}{2}}
    \CC{ M^\vee_{\dpe}}{V_\msf e^\vee}{M^\vee_{\dme}}{m_{\dpe, \pme}}{w_\msf e}{m_{\dme, \pme}}_{i_{\msf e},A \sigma_{\pme}}
    \frac{1}{(d_{M_{\dme}})^\frac{1}{2}} \,
    \CC{M^\vee_{\dpe}}{V_\msf e^\vee}{M^\vee_{\dme}}{m_{\dpe, \ppe}}{v_\msf e}{m_{\dme, \ppe}}_{i_{\msf e},A\sigma_{\ppe}}^* \, .
\end{split}
\end{equation}
By analogy with sec.~\ref{sec:VecG-FT}, we now distinguish two situations associated with the two types of 3-valent vertices in $\Std$. Given any vertex $\msf v \in \msf V(\Std)$, we still denote by $\msf e_{\msf v,1}$ and $\msf e_{\msf v,2}$ the two incoming or outgoing edges, and by $\msf e_{\msf v,3}$ the remaining one. On the one hand, whenever there are two incoming edges at $\msf v$, one can reorganise the results of the summations over the gauge field degrees of freedom so as to obtain the following contribution associated with $\msf v$:  
\begin{equation} 
\begin{split}
    \label{eq:pre6J1}
    \frac{1}{(d_{M_{\dme_{\msf v,1}}})^\frac{1}{2}}
    \CC{ M^\vee_{\dpe_{\msf v,1}} }{ V^\vee_{\msf e_{\msf v,1}} }{ M^\vee_{\dme_{\msf v,1}} }{ m_{\dpe_{\msf v,1},\msf v} }{ v_{\msf e_{\msf v,1}} }{ m_{\dme_{\msf v,1},\msf v} }_{i_{\msf e_{\msf v,1}},A\sigma_\msf v}^* \, 
    &\frac{1}{(d_{M_{\dme_{\msf v,2}}})^\frac{1}{2}}
    \CC{ M^\vee_{\dme_{\msf v,1}} }{ V^\vee_{\msf e_{\msf v,2}} }{ M^\vee_{\dme_{\msf v,2}} }{ m_{\dme_{\msf v,1},\msf v} }{ v_{\msf e_{\msf v,2}} }{ m_{\dme_{\msf v,2},\msf v} }_{i_{\msf e_{\msf v,2}},A\sigma_\msf v}^*
    \\
    \cdot \; &\frac{1}{(d_{M_{\dme_{\msf v,2}}})^\frac{1}{2}}
    \CC{ M^\vee_{\dpe_{\msf v,1}} }{ V^\vee_{\msf e_{\msf v,3}} }{ M^\vee_{\dme_{\msf v,2}} }{ m_{\dpe_{\msf v,1},\msf v} }{ w_{\msf e_{\msf v,3}} }{ m_{\dme_{\msf v,2},\msf v} }_{i_{\msf e_{\msf v,3}},A\sigma_\msf v}\, ,
\end{split}
\end{equation}
where we used the fact that by convention $\dpe_{\msf v,1} \equiv \dpe_{\msf v,3}$, $\dme_{\msf v,2} \equiv \dme_{\msf v,3}$ and $\dme_{\msf v,1} \equiv \dpe_{\msf v,2}$. On the other hand, whenever there are two outgoing edges at $\msf v$, one obtains
\begin{equation} 
\begin{split}
    \label{eq:pre6J2}
    \frac{1}{(d_{M_{\dme_{\msf v,1}}})^\frac{1}{2}}
    \CC{ M^\vee_{\dpe_{\msf v,1}} }{ V^\vee_{\msf e_{\msf v,1}} }{ M^\vee_{\dme_{\msf v,1}} }{ m_{\dpe_{\msf v,1},\msf v} }{ w_{\msf e_{\msf v,1}} }{ m_{\dme_{\msf v,1},\msf v} }_{i_{\msf e_{\msf v,1}},A\sigma_\msf v} \,
    &\frac{1}{(d_{M_{\dme_{\msf v,2}}})^\frac{1}{2}}
    \CC{ M^\vee_{\dme_{\msf v,1}} }{ V^\vee_{\msf e_{\msf v,2}} }{ M^\vee_{\dme_{\msf v,2}} }{ m_{\dme_{\msf v,1},\msf v} }{ w_{\msf e_{\msf v,2}} }{ m_{\dme_{\msf v,2},\msf v} }_{i_{\msf e_{\msf v,2}},A\sigma_\msf v}
    \\
    \cdot \; &\frac{1}{(d_{M_{\dme_{\msf v,2}}})^\frac{1}{2}}
    \CC{ M^\vee_{\dpe_{\msf v,1}} }{ V^\vee_{\msf e_{\msf v,3}} }{ M^\vee_{\dme_{\msf v,2}} }{ m_{\dpe_{\msf v,1},\msf v} }{ v_{\msf e_{\msf v,3}} }{ m_{\dme_{\msf v,2},\msf v} }_{i_{\msf e_{\msf v,3}},A\sigma_\msf v}^* \, ,
\end{split}
\end{equation}
where we are using the same identifications as above. At this stage, one can use the invariance property \eqref{eq:interModCG} in order to rewrite the above expressions in terms of plain Clebsch--Gordan coefficients in $\Rep(A)$. Focusing for now on eq.~\eqref{eq:pre6J1}, one finds
\begin{align*} 
    \nn
    \frac{1}{(d_{M_{\dme_{\msf v,1}}})^\frac{1}{2}}
    \CC{ M^\vee_{\dpe_{\msf v,1}} }{ V^\vee_{\msf e_{\msf v,1}} }{ M^\vee_{\dme_{\msf v,1}} }{ m_{\dpe_{\msf v,1},\msf v} }{ u_{\msf e_{\msf v,1}} }{ m_{\dme_{\msf v,1},\msf v} }_{i_{\msf e_{\msf v,1}}}^*
    \!\!\! \rho_{\msf e_{\msf v,1}}^\vee(\sigma_\msf v)_{u_{\msf e_{\msf v,1}}}^{v_{\msf e_{\msf v,1}}}
    \,
    \frac{1}{(d_{M_{\dme_{\msf v,2}}})^\frac{1}{2}}
    &\CC{ M^\vee_{\dme_{\msf v,1}} }{ V^\vee_{\msf e_{\msf v,2}} }{ M^\vee_{\dme_{\msf v,2}} }{ m_{\dme_{\msf v,1},\msf v} }{ u_{\msf e_{\msf v,2}} }{ m_{\dme_{\msf v,2},\msf v} }_{i_{\msf e_{\msf v,2}}}^*
    \!\!\! \rho_{\msf e_{\msf v,2}}^\vee(\sigma_\msf v)_{u_{\msf e_{\msf v,2}}}^{v_{\msf e_{\msf v,2}}}
    \\
    \label{eq:pre6J3}
    \cdot \; \frac{1}{(d_{M_{\dme_{\msf v,2}}})^\frac{1}{2}}
    &\CC{ M^\vee_{\dpe_{\msf v,1}} }{ V^\vee_{\msf e_{\msf v,3}} }{ M^\vee_{\dme_{\msf v,2}} }{ m_{\dpe_{\msf v,1},\msf v} }{ u_{\msf e_{\msf v,3}} }{ m_{\dme_{\msf v,2},\msf v} }_{i_{\msf e_{\msf v,3}}}
    \!\!\! \rho_{\msf e_{\msf v,3}}(\sigma_\msf v)_{u_{\msf e_{\msf v,1}}}^{w_{\msf e_{\msf v,1}}}\, ,
\end{align*}
which is true for any $\sigma_\msf v \in G$. Since the previous expression is valid for any $\sigma_\msf v$, one can freely sum over $\sigma_\msf v \in G$, provided that we divide the partition function by $|G|$. But,
\begin{equation}
\begin{split}
    \label{eq:CC*3}
    &\frac{1}{|G|} \sum_{\si_{\msf v} \in G}
    (d_{V_{\msf e_{\msf v,1}}} d_{V_{\msf e_{\msf v,2}}})^\frac{1}{2} \, 
    \rho^\vee_{\msf e_{\msf v, 1}} (\si_{\msf v})_{u_{\msf e_{\msf v, 1}}}^{v_{\msf e_{\msf v, 1}}} \,
    \rho^\vee_{\msf e_{\msf v, 2}} (\si_{\msf v})_{u_{\msf e_{\msf v, 2}}}^{v_{\msf e_{\msf v, 2}}} \,
    (d_{V_{\msf e_{\msf v, 3}}})^\frac{1}{2} \, 
    \rho_{\msf e_{\msf v, 3}}(\si_{\msf v})_{u_{\msf e_{\msf v, 3}}}^{w_{\msf e_{\msf v, 3}}}
    \\
    & \q = 
    \sum_{i_\msf v}
    \Big(\frac{d_{V_{\msf e_{\msf v,1}}} d_{V_{\msf e_{\msf v,2}}}}{d_{V_{\msf e_{\msf v,3}}} }\Big)^{\frac{1}{4}} 
    \CC{V^\vee_{\msf e_{\msf v, 1}}}{V^\vee_{\msf e_{\msf v, 2}}}{V^\vee_{\msf e_{\msf v, 3}}}{u_{\msf e_{\msf v, 1}}}{u_{\msf e_{\msf v, 2}}}{u_{\msf e_{\msf v, 3}}}_{i_\msf v}
    \Big(\frac{d_{V_{\msf e_{\msf v,1}}} d_{V_{\msf e_{\msf v,2}}}}{d_{V_{\msf e_{\msf v,3}}} }\Big)^{\frac{1}{4}} 
    \CC{V^\vee_{\msf e_{\msf v, 1}}}{V^\vee_{\msf e_{\msf v, 2}}}{V^\vee_{\msf e_{\msf v, 3}}}{v_{\msf e_{\msf v, 1}}}{v_{\msf e_{\msf v, 2}}}{w_{\msf e_{\msf v, 3}}}_{i_\msf v}^* .
\end{split}
\end{equation}
Assembling the various Clebsch--Gordan coefficients associated with the vertex $\msf v$ together with the corresponding multiplicative factors of quantum dimensions, one finds
\begin{equation}
    \label{eq:6JRep(A)}
\begin{split}
    &
    \frac{1}{(d_{M_{\dme_{\msf v,1}}})^\frac{1}{2}}
    \CC{ M^\vee_{\dpe_{\msf v,1}} }{ V^\vee_{\msf e_{\msf v,1}} }{ M^\vee_{\dme_{\msf v,1}} }{ m_{\dpe_{\msf v,1},\msf v} }{ u_{\msf e_{\msf v,1}} }{ m_{\dme_{\msf v,1},\msf v} }_{i_{\msf e_{\msf v,1}}}^*
    \frac{1}{(d_{M_{\dme_{\msf v,2}}})^\frac{1}{2}}
    \CC{ M^\vee_{\dme_{\msf v,1}} }{ V^\vee_{\msf e_{\msf v,2}} }{ M^\vee_{\dme_{\msf v,2}} }{ m_{\dme_{\msf v,1},\msf v} }{ u_{\msf e_{\msf v,2}} }{ m_{\dme_{\msf v,2},\msf v} }_{i_{\msf e_{\msf v,2}}}^*
    \\
    \cdot &\Big(\frac{d_{V_{\msf e_{\msf v,1}}} d_{V_{\msf e_{\msf v,2}}}}{d_{V_{\msf e_{\msf v,3}}} }\Big)^{\frac{1}{4}} 
    \CC{V^\vee_{\msf e_{\msf v, 1}}}{V^\vee_{\msf e_{\msf v, 2}}}{V^\vee_{\msf e_{\msf v, 3}}}{u_{\msf e_{\msf v, 1}}}{u_{\msf e_{\msf v, 2}}}{u_{\msf e_{\msf v, 3}}}_{i_\msf v}
    \frac{1}{(d_{M_{\dme_{\msf v,2}}})^\frac{1}{2}}
    \CC{ M^\vee_{\dpe_{\msf v,1}} }{ V^\vee_{\msf e_{\msf v,3}} }{ M^\vee_{\dme_{\msf v,2}} }{ m_{\dpe_{\msf v,1},\msf v} }{ u_{\msf e_{\msf v,3}} }{ m_{\dme_{\msf v,2},\msf v} }_{i_{\msf e_{\msf v,1}}}
    \\
    & \q = 
    \Big(\frac{d_{V_{\msf e_{\msf v,1}}} d_{V_{\msf e_{\msf v,2}}}}{d_{V_{\msf e_{\msf v,3}}} }\Big)^{\frac{1}{4}}
    \frac{1}{(d_{M_{\dme_{\msf v,1}}})^\frac{1}{2}}
    \bigg(\F{\cat}^{ M^\vee_{\dpe_{\msf v,1}} V^\vee_{\msf e_{\msf v,1}} V^\vee_{\msf e_{\msf v,2}} }_{M^\vee_{\dme_{\msf v,2}}} \bigg)_{M^\vee_{\dme_{\msf v,1}},i_{\msf e_{\msf v,1}}i_{\msf e_{\msf v,2}}}^{V^\vee_{\msf e_{\msf v,3}},i_\msf v i_{\msf e_{\msf v,3}}} \, .
\end{split}
\end{equation}
where we used the definition \eqref{eq:def6J} of the $\F{\cat}$-symbols of the $\Rep(G)$-module category $\Rep(A)$.
Starting from eq.~\eqref{eq:pre6J2} yields a similar result in terms of $\Fbar{\cat}$-symbols. These quantities precisely correspond to the entries of tensors of the form \eqref{eq:PEPS} entering the definition of the topological state $|\Std,\Rep(G),\Rep(A) \ra$. More specifically, to every vertex, we assign a tensor whose entries are of the form \eqref{eq:6JRep(A)}. This set of tensors are contracted along multiplicity indices of the type $i_\msf e$, resulting in the topological state $|\Std,\Rep(G),\Rep(A) \ra$. Moreover, notice that expression \eqref{eq:6JRep(A)} does not involve Clebsch--Gordan coefficients containing indices of the type $v_\msf e$ and $w_\msf e$, which are the type of indices appearing with the matrices $\theta^\vee_\msf e(\rho_\msf e)$. Tensors evaluating to these Clebsch--Gordan coefficients are contracted to each other via the matrices $\theta^\vee_\msf e(\rho_\msf e)$ so as to define the state $|\Std,\Rep(G);\theta,\vartheta^{\rm triv.} \ra$. Finally, contracting these tensor network states along multiplicity indices of the type $i_\msf v$ precisely recovers the inner product in $\mc H_{\Rep(G)}(\Std)$ of these two states.  Bringing everything together, this establishes the equality
\begin{equation}
\begin{split}
    \mc Z^{(\Vect_G)^\vee_{\mc M(A)}}(\Sigma_\triangle^\vee ; \theta) 
    = \la \Sigma_\triangle^\vee, \Rep(G), \Rep(A) \, | \,\Sigma_\triangle^\vee, \Rep(G) ; \theta^\vee, \vartheta^{\rm triv.} \ra \, ,
\end{split}
\end{equation}
which, together with eq.~\eqref{eq:gauged_ZIsing0_2}, allows us to establish the Kramers--Wannier duality below. Note that in the spirit of sec.~\ref{sec:FT_lines} we could have performed this computation in the presence of topological lines whose actions were obtained in sec.~\ref{sec:lines_gauging}.

\subsection{Non-abelian Kramers--Wannier duality}

In the previous sections, we demonstrated how the Fourier transform relates two $\Vect_G$-symmetric theories realised as boundary theories of topological field theories with input data $\Vect_G$ and $\Rep(G)$, respectively; we explained how to independently gauge a subsymmetry in both descriptions; we subsequently related the partitions functions of the resulting $(\Vect_G)^\vee_{\mc M(A)}$-symmetric theories. Recall that we define a Kramers--Wannier dual of a theory as the combination of performing a Fourier transform and gauging a subsymmetry. Establishing equality of a partition function and its Kramers--Wannier dual thus requires equating partition functions before and after gauging. As anticipated, this requires elucidating the interplay between gauging and topological sectors.

Given a spherical fusion category $\mc C$ and a $\mc C$-module category $\mc M$, we explained in sec.~\ref{sec:lines} how to construct a basis of the Hilbert space $\mc Z_\mc C(\Sigma_\Upsilon)$. This basis is indexed by simple objects in the Drinfel'd center $\ms Z(\mc C^\vee_\mc M)$ of the symmetry category $\mc C^\vee_\mc M$, which provide the bulk topological operators of $\mc Z_\mc C$, and we denoted the basis vector associated with the simple object $(Z,R_{Z,-})$ by $| \Sigma_\Upsilon,\mc C,\mc M,(Z,R_{Z,-}) \ra$. In sec.~\ref{sec:non_topo}, we used these basis states to write the partition function of $\mc C^\vee_\mc M$-symmetric theories in non-trivial sectors. Whenever we gauge part of the symmetry via a change of brane boundary condition $\mc M \to \mc N$, we are considering a different basis of the Hilbert space $\mc Z_\mc C(\Sigma_\Upsilon)$, namely that associated with $\mc N$. But $| \Sigma_\Upsilon,\mc C,\mc M,(Z,R_{Z,-}) \ra$ can always be expressed in this new basis. This confirms that one can always find a superposition of sectors in the gauged theory so that the partition function coincides with that associated with the topological state $| \Sigma_\Upsilon,\mc C,\mc M,(Z,R_{Z,-}) \ra$. The explicit basis transformation is encoded into the equivalence $\ms Z(\mc C^\vee_\mc M) \simeq \ms Z(\mc C^\vee_\mc N)$, which follows from the Morita equivalence between $\mc C^\vee_\mc M$ and $\mc C^\vee_\mc N$. Note that figuring this change of basis out may be challenging in practice, as it requires finding a common parameterisation of simple objects in $\ms Z(\mc C^\vee_\mc M)$ and $\ms Z (\mc C^\vee_\mc N)$ while $\mc C^\vee_\mc M$ and $\mc C^\vee_\mc N$ may not be monoidally equivalent. Without loss of generality, let us suppose that $\mc N = \mc C$. We can find this common parameterisation invoking $\mc Z(\mc C^\vee_\mc M) \xrightarrow{\sim} (\mc C^\vee_\mc M \boxtimes \mc C)^\vee_\mc M \xleftarrow{\sim} \ms Z(\mc C)$, together with the fact that $(\mc C^\vee_\mc M \boxtimes \mc C)^\vee_\mc M$ is symmetric in $\mc C$ and $\mc C^\vee_\mc M$. 

Let us now specialise to the case of $G$-Ising models,  and consider gauging the whole symmetry. In this case, the fact that a superposition of sectors in the gauged theory can be found so that the resulting partition function coincides with the initial one can be inferred from the notion that gauging amounts to a Fourier transform on the moduli space of flat connections. As discussed in sec.~\ref{sec:GIsing}, simple objects in $\ms Z(\Vect_G)$ can be conveniently labelled by pairs $([c_1], V_{c_1})$ consisting of a conjugacy class and an irreducible representation of its centraliser. With the notations of eq.~\eqref{eq:linesC}, the partition function $\mc Z^{\Vect_G}(\Std;\theta)$ of the $G$-Ising model in the trivial sector, i.e. without line insertions, is associated with the object $\bigoplus_{V \in \what G}d_V \cdot ([\mathbb 1],V)$ of $\ms Z(\Vect_G)$ so that $\mc Z^{\Vect_G}(\Std;\theta) \equiv \mc Z^{\Vect_G}(\Std;\theta)([\mathbb 1,\mathbb 1])= \sum_{V \in \what G} d_V \, \mc Z^{\Vect_G}(\Std;\theta)([\mathbb 1],V)$. In contrast, the partition function of the theory in the sector labelled by the identity in $\ms Z(\Vect_G)$ is provided by
$\mc Z^{\Vect_G}(\Std;\theta)([\mathbb 1],\ub 0) = \frac{1}{|G|}\sum_{h \in G }\mc Z^{\Vect_G}(\Std;\theta)([\mathbb 1,h])$. For instance, the network of lines producing the partition function $\mc Z^{\Vect_G}(\Std;\theta)([\mathbb 1],V)$ explicitly reads
\begin{equation}
    \frac{1}{|G|} \sum_{g \in G} \tr_V\big(\eta(g) \big) \; 
    \networkLines{}{g} \, ,
\end{equation}
where $\eta : G \to \End_\mathbb C(V)$. After gauging the $\Vect_G$ symmetry, the dual symmetry is found to be $(\Vect_G)^\vee_\Vect \simeq \Rep(G)$. The same pair of labels $([\mathbb 1],V)$ parameterises a simple object in $\mc Z(\Rep(G))$, which now encodes a sector in the gauged theory with respect to $\Rep(G)$. The network of lines producing the partition function $\mc Z^{(\Vect_G)^\vee_\Vect}(\Std;\theta)([\mathbb 1],V)$ is given by
\begin{equation}
    \frac{1}{|G|} \sum_{W \in \what G} \;
    d_W \networkLines{V}{W} \, ,
\end{equation}
where, in this case, the half-braiding isomorphism $R_{V,W} : V \otimes W \xrightarrow{\sim} W \otimes V$ is simply provided by the operator that permutes the order of vector spaces in the tensor product. A treatment of the general case can be found in ref.~\cite{Lootens:2022avn}. We can readily confirm that 
\begin{equation}
    |\Sigma_\triangle,\Vect_G,\Vect_G,([\mathbb 1],V) \ra = | \Sigma_\triangle, \Vect_G,\Vect,([\mathbb 1],V) \ra \, ,
\end{equation}
from which follows that $\mc Z^{\Vect_G}(\Std;\theta)([\mathbb 1],V) = \mc Z^{(\Vect_G)_\Vect^\vee}(\Std;\theta)([\mathbb 1],V)$. 
In particular, considering instead the object $\bigoplus_{V \in \what G}d_V \cdot ([\mathbb 1],V)$ relates the partition function of the $G$-Ising model without line insertions to that of the gauged model with the sum of all possible line insertions. 
More generally, one has $\mc Z^{\Vect_G}(\Std;\theta)([ c_1],V_{c_1}) = \mc Z^{(\Vect_G)_\Vect^\vee}(\Std;\theta)([ c_1],V_{c_1})$. 
Subsequently performing the Fourier transform allows us to equate the partition function of the $G$-Ising model in a given sector and that of its Kramers--Wannier dual in a dual sector:
\begin{equation}
\begin{split}
    &\la \Sigma_\triangle,\Vect_G,\Vect_G, ([c_1],V_{c_1}) \, | \, \Sigma_\triangle, \Vect_G ; \theta, \vartheta^{\rm triv.} \ra
    \\
    & \q \propto \, 
    \la \Std, \Rep(G), \Rep(G),([c_1],V_{c_1}) \, | \, \Std, \Rep(G), \theta^\vee, \vartheta^{\rm triv.} \ra \, .
\end{split}
\end{equation}
Similar relations hold whenever we only gauge a subsymmetry $A$ instead, in which case simple objects in $\ms Z((\Vect_G)^\vee_{\mc M(A)})$ can still be parametrised by pairs $([c_1],V_{c_1})$. Finally, note that the same reasoning applies to open versions of the bulk topological operators, whereby the same object in $\ms Z(\Vect_G) \cong \ms Z((\Vect_G)^\vee_{\mc M(A)})$ can give rise to an order or a  disorder operator with respect to the symmetry $(\Vect_G)^\vee_{\mc M(A)}$ depending on the choice of $A$.

\newpage
\bigskip
\titleformat{name=\section}[display]
{\normalfont}
{\footnotesize\centering {APPENDIX \thesection}}
{0pt}
{\large\bfseries\centering}
\appendix

{}

\newpage

\renewcommand*\refname{\hfill References \hfill}
\bibliographystyle{alpha}
\bibliography{ref}

\end{document}